\documentclass[preprint,aps,amsmath,nofootinbib]{revtex4-1}
\usepackage{graphicx,array,dcolumn}
\usepackage{calc,tabularx, epsfig,mathrsfs}
\usepackage{hyperref}

\usepackage{amsmath,verbatim,enumerate}
\usepackage{amssymb}
\usepackage{wasysym}
\usepackage{xcolor}
\usepackage{subfigure}
\usepackage{graphicx}
\usepackage{tcolorbox}
\usepackage{multirow}
\usepackage{xspace}
\usepackage{slashed}
\usepackage{mathtools}
\usepackage{float}
\usepackage{cancel}
\usepackage{pgfplots}
\pgfplotsset{compat=1.18}
\allowdisplaybreaks[1]
\newlength{\figurewidth}
\newcommand{\beq}{\begin{equation}}
\newcommand{\eeq}{\end{equation}}
\newcommand{\bea}{\begin{eqnarray}}
\newcommand{\eea}{\end{eqnarray}}
\newcommand{\ba}{\begin{array}}
\newcommand{\ea}{\end{array}}

\newcommand{\Bl}{\biggl}
\newcommand{\Br}{\biggr}
\newcommand{\bl}{\bigl}
\newcommand{\br}{\bigr}
\newcommand{\bt}{\beta}
\newcommand{\g}{\gamma}
\newcommand{\ep}{\epsilon}

\newcommand{\D}{\Delta}

\newcommand{\om}{\omega}

\makeatother
\begin{document}
%
\title{
Complex Saddles of Charged-AdS Gravitational partition function
}
\setlength{\figurewidth}{\columnwidth}
%
\author{Manishankar Ailiga}
\email{manishankara@iisc.ac.in}

\author{Shubhashis Mallik} 
\email{shubhashism@iisc.ac.in}

\author{Gaurav Narain}
\email{gnarain@iisc.ac.in}

\affiliation{
Center for High Energy Physics, Indian Institute of Science,
C V Raman Road, Bangalore 560012, India.
\vspace{10mm}
}

\vspace{20mm}

\begin{abstract}
In this paper, we consider the Euclidean partition function of uncharged and charged $AdS_{d+1}$ black hole geometries in  canonical and grand canonical ensemble for $d\geq3$. It is seen that the partition function can be reduced to a one-dimensional integral, which can be investigated using methods of Picard-Lefschetz. The saddles of the system correspond to either naked-singular geometry, thermal-AdS, small-, intermediate- or large-sized black hole for different ranges of parameter space. These are solutions of Einstein's equation, which are dominant saddles of the partition function in various regimes of parameter space. A naive analysis of the partition function involving these saddles would lead to conflicts with the standard understanding of black hole thermodynamics and also with AdS/CFT. However, when the partition function is analysed using Picard-Lefschetz, it is seen that naked-singular geometries turn out to be irrelevant and therefore do not contribute. This also aligns well with the Cosmic Censorship hypothesis. Depending on the ensemble, saddles corresponding to negative specific heat are either small- or intermediate-sized black holes. Although they are {\it relevant} in the partition function but are sub-dominant. They drop out under homology averaging. Only saddles corresponding to non-negative specific heat contribute to the Euclidean partition function. Finally, we analyze the allowability of these complex geometries using the KSW criterion.

\vspace{55mm}

\end{abstract}
\vspace{5mm}


\maketitle

\tableofcontents


\section{Introduction}

\label{intro}

Thermal partition function given by $Z(\bt)$ which is a weighted sum of energy eigenstates, is wonderfully described by Euclidean path-integral for non-gravitational systems. If the system has a Hamiltonian $H$ that is bounded from below then one has
\beq
\label{eq:zbeta_H-gen}
Z(\bt) = {\rm Tr} \,e^{-\bt H} \, ,
\eeq
where the trace ${\rm Tr}$ is defined as the summation over all the energy eigenstates. In the case of gravity, in a sense the same is true as pointed out long ago by Gibbons and Hawking \cite{Gibbons:1976ue}, where the gravitational path-integral $Z(\bt)$ is defined as path-integral over some class of metrics with Euclidean signature boundary $S^1 \times \mathbb{Y}$ where $S^1$ has proper length $\bt$. It is well-known that path-integrals get dominant contribution from saddle configurations, which are the configurations satisfying the onshell equation of the action describing the system. This thing works well in non-gravitational cases where the saddles correspond to configuration with least energy. In gravitational systems, things are not so straightforward. 

Euclidean gravitational partition function suffers from the conformal factor problem, where the path-integral over the conformal factor is unbounded from below \cite{Gibbons:1978ac}. A simple treatment proposed earlier was to rotate the contour of integration of the conformal-factor, which although is adhoc but does the task. This procedure has been utilized by several authors giving physically satisfying results \cite{Allen:1984bp, Prestidge:1999uq, Kol:2006ga, Headrick:2006ti, Monteiro:2008wr, Monteiro:2009tc, Marolf:2018ldl} where the choice of contour is ad hoc lacking justification from first principles. 

A better proposal is to deal with the Lorentzian path-integral where one doesn't encounter the conformal factor problem directly in the sense that the path-integral is defined over Lorentzian geometries with the contour of integration taken to be defined by real Lorentzian-signature metrics. The notion is then carefully scrutinize the possible deformation of integration contour in the complex plane to come up with an equivalent ``Euclidean'' path-integral which right from beginning comes along with a specific integration contour where meaningful computation can be performed. This approach has been considered by various authors in both continuum and discrete formulation of path-integral of gravity  \cite{Schleich:1987fm, Mazur:1989by, Hartle:2020glw, Giddings:1989ny, Marolf:1996gb, Dasgupta:2001ue, Ambjorn:2002gr, Feldbrugge:2017kzv, Feldbrugge:2017fcc, Ailiga:2023wzl, Ailiga:2024wdx, Ailiga:2025fny}. 

The benefit of using a Lorentzian formulation of path-integral for gravity is that it is not ruled out for obvious reason like unboundedness happening in Euclidean gravity. This is because the Lorentzian gravitational action $I_{\rm grav}$ for smooth real Lorentzian metrics is purely real, hence implying that integrand of the path-integral $\exp(i I_{\rm grav})$ is a pure phase leading to an oscillatory integral! Is there a way to define the Euclidean path-integral $Z(\bt, \cdots)$ (where the $(\cdots)$ represent charges, angular momentum, etc.) via Lorentzian path-integral by some transformation? This was investigated in \cite{Marolf:2022ybi}, where the author proposes to consider Lorentzian path-integral as path-integral over all Lorentzian metrics including co-dimension-2 singularities, and uses a smearing function $f_\bt(T)$ to convert the Lorentzian path-integral to a ``Euclidean'' one. The smearing function is defined by
\beq
\label{eq:sm_func}
\int_{-\infty}^\infty {\rm d} T 
f_\bt(T) \,\, e^{-i T \bar{\om}}
= e^{-\bt \bar{\om}} \, ,
\eeq
for all $\bar{\om}$ above the ground state of Hamiltonian. This immediately gives that the Euclidean partition function can be written as
\beq
\label{eq:Tr_eucH_tr}
Z(\bt) = {\rm Tr} \, e^{-\bt H} 
= \int_{-\infty}^\infty {\rm d}T
f_\bt(T) \,\, {\rm Tr} \, e^{-i T H} \, ,
\eeq
where $Z_L(T)= {\rm Tr} \, e^{-i T H}$ can be thought of as one parameter family of path-integral over Lorentzian metrics \cite{Marolf:2022ybi}. If the Euclidean path-integral involves Black hole as saddles, then the Lorentzian path-integral must allow inclusion of co-dimension-2 singularities in the path-integral. If these co-dimension-2 surfaces are denoted by $\g$, then 
\beq
\label{eq:ZLT_s}
Z_L(T) = \int {\rm d} S \,\, Z_L(T; S) \, ,
\eeq
where $S = {\rm Area}(\g)/4G_N$ and $Z_L(T;S)$ is the path-integral over Lorentzian metrics of fixed $S$ (or fixed area of the co-dimension-2 surface $\g$). Saddle configurations of $Z_L(T; S)$ then become constrained saddles of the $Z_L(T)$. These constrained saddles can become relevant depending on the integration contour and can contribute non-trivially to the path integral once the integration contour is deformed using the Picard-Lefschetz methodology \cite{Marolf:2022ybi, Chen:2025leq} (see \cite{Witten:2010cx, Feldbrugge:2017kzv, Narain:2021bff, Shoji:2025riv,Tanizaki:2014xba,Witten:2010zr} for discussion on Picard-Lefschetz technique). 

In the saddle point approximation, $Z_L(T; S)$ gets dominant contribution from the co-dimension-2 surface which is the saddle of the action $I_{\rm grav}$. These correspond to blackhole geometries of fixed area. This implies
\beq
\label{eq:ZLTS_sad}
Z_L(T;S) \sim e^{i I_{\rm grav}}
= e^{S - i T E(S)} \, ,
\eeq
where $E(S)$ is the energy for the BH saddle of fixed $S$ \cite{Marolf:2022ybi, Chen:2025leq}. Plugging this back into eq. (\ref{eq:ZLT_s}), we get an expression for the contribution of the black hole saddle in the Lorentzian path-integral to be
\beq
\label{eq:ZLT_exp_sad}
Z_L(T) = \int_0^\infty {\rm d} S \,\,
e^{S - i T E(S)} \, .
\eeq
Using the smearing function this can be integral transformed following eq. (\ref{eq:Tr_eucH_tr}) to obtain the black hole contribution to the Euclidean path-integral to be
\beq
\label{eq:EPI_grav}
Z(\bt) = \int_0^\infty {\rm d}S \, {\rm d}T \,\, 
f_\bt(T) \, \, e^{S - i T E(S)} 
= \int_0^\infty {\rm d}S \, e^{S - \bt E(S)}\, ,
\eeq
where it is understood that the $T$ integration has to be performed first before the integration over $S$. The same expression can be obtained from the trace, if one notices that $e^S$ can be identified as the density of states corresponding to a system with energy $E(S)$. In the case when there is charge the above becomes two dimensional integral 
\begin{equation}
    \label{eq:grand_partition_1}
    Z(\bt,\Phi) 
= \int_0^\infty {\rm d}S\int d Q \, e^{S - \bt E_\Phi(S,Q)}\, ,\quad  
{\rm with} 
\quad
E_\Phi(S,Q)=E(S)-\Phi Q \, .
\end{equation}
The essence of this procedure is that we do not deal with the conformal factor problem of the Euclidean path-integral, instead we use the Lorentzian path-integral to define a Euclidean one via an integral transform, which, for the case of black hole geometries, reduces to the known partition function. 

Once we have comfortably defined the Euclidean partition function $Z(\bt)$ for the case of black hole geometries, we proceed to study it in detail for the case of AdS-Schwarzschild and charged black holes in arbitrary dimensions. Our starting point in this study is eq. (\ref{eq:EPI_grav})  and eq. (\ref{eq:grand_partition_1}). These are either one or two-dimensional integral which can be evaluated by analysing them in the complex plane. As this targets geometries which correspond to black hole solutions, it is worth exploring what are the possible saddles of the system and what do they correspond to. Furthermore, it is crucial to investigate which saddles give dominant contribution to the partition function for various values of parameter $\bt,Q,\Phi$ and whether they are compatible with the known results of AdS/CFT for various dimensions. 

It may happen that certain saddles, which are dominant in the partition function for various ranges of parameters, are incompatible with the known wisdom of AdS/CFT. In such cases, it is worth asking whether such saddles become ``relevant'' in the partition function when the integrations in eqs. (\ref{eq:EPI_grav}) and (\ref{eq:grand_partition_1}) are performed in the complex plane using Picard-Lefschetz methods. An interesting situation seen in the recent work \cite{Mahajan:2025bzo, Singhi:2025rfy} is that there are saddles that are at odds with the usual expectations from AdS/CFT, do not contribute to the partition function. They are ``not" relevant according to the PL method. This paper investigates such situations in more detail for uncharged and charged black holes in arbitrary $d+1$-dimensions. 

The outline of the paper is as follows: In Sec. \ref{sec:Sch_ads_d}, we revisit the puzzle and findings reported in \cite{Mahajan:2025bzo} for AdS Schwarzschild black hole by generalizing to arbitrary dimensions. Here, we discuss the implementation of Picard Lefschetz and homology averaging of thimbles, along with thermodynamic stability from the viewpoint of PL analysis. In Sec. \ref{sec:charge_ads}, we extend this study to the charged AdS Schwarzschild black holes both in grand canonical ensemble (fixed potential) and the canonical ensemble (fixed charge). Sec. \ref{d_KSW} is devoted to the analysis of weaker verison of the KSW criterion and its compatibility with the puzzle raised in the earlier sections. Finally, summary of our findings and conclusions are presented in section \ref{conc}.

\section{ AdS Schwarzschild black hole in $d+1$ dimension}
\label{sec:Sch_ads_d}

We first consider the case of $AdS_{d+1}$ Euclidean Schwarzschild black hole with the boundary $S^1_\beta\times S^{d-1}$ ($d\geq 3$) at a given inverse temperature $\beta$. Such a black hole has a line element 
\begin{equation}
\label{eq:Ads_sch_bh}
\begin{split}
    &ds^2=f(r) d\tau^2+\frac{dr^2}{f(r)}+r^2d\Omega_{d-1},\hspace{5mm}\tau\sim\tau+\beta, \,\,\,\\
   & \,\,f(r)=1+\frac{r^2}{l^2}-\frac{\mu}{r^{d-2}},\,\,\mu> 0.
\end{split}    
\end{equation}
This is solution to equation of motion following from Einstein-Hilbert action with appropriate surface terms and counterterms to cancel the large volume divergence \cite{Skenderis:2002wp}. For $\mu=0$, it describes a thermal AdS solution without any horizon. However, when $\mu>0$, we get a horizon at $r_+$, which is the largest positive real root of $f(r_+)=0$ and $r_+\leq r<\infty$. Here, $r\geq r_+$ ensures the existence of a timelike killing vector everywhere outside the horizon, and regularity at the horizon enforces $\beta$ to be $4\pi/f'(r_+)$. The entropy $S$ and energy $E$ as function of $r_+$ are given by
\begin{equation}
    \label{eq:entropy_and_mass}
    S(r_+)=\frac{\omega r_+^{d-1}}{4G},\hspace{4mm} E(r_+)=\frac{\omega(d-1)}{16\pi G}\left(r_+^{d-2}+\frac{r_+^d}{l^2}\right) \, ,
\end{equation}
where $\omega$ is the area of $(d-1)$ dimensional unit sphere, and given by $ \omega=2\pi^{d/2}/\Gamma(d/2)$. In eq. (\ref{eq:entropy_and_mass}), we omitted the zero-point Casimir energy $E_0(d)$ of empty AdS and set $E(r_+=0)=0$, as it plays no role in what follows. The exponent in the integrand in eq. (\ref{eq:EPI_grav}) takes the form $\mathcal{I}(r_+)=S(r_+)-\beta E(r_+)$. The measure ${\rm d}S$ becomes a measure over $r_+$ with range of integration $0\leq r_+ < \infty$ (the Jacobian generated in the transformation can be ignored as we are working in the saddle-point approximation ). The Euclidean partition function mentioned in eq. (\ref{eq:EPI_grav}) reduces to
\begin{equation}
\label{eq:can_partion_ads_bh}
Z(\beta) \approx \int_0^\infty {\rm d} r_+
\exp\left[\frac{\omega r_+^{d-1}}{4G}-\beta\frac{\omega(d-1)}{16\pi G}\left(r_+^{d-2}+\frac{r_+^d}{l^2}\right)\right] 
= \int_0^\infty {\rm d} r_+ e^{\mathcal{I}(r_+)} \, ,
\end{equation}
where $\beta$ is any arbitrary positive parameter.
A similar setup is also considered in the literature \cite{Mahajan:2025bzo}, where the authors specifically study in $d=4$ and briefly comment on arbitrary dimensions. Studies along these lines were also taken in \cite{DiTucci:2020weq} in $d=3$. The saddle configurations can be determined by considering the exponent in eq. (\ref{eq:can_partion_ads_bh}) and extremizing the $\mathcal{I}(r_+)$. This gives the saddle-point equation
\begin{equation}
    \label{eq:saddle_can_partion}
    r_+^{d-3}\left[r_+^2-\frac{4\pi  l^2}{d\beta}r_++\frac{l^2(d-2)}{d}\right]=0.
\end{equation}
This is an algebraic equations whose roots can be found easily. These are given by
\begin{equation}
\label{eq:saddlesadsach}
\begin{split}
   & r_+=0,\cdots ,0\,\,\, (d-3) \text{times}\\
    &r_+^{\pm}=\frac{2l^2\pi}{d\beta}\pm\frac{l\sqrt{4l^2\pi^2-(d-2)d\beta^2}}{d\beta} \, .
    \end{split}
\end{equation}
The root appearing at $r_+=0$ correspond to thermal $AdS$ solution which exists for all $\beta$. The other solutions $r_+^\pm$ correspond to real black hole saddle geometries for $\beta<\beta_{\rm max}$, with 
\begin{equation}
    \label{eq:betamax}
    \beta_{\rm max}=\frac{2l\pi}{\sqrt{d(d-2)}} \, .
\end{equation}
For $\bt>\bt_{\rm max}$, these $r_+^\pm$ saddles becomes complex. For $\bt>\bt_{\rm max}$, $r_+=0$ (thermal AdS) is the only real saddle. When $\beta=\beta_{\rm max}$ both saddles merge. For later convenience, we will write the above saddle configurations in terms of $\bt_{\rm max}$. These are given by
\begin{equation}
\label{eq:roots_canonical}
r_+^{\pm}=\frac{2l^2\pi}{d\beta}\left(1\pm\sqrt{1-\frac{\beta^2}{\beta_{\rm max}^2}}\right) 
= \frac{2l^2\pi}{d\beta} 
\bl(
1 \pm \tanh \theta
\br)\, ,
\end{equation}
where we define 
\beq
\label{eq:tanhTH}
\tanh \theta = \sqrt{1-\frac{\beta^2}{\beta_{\rm max}^2}}\, .
\eeq
It is worth highlighting that the regularity condition is satisfied for all $\beta$ and is not restricted only to $\beta<\beta_{\rm max}$ where we have real black hole solutions.

If one computes the Kretschmann scalar (see appendix \ref{sec:curvature_scalar}), then for such configurations we have 
\begin{equation}
\label{eq:K_explicit_1}
\begin{split}
K(r)=\frac{d(d-1)^2 (d-2) \mu^2}{r^{2d}}+\frac{2 d (d+1)}{l^4}\, ,
\quad
{\rm where}
\quad
\mu^2=r_+^{2(d-2)}\biggl(1+\frac{r_+^2}{l^2}\biggr)^2 \, .
\end{split}
\end{equation}
From eq (\ref{eq:K_explicit_1}), it is evident that at $r=0$, $K(r=0)$ blows up in $d\geq 3$, signifying the presence of curvature singularity irrespective of whether $r_+$ is real (for $\beta<\beta_{\rm max}$) or complex (which happens for $\bt>\bt_{\rm max}$).
The real saddles describe the Euclidean cigar geometries, while in the Lorentzian continuation, they correspond to a black-hole with a singularity hidden behind the horizon. On the other hand, for complex saddles things are different. Their Lorentzian continuation correspond to naked singularities. Such naked singularities are also solutions of Einstein's equations. 

In the following subsections, we will study the partition function for various regimes of $\bt$, analysing the saddle configurations, their dominance and relevance, and compute the partition function using the methods of Picard-Lefschetz. To do so, we would need to study the argument of the exponent in eq. (\ref{eq:can_partion_ads_bh}) at the various saddles. $\mathcal{I}(r_+)$ at various saddles is given by
\begin{equation}
\label{eq:exponent_at_saddle}
\begin{split}
&\mathcal{I}(r_+=0)=0 \, , \\
&\mathcal{I}(r_+=r_+^\pm)
=\frac{\omega l^{d-1}}{4G}
\frac{(d-2)^{(d-3)/2}}{d^{(d+1)/2}}
\exp\{{\pm (d-1)\theta }\}
\bl[-1 \pm (d-1) \tanh \theta \br] \, .
\end{split}
\end{equation}
From here we see that for thermal-AdS saddle $\mathcal{I}(r_+=0)=0$. For $\bt\leq \bt_{\rm max}$, $\mathcal{I}(r_+)$ at other two saddles $r_+^\pm$ is purely real as $\theta$ is purely real. For $\bt> \bt_{\rm max}$, $\theta \to i\tilde{\theta}$, becomes purely imaginary quantity with $\tilde{\theta}$ being real, leading to complex $\mathcal{I}(r_+)$ at the two complex saddles which correspond to naked-singularity.

\subsection{$Z(\bt)$ for $\beta>\beta_{\rm max}$}
\label{sec:cbh}

In this regime, we have one real saddle and two complex saddle configurations. The real saddle corresponds to thermal AdS geometry, while the complex-conjugate saddle correspond to naked singular geometries. As noticed in \cite{Mahajan:2025bzo}, these complex saddles give dominant contribution in the partition function for certain range of values of $\beta$, as can be seen by computing $\mathcal{I}(r_+)$. Naively, it will be expected that being dominant saddles, they are the ones whose contribution will be most important in the semi-classical limit. However, the situation changes if such saddles, despite being dominant becomes {\it irrelevant} in the computation of the partition function using Picard-Lefschetz methods. This situation was analysed for the case of $d=4$ in \cite{Mahajan:2025bzo}. In our paper we examine the situation in arbitrary dimensions. 

\begin{figure}
    \centering
\includegraphics[width=0.74\linewidth]{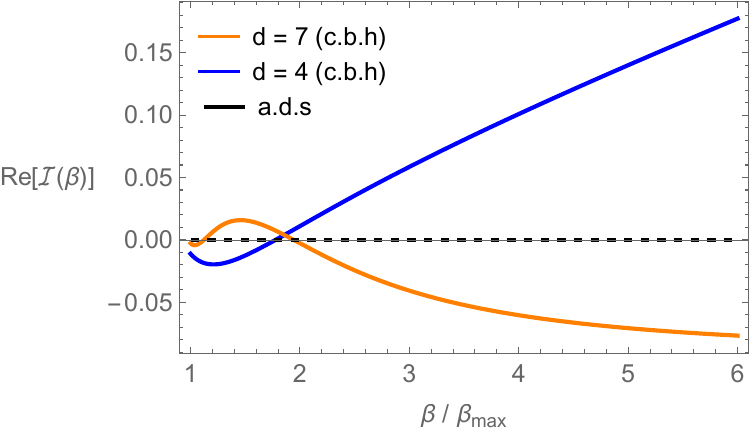}
\caption{Real part of $\mathcal{I}$ (see eq. (\ref{eq:re_Ir+})) as a function of $\beta$ ($>\bt_{\rm max}$) at various saddles for $d=4$ and $d=7$. ${\rm Re}[\mathcal{I}(r_+^\pm)]>0$ for complex saddles (naked-singular geometries) imply that for these values of $\bt$ naked singularities are dominant over the Thermal-AdS saddle.}
\label{fig:dominance_check}
\end{figure}

In order to compute the partition function for $\bt>\bt_{\rm max}$, we start by analysing ${\rm Re}(\mathcal{I}(r_+))$ at various saddles. It will indicate the saddles which will become dominant as $\bt$ varies. It is seen from eq. (\ref{eq:exponent_at_saddle}) that for arbitrary $d$, $\mathcal{I}(r_+)$ for thermal AdS is always zero. For $\bt>\bt_{\rm max}$, the other two saddles $r_+^\pm$ are complex conjugate and correspond to naked singularities. For them $\mathcal{I}(r_+)$ is complex as seen from eq. (\ref{eq:exponent_at_saddle}) . 
To analyse the dominance, one needs to extract the real part of ${\rm Re}[\mathcal{I}(r_+)]$ for each of these complex saddles. If the ${\rm Re}[\mathcal{I}(r_+)]>0$ for certain values of $\beta$, then these naked singularities become dominant over the thermal-AdS saddle, which will be in conflict with the results of AdS/CFT. For $\bt > \bt_{\rm max}$, the ${\rm Re}[\mathcal{I}(r_+^\pm)]$ is given by
\bea
\label{eq:re_Ir+}
{\rm Re}[\mathcal{I}(r_+^\pm)]
=&& -\frac{\omega l^{d-1}}{4G \bt_{\rm max}}
\frac{(d-2)^{(d-3)/2}}{d^{(d+1)/2}}
\bl[
\bt_{\rm max} \cos((d-1)\tilde{\theta})
\notag \\
&&
+ (d-1)\sqrt{\beta^2-\bt_{\rm max}^2} 
\sin((d-1)\tilde{\theta})
\br] \, .
\eea
It is easy to see from here that for certain values of $\bt$, say for example large-$\bt$ ($\bt\to\infty$), $\cos((d-1)\theta)=0$ or $\pm 1$, while $\sin((d-1)\theta)=\pm 1$ or $0$ (depending on the dimension $d$), respectively. In the case when it is $-1$ (for example, $d=4$), then naked singularities are dominant in the partition function! More generically, for large-$\bt$, when $d=4n, n\in (1,2,3,..)$, ${\rm Re}[\mathcal{I}(r_+^\pm)] \sim +(\beta/\beta_{\rm max})>0$ thereby claiming dominance over thermal-AdS. This was already noticed in \cite{Mahajan:2025bzo} for $AdS_5(n=1)$.  When $d=4n-1$, for large-$\bt$, ${\rm Re}[\mathcal{I}(r_+^\pm)]<0$ (independent of $\bt$) and thermal-AdS remains a dominant saddle. However, for $d=4n+1$, ${\rm Re}[\mathcal{I}(r_+^\pm)]>0$ and is independent of $\beta$, thereby again dominating over the thermal-AdS saddle. For $d=4n+2$, ${\rm Re}[\mathcal{I}(r_+^\pm)] \sim -(\beta/\beta_{\rm max})$ for large-$\bt$ thereby making thermal-AdS dominant saddle. We summarize this in the table \ref{tab:dominant_saddle}.
\begin{table}[hbt!]
    \centering
    \begin{tabular}{|c|c|}
    \hline
       $ d$ \,\,&\,\, dominant saddle\,\, \\
        \hline
        \,\,$4n-1$\,\, & \,\,Thermal-AdS\,\,\\
         \hline
       \,\, $4n$\,\, &\,\, Naked-singularity \,\,\\
         \hline
       \,\,  $4n+1$ \,\,&\,\, Naked-singularity \,\,\\
         \hline
        \,\, $4n+2$\,\,& \,\, Thermal-AdS\,\,\\
         \hline
    \end{tabular}
    \caption{ Dominant saddles of the partition function in the large $\beta (\gg \beta_{\rm max})$ limit for various dimensions d with $n \in (1,2,3,\cdots)$.}
    \label{tab:dominant_saddle}
\end{table}
On doing a numerical plot of ${\rm Re}[\mathcal{I}(r_+^\pm)]$ as a function of arbitrary $\bt>\bt_{\rm max}$, we notice from fig. (\ref{fig:dominance_check}) the values of $\bt$ for which naked singularities give dominant contribution in the partition function.

From this discussion, it is clear that for certain values of $\bt$ for the case of $\bt>\bt_{\rm max}$, naked singular geometries, which are complex saddles, are dominant. Naively, one would expect that they would control the nature of the partition function in the semiclassical limit, as they are most dominant, though they correspond to complex metrics ($r_+$ being complex implies $\mu$ is complex). Occurrences of complex metrics are not new in the literature. They have shown up in studies of quantum cosmology, black holes \cite{Feldbrugge:2017kzv, Maldacena:2024uhs, Ivo:2024ill, Ailiga:2024mmt, Ailiga:2023wzl, DiTucci:2020weq, Lehners:2021mah, Jonas:2022uqb, Nishimura:2024one, Hertog:2015nia, Cabo-Bizet:2020ewf} and real-time holography as well \cite{Skenderis:2008dh, Glorioso:2018mmw, Jana:2020vyx}, where they have lead to physically meaningful results. However, in our case if they are to be considered, then our findings will be in tension with the understanding of black hole thermodynamics. Furthermore, if these additional saddles are considered, new phase transitions will occur among these complex saddles. This will also be in conflict with AdS-CFT, where it is well understood that thermal AdS reproduces results consistent with the dual CFT in the low temperature regime \cite{Maldacena:1997re,Witten:1998zw, Witten:1998qj, Aharony:2003sx}.

This is the puzzle first raised in the note \cite{Mahajan:2025bzo}, where naive computation of the partition function based on saddle dominance lead to tension with AdS/CFT. However, if the partition function is computed carefully, making use of Picard-Lefschetz methods and tools from Morse theory, then the puzzle is resolved on its own, as noticed in \cite{Mahajan:2025bzo}. In this paper, we study this situation for generic dimensions and analytically show that complex saddles corresponding to naked-singular geometries become {\it irrelevant} in the path-integral and hence don't contribute.

In the next subsequent section, we analyse the partition function stated in eq. (\ref{eq:can_partion_ads_bh}) using Picard-Lefschetz methods for the case of $\bt>\bt_{\rm max}$.

\subsubsection{Complex saddles and their relevance for $\bt>\bt_{\rm max}$}
\label{sssec:csir}

After defining the starting integration contour $\mathbb{D}\in [0,\infty)$ and determining all the saddles of eq. (\ref{eq:can_partion_ads_bh}), we proceed to find the {\it relevant} saddles which contribute to the integral utilizing the Picard-Lefschetz method. According to this method, we analytically continue the real $r_+$ to a complex variable $r_+=r_{+}^x+i\,r_+^y$. When $r_+$ is complex, the exponent $\mathcal{I}(r_+)$ becomes a holomorphic function of $r_+$ with real ($\mathfrak{h}$) and imaginary ($\mathcal{H}$) part: $\mathcal{I}(r_+)=\mathfrak{h}(r_+)+i\mathcal{H}(r_+)$. The direction of steepest descent/ascent ($\mathcal{J}_\sigma/\mathcal{K}_\sigma$) in the complex $r_+$ plane are given by the equation
\begin{equation}
\label{eq:floweq}
\frac{\partial r_+^x}{\partial \lambda}=\mp\frac{\partial \mathfrak{h}(r_+^x,r_+^y)}{\partial r_+^x}
\hspace{3mm}
{\rm and}
\hspace{3mm}
\frac{\partial r_+^y}{\partial \lambda}=\mp\frac{\partial\mathfrak{h}(r_+^x,r_+^y)}{\partial r_+^y},
\end{equation}
where $\lambda$ is the real parameter along the curve. In the above equation, $-$ sign corresponds to the steepest descent and $+$ sign corresponds to the steepest ascent. Along these curves $\mathcal{H}$ remains constant. i.e., $d\mathcal{H}/d\lambda=0$. At the saddles, one satisfies $\partial r_+^x/\partial \lambda=\partial r_+^y/\partial \lambda=0$. Note that as $\mathcal{I}(r_+)$ is purely real when $r_+$ is real, the Morse function satisfies $\mathfrak{h}(r_+^x,-r_+^y)=\mathfrak{h}(r_+^x,r_+^y)$ and hence the flowlines are invariant under $(r_+^x,r_+^y)\rightarrow(r_+^x,-r_+^y)$. In the complex $r_+$-plane, the flowlines are symmetric around the real axis ($r_+^y=0$), see appendix \ref{sec:symmetry} for more details. The real integration contour ($\mathbb{D}$) can be deformed to a new contour ($\mathcal{C}$), which is the summation of steepest contours ($\mathcal{J}_\sigma$)
\begin{equation}
    \label{eq:PL}
\mathcal{C}=\sum_{\sigma\in\text{saddle}} n_\sigma \mathcal{J}_\sigma,\quad n_\sigma= \text{Int}(\mathbb{D},\mathcal{K}_\sigma),
\end{equation}
where $\text{Int}(\, ,\,)$ determines the intersection between two curves. The intersection number, $n_\sigma$ can take values $0,\pm 1$. The sign depends upon the orientation of the thimbles. $n_\sigma=0$ will imply the saddle is {\it irrelevant} and doesn't contribute to the integral. Any non-zero value of $n_\sigma$ will imply the saddle is {\it relevant} and hence contribute to the integral.

\begin{figure}
    \centering
    \includegraphics[width=11cm, height=9cm]{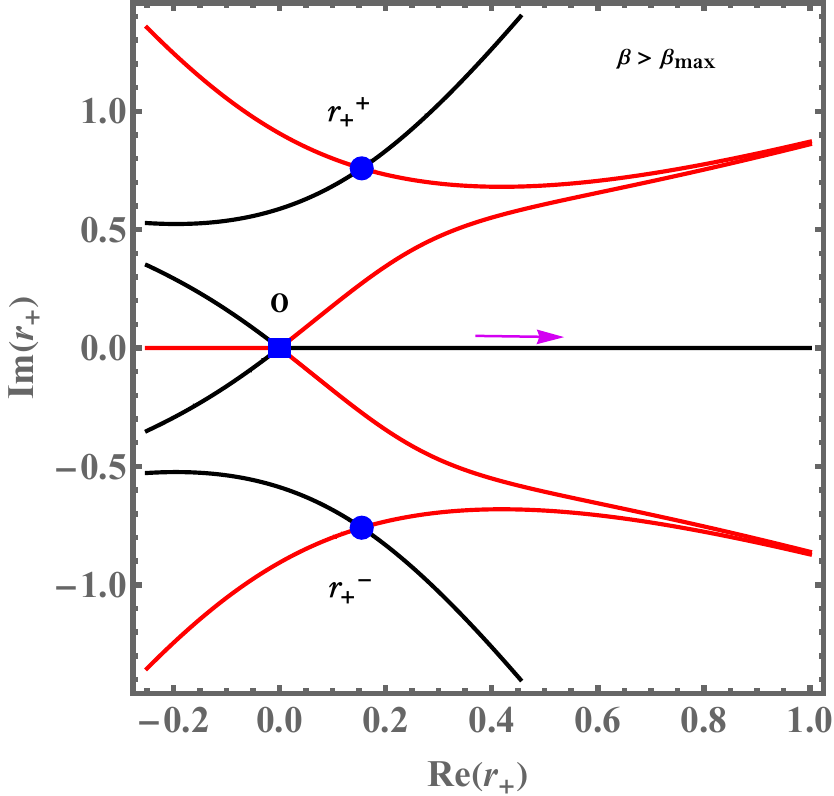}
    \caption{ Steepest descent/ascent lines are plotted in black/red for $AdS_6$, for $\beta=5\beta_{\rm max}$, $l=1$. Thermal-AdS is at $r_+=0$ and $r_+^\pm$ are naked singularities. The purple arrow is the direction of integration cycle. As the steepest ascent curves from these complex saddles don't intersect the real line, hence they don't contribute to the partition function.}
    \label{fig:bt_gr_bmax_AdS6}
\end{figure}

Hence, in order to determine the $n_\sigma$ for complex saddles ($r_+^\pm$), one needs to see if the steepest ascent flow-lines from these saddles intersect the real integration contour ($\mathbb{D}$). To find it, we note that to start with, $\mathcal{I}(r_+)$ is purely real along the integration contour, and hence, $(\mathcal{I}(r_+))^*=\mathcal{I}(r_+^*)$, where $(.)^*$ is complex conjugation. A necessary condition (not sufficient) for the thimbles ($\mathcal{K}_\sigma$) from these complex saddles to intersect the real axis is that $\mathcal{H}(r_+^\pm)=0$. It is simply because, on the real axis $\mathcal{H}=0$ and $\mathcal{H}$ is conserved along the flow-lines. The expression of $\mathcal{H}(r_+^\pm)$ can be read out from eq. (\ref{eq:exponent_at_saddle}) and it is given by,
\begin{equation}
\label{eq:mathcal_H_at_saddle}
\mathcal{H}(r_+^\pm)=\pm \frac{\omega}{4G}\frac{1}{d}\left(\frac{l^2(d-2)}{d}\right)^{(d-1)/2}\left\{\sin((d-1)u)-\frac{(d-1)}{(d-2)}\frac{\beta}{\beta_{\rm max}}\sin((d-2)u)\right\},
\end{equation}
where, $u=\arctan(\sqrt{\beta^2/\beta_{\rm max}^2-1})$.
To see if $\mathcal{H}(r_+^\pm)=0$ directly from eq. (\ref{eq:mathcal_H_at_saddle}) is not possible analytically for arbitrary $\beta>\beta_{\rm max}$ and $d$. However, in the regime of $\beta\gg\beta_{\rm max}$, we can write 
\begin{equation}
\label{eq:large_beta}
\begin{split}
\biggl.\mathcal{H}(r_+^\pm)\biggr|_{\beta\gg\beta_{\rm max}}=\pm \frac{\omega}{4G}\frac{1}{d}\left(\frac{l^2(d-2)}{d}\right)^{(d-1)/2}\left\{\frac{(d-1) \sin \left(\frac{\pi  d}{2}\right)}{d-2}\frac{\beta}{\beta_{\rm max}}\right.\\
\left.-d \cos \left(\frac{\pi  d}{2}\right)+\mathcal{O}\left(\frac{\beta_{\rm max}}{\beta}\right)\right\}
\end{split},
\end{equation}
which can never vanish for arbitrary dimensions. It is easy to convince oneself that the same will hold for an arbitrary $\beta>\beta_{\rm max}$, by observing that the integral along the real axis is absolutely convergent, and hence the steepest ascent flow-line of any saddle not lying on the real line can never intersect the real line. This implies, $n_\sigma$ for complex saddles is always zero making such complex saddles {\it irrelevant} in the computation of partition function (see fig. \ref{fig:bt_gr_bmax_AdS6}). Hence, the partition function receives contributions only from the thermal-AdS for $\beta>\beta_{\rm max}$ thereby giving
\begin{equation}
\label{eq:partition_function}
Z(\beta>\beta_{\rm max}) \Br \rvert_{\rm thermal-AdS} \sim \exp\bl[\mathcal{I}(r_+) \br] \Br \rvert_{\rm thermal-AdS} = 1\, .
\end{equation}
This result reveals two important facts: Firstly, one recovers the AdS-CFT conjecture and the standard thermodynamic perspective, according to which thermal AdS should only contribute to the partition function. Secondly, since the complex saddles are naked singularities and don't contribute to the partition function for any value of $\bt>\bt_{\rm max}$, despite being dominant saddles for certain regimes of $\bt$, this shows that the results are compatible with the cosmic censorship hypothesis \cite{Penrose:1964wq}, which states that naked singularities are unphysical. Indeed, Picard-Lefschetz cleanly discards these complex saddles even though they are solutions of Einstein's equation and are even dominant saddles for certain regimes of $\bt$.  

%

\subsection{$Z(\bt)$ for $\beta\leq\beta_{\rm max}$}
\label{sec:real_saddle}

In the high temperature regime (when $\beta\leq \beta_{\rm max}$), all the saddles are purely real which can be seen from eq. (\ref{eq:roots_canonical}), Consequently, $\theta$ is also real. This gives
\beq
\label{eq:sadrel}
0=r_+^{\rm thermal-AdS} < r_+^- \leq r_+^+ \, .
\eeq
This implies that the action $\mathcal{I}(r_+)$ at various saddles is also real as can be noticed from eq. (\ref{eq:exponent_at_saddle}). As range of $\bt$ is $0\leq \bt \leq \bt_{\rm max}$, this implies that $0\leq \theta< \infty$ ($\theta\to\infty$ correspond to $\beta\rightarrow 0$). From eq. (\ref{eq:exponent_at_saddle}) it can be immediately inferred that we have $\mathcal{I}(r_+^-) <0$. The action at saddle $r_+^+$ is always greater than the action at saddle $r_+^-$: $\mathcal{I}(r_+^-) <\mathcal{I}(r_+^+)$. $\mathcal{I}(r_+^+)$ changes sign depending on $\bt$ and dimensions. If $\tanh \theta <1/(d-1)$, then $\mathcal{I}(r_+^+)<0$; and for $\tanh \theta >1/(d-1)$ one has $\mathcal{I}(r_+^+)>0$. In the language of $\bt$ this means that for $\bt< 2 l \pi /(d-1)$, we have $\mathcal{I}(r_+^+)<0$, while for $\bt> 2 l \pi /(d-1)$ the action $\mathcal{I}(r_+^+)>0$. This is the Hawking-Page transition \cite{Hawking:1982dh}. Summarizing it, we have 
\bea
\label{eq:Irsad_+-_low}
&& \mathcal{I}(r_+^-) <0 \, \hspace{5mm}
\forall \,\,  0\leq \bt\leq \bt_{\rm max} \, ,
\notag \\
&&
\mathcal{I}(r_+^+) \leq 0 
\hspace{5mm} \forall 
\,\, 0\leq \bt \leq \frac{2 l \pi}{d-1} \, ,
\notag \\
&&
\mathcal{I}(r_+^+) >0 
\hspace{5mm} 
\forall \,\,  \frac{2 l \pi}{d-1} \leq \bt \leq \bt_{\rm max} \, .
\eea
In the next subsequent section, we analyse the partition function stated in eq. (\ref{eq:can_partion_ads_bh}) using Picard-Lefschetz methods for the case of $\bt<\bt_{\rm max}$. 
\subsubsection{Real saddles and their relevance for $\bt<\bt_{\rm max}$}
\label{sec:real_saddle_relevance}
\begin{figure}[htbp]
\centering
\subfigure[\,\,\, $\ep=\pi/10$ \label{fig:15D}]{
    \includegraphics[width=0.47\textwidth]{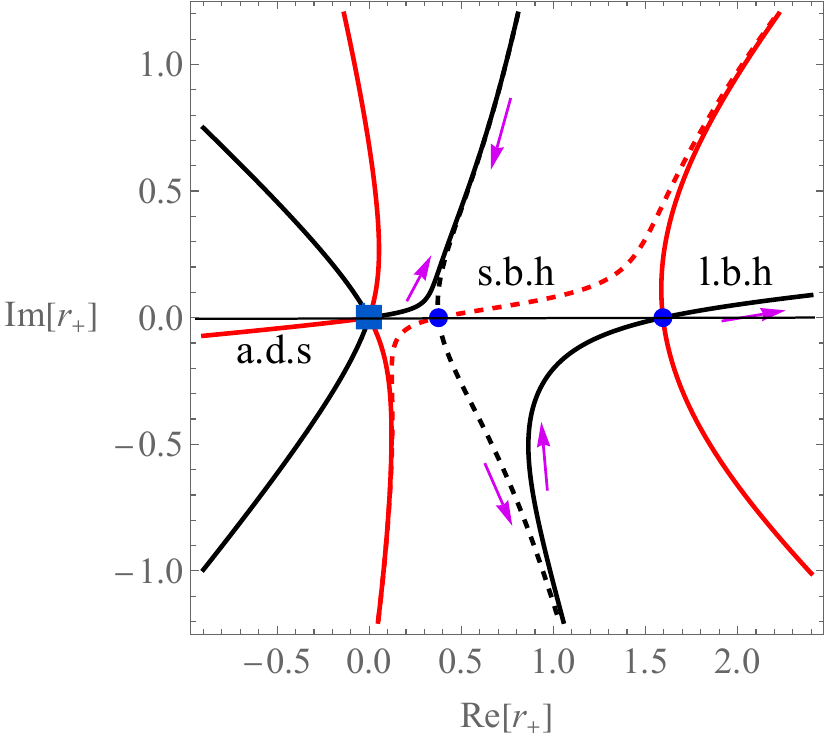}
}
\subfigure[$\,\,\ep=-\pi/10$\label{fig:26D}]{
\includegraphics[width=0.41\textwidth]{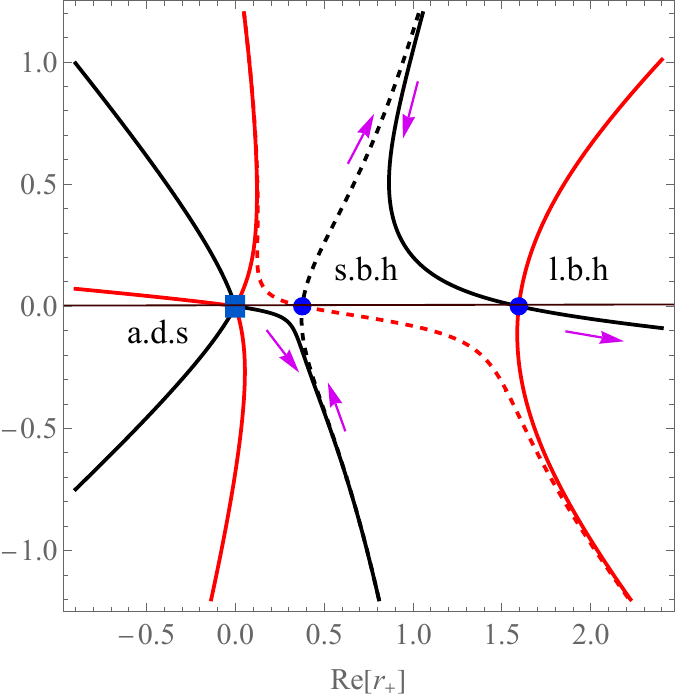}
}
\caption{Steepest descent-ascent contour for $AdS_6$ in the high temperature regime ($\beta<\beta_{\rm max}$) with rotating $G=|G|e^{i\ep},|G|=1,l=1$. Steepest descent ($\mathcal{J}_\sigma$) curves are plotted in black, and steepest ascent ($\mathcal{K}_\sigma$) curves are plotted in red. The arrows indicate the direction of the integration cycle. Steepest ascent from all the saddles intersect the real line (for either sign) and hence contribute to the partition function. Note that due to the degeneracy of the ads saddle in $AdS_6$, there are three descent/ascent lines emanating from the saddle. However, only the branch lying on the positive axis participates in the Stokes jump.}
\label{fig:stokes_jump}
\end{figure}

For the real saddles, the action $\mathcal{I}(r_+)$ being always real implies that their imaginary part trivially vanishes. This means that the steepest ascent emanating from them always intersects our original integration contour $\mathbb{D}$, which lies in the range $0\leq r_+ < \infty$. If we now look at the imaginary part of ${\rm Im}(\mathcal{I}(r_+^+))$, which we define as $\mathcal{H}$, we notice that $\mathcal{H}=0$ for all the saddles, the saddles lie on a Stokes ray. In such situations where the steepest ascent thimble of a saddle intersects the steepest descent thimble of another saddle the Picard-Lefschetz method breaks down and the associated flowline is known as Stokes ray. To find out $n_\sigma$ unambiguously and hence the {\it relevance} of saddles, one needs to resolve such degeneracy. As the Stokes ray lies on the real axis ($r_+^y=0$), which is the axis of symmetry of the flowlines, to resolve the Stokes ray, one needs to break the symmetry; see appendix \ref{sec:symmetry} for more details. There are various ways one can break the symmetry, for example, by complexifying $\beta$, as in \cite{Maldacena:2019cbz}. It can also be achieved by complexifying Newton's constant ($G=|G|e^{i\ep}, |\ep|\ll 1$)\cite{Ailiga:2025fny,Mahajan:2025bzo,Ivo:2025yek}. Choosing either sign of $\ep$, we find that all three saddles are relevant by the Picard-Lefschetz, see fig. (\ref{fig:stokes_jump}). We also note that such complexification of parameters often modifies the asymptotic convergence, and it could be a threat to the convergence of the integral in general minisuperspace (as seen in \cite{Kanazawa:2014qma, Ailiga:2025fny}). However, in the present case, as long as $|\ep|\ll 1$, the integral in eq. (\ref{eq:can_partion_ads_bh}) remains convergent and well-defined.

Although the contributing saddles are the same, as one tunes $\ep$ from $0^+$ to $0^-$, one observes a homological jump in the thimbles/intersection numbers \cite{Witten:2010cx,Behtash:2017rqj,Honda:2024aro,Kanazawa:2014qma}, and as a result, the deformed integration contours are different for $\ep>0$ and $\ep<0$, see fig. \ref{fig:stokes_jump}. The resultant homological jump can be expressed as 
\begin{equation}
    \label{eq:jump_in_thimble_s_sigma}
\mathcal{J}_i\rightarrow\mathcal{J}_i+\sum_{j\neq i}\sigma_{ij}\mathcal{J}_j,\quad n_i\rightarrow n_i+\sum_{j\neq i}\eta_{ij}n_j,
\end{equation}
where $i$ refers to the saddles participating in the jump.
$\sigma_{ij}$ and $\eta_{ij}$ are identically zero whenever there is no Stokes jump. Also, $\sigma_{ii}=\eta_{ii}=0$, and they are elements of nilpotent matrices. Whenever Stokes jump happens, smoothness across the jump requires
\begin{equation}
\label{eq:continuity_across_stokes_ray}
\sigma_{ij}+\eta_{ji}+\sum_{m}\eta_{mi}\sigma_{mj}=0,\quad \forall\,\, i\neq j.
\end{equation}
To be concrete, let's consider the example of the AdS Schwarzschild black hole. In this case, we observed Stokes ray among thermal a.d.s, small black hole (s.b.h) and large black hole (l.b.h) saddles when $\beta<\beta_{\rm max}$. 
To proceed, let us fix the orientation of the thimbles: we take the orientation to be positive ($+1$) from left to right for $\mathcal{J}(\rm a.d.s)$ and $\mathcal{J}(\rm l.b.h)$, while from down to up for $\mathcal{J}(\rm s.b.h)$. For the reverse orientation, we assign $(-1)$.
With this convention, the change in orientation of the thimble as we tune $\ep$ from positive to negative (see fig. \ref{fig:stokes_jump}), can be encoded in the matrix form as

\begin{equation}
\label{eq:stokes_jump_matrix_J}
    \left(\begin{array}{c}
         \mathcal{J}(\rm a.d.s)  \\
       \mathcal{J}(\rm s.b.h)\\
       \mathcal{J}(\rm l.b.h)
    \end{array}\right)\rightarrow \left(\begin{array}{ccc}
         1&  1&0\\
        0 & 1 &0\\
        0 &1 &1
    \end{array}\right)\left(\begin{array}{c}
         \mathcal{J}(\rm a.d.s)  \\
       \mathcal{J}(\rm s.b.h)\\
       \mathcal{J}(\rm l.b.h)
    \end{array}\right) \hspace{3mm}\text{for $\ep^+\rightarrow\ep^-$}.
\end{equation}
The matrix in the above equation encodes all the elementary jumps of each thimble. The jump in eq. (\ref{eq:stokes_jump_matrix_J}) is also expected by realizing that  Morse function $\mathfrak{h}(\rm s.b.h)$ is always subdominant than $\mathfrak{h}(\rm a.d.s)$ and $\mathfrak{h}(\rm l.b.h)$. However, the dominance between $\mathfrak{h}(\rm a.d.s)$ and $\mathfrak{h}(\rm l.b.h)$ changes at Hawking-Page transition. Since the integral over $\mathcal{J}(\rm s.b.h)$ is exponentially smaller than both $\mathcal{J}(\rm a.d.s)$, and $\mathcal{J}(\rm l.b.h)$, the thimble $\mathcal{J}(\rm s.b.h)$ can't jump multiple of $\mathcal{J}(\rm a.d.s)$, and $\mathcal{J}(\rm l.b.h)$. Also, the jump matrix remains unchanged across the transition. 
Comparing the transformation in eq. (\ref{eq:stokes_jump_matrix_J}) with eq. (\ref{eq:jump_in_thimble_s_sigma}), we get the jump matrix for the thimbles ($\mathcal{J}_\sigma$)
\begin{equation}
    \label{eq:sigma_matrix}
[\sigma]_{ij}=\left(\begin{array}{ccc}
         0&  1&0\\
        0 & 0 &0\\
        0 &1 &0
    \end{array}\right),\quad [\sigma^2]_{ij}=[0]_{ij}.
\end{equation}
Clearly, the jump matrix is a nilpotent matrix of order 2.

To ensure the smoothness across the jump, i.e., to ensure $n_\sigma(\rm a.d.s)\mathcal{J}(a.d.s)+n_\sigma(\rm s.b.h)$ $\mathcal{J}(\rm s.b.h)+n_\sigma(\rm l.b.h)\mathcal{J}(l.b.h)$ remains continuous across the jump, one requires the condition eq. (\ref{eq:continuity_across_stokes_ray}) to hold. From it, one can compute the jump matrix for the intersection numbers ($n_\sigma$), which reads 
\begin{equation}
    \label{eq:eta_matrix}
[\eta]_{ij}=\left(\begin{array}{ccc}
         0&  0&0\\
        -1 & 0 &-1\\
        0 &0 &0
    \end{array}\right),\quad [\eta^2]_{ij}=[0]_{ij}.
\end{equation}
Similarly, the jump matrix ($\eta$) is also a nilpotent matrix of order 2. With the $\eta$ matrix as given in eq. (\ref{eq:eta_matrix}), we conclude that $n_\sigma$ corresponds to the a.d.s and l.b.h saddles don't change. Only the $n_\sigma$ corresponds to the s.b.h changes to account for all the changes in $\mathcal{J}_\sigma$ across the Stokes ray. Explicitly, one obtains
\begin{equation}
    \label{eq:change_n_sigma}
    \begin{split}
  &  n_\sigma(\rm a.d.s)\rightarrow n_\sigma(a.d.s),\quad n_\sigma(\rm l.b.h)\rightarrow n_\sigma(l.b.h),\\
   & n_\sigma(\rm s.b.h)\rightarrow -n_\sigma(\rm a.d.s)-n_\sigma(\rm l.b.h)+n_\sigma(\rm s.b.h).
    \end{split}
\end{equation}

With the given orientation of the thimbles, the real line  $\mathbb{R}\in(-\infty,+\infty)$ can be deformed in two homologically different ways for $\ep>0$ and $\ep<0$, as (in $d=4$),
\begin{equation}
    \label{eq:homology_jump}
    \begin{split}
\mathbb{\mathbb{R}}=&\mathcal{J}(\rm  a.d.s)-\mathcal{J}(\rm s.b.h)+\mathcal{J}(\rm l.b.h) \hspace{3mm} \text{for $\ep >0$} \\
    =&\mathcal{J}(\rm a.d.s)+\mathcal{J}(\rm s.b.h)+\mathcal{J}(\rm l.b.h) \hspace{3mm} \text{for $\ep <0$}.
    \end{split}
\end{equation}

Remembering that the integration contour was initially defined from $0$ to $\infty$, one should put $\frac{1}{2}\mathcal{J}(\rm a.d.s)$ in eq. (\ref{eq:homology_jump}) to account only half of the thimble, \cite{Mahajan:2025bzo,Singhi:2025rfy}.
For $d>4$, we encounter a degeneracy of order $(d-3)$ for the thermal a.d.s saddle. The presence of degeneracy for a.d.s saddle wouldn't spoil the above analysis, but demands additional care. The appearance of degeneracy implies the existence of multiple steepest-descent thimbles ($\mathcal{J}(\rm a.d.s)$) emanating from the saddle (precisely $(d-2)$ directions, as explained below in sec. \ref{sec:stability_direction}). Only the part of the thimble lying on the positive-real axis participates in the Stokes jump, and the other thimbles show no qualitative change (see fig. \ref{fig:stokes_jump}). However, eq. (\ref{eq:homology_jump}) is still applicable, as the contribution from the extra thimbles automatically drops out while moving along the flowline. Hence, one can use the same argument and take $\frac{1}{2}\mathcal{J}(\rm a.d.s)$.
Since $n_\sigma$ is computed using $\rm Int(\mathbb{D},\mathcal{K}_\sigma)$ and $n_\sigma$ remains unaffected for a.d.s and l.b.h saddles in the homology jump, $\mathcal{K}_\sigma$'s corresponding to these saddles don't change.
Now, since the integration over each of the thimbles $\mathcal{J}$'s are convergent, in the saddle point approximation, one can compute the partition function eq. (\ref{eq:can_partion_ads_bh}) in the high temperature limit,
\begin{equation}
\label{eq:partition_function_high_temp}
    \begin{split}
    Z(\beta<\beta_{\rm max})&=\left[\frac{1}{2}\int_{\mathcal{J}(\rm a.d.s)}dr_+-\int_{\mathcal{J}(\rm s.b.h)}dr_++\int_{\mathcal{J}(\rm l.b.h)}dr_+\right] \exp(\mathcal{I}(r_+))\hspace{7mm} \text{for $\ep >0$}\\
    &= \left[\frac{1}{2}\int_{\mathcal{J}(\rm a.d.s)}dr_++\int_{\mathcal{J}(\rm s.b.h)}dr_++\int_{\mathcal{J}(\rm l.b.h)}dr_+\right] \exp(\mathcal{I}(r_+)) \hspace{7mm} \text{for $\ep <0$}.
    \end{split}
\end{equation}
 From eq. (\ref{eq:partition_function_high_temp}), we note that the partition function is different for $\ep>0$ and $\ep<0$ due to the sign difference in the second term and also not real. However, in the limit of vanishing $\ep$ (Real $G$), one gets the real answer from either side. These can be done systematically by taking the ``homology average" of these two contours. By taking such averaging, one drops the contribution from the subdominant saddle \cite{Mahajan:2025bzo}, and gets
 \begin{equation}
\label{eq:partition_function_1}
Z(\beta<\beta_{\rm max})=\frac{1}{2}\int_{\mathcal{J}(\rm a.d.s)}dr_+ \exp(\mathcal{I}(r_+))+\int_{\mathcal{J}(\rm l.b.h)}dr_+\exp(\mathcal{I}(r_+)) \, ,
 \end{equation}
\begin{figure}[htbp]
\centering
\subfigure[\,\,\, $\ep=\pi/10$ \label{fig:cric_pos}]{
    \includegraphics[width=0.54\textwidth]{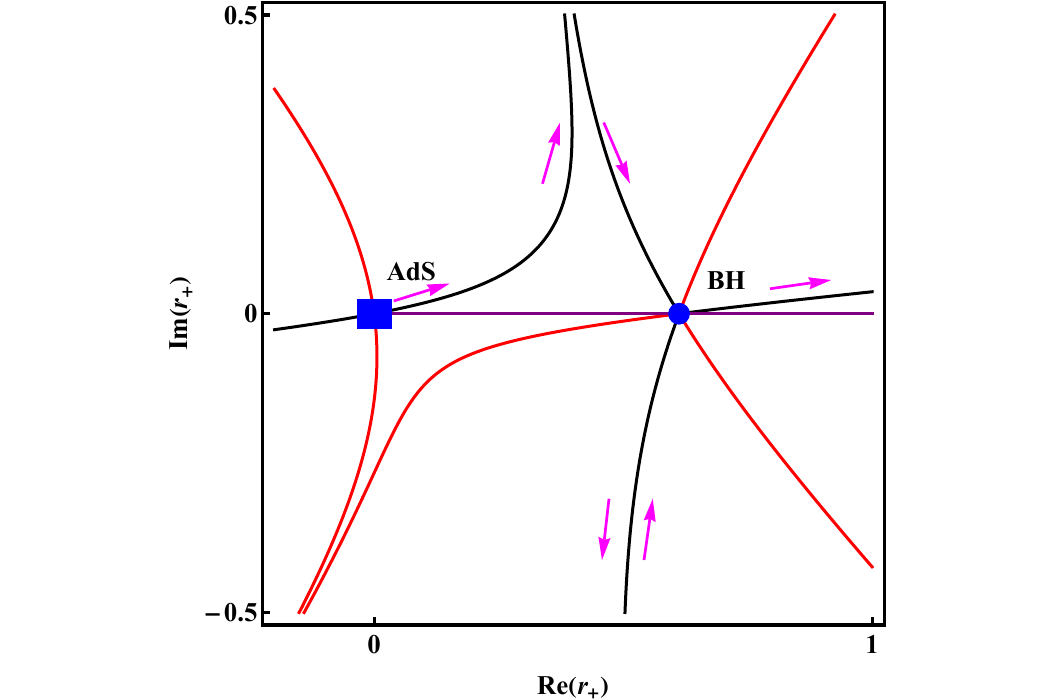}
}
\hspace{-2.6cm}
\subfigure[$\,\,\ep=-\pi/10$\label{fig:cric_neg}]{
\includegraphics[width=0.54\textwidth]{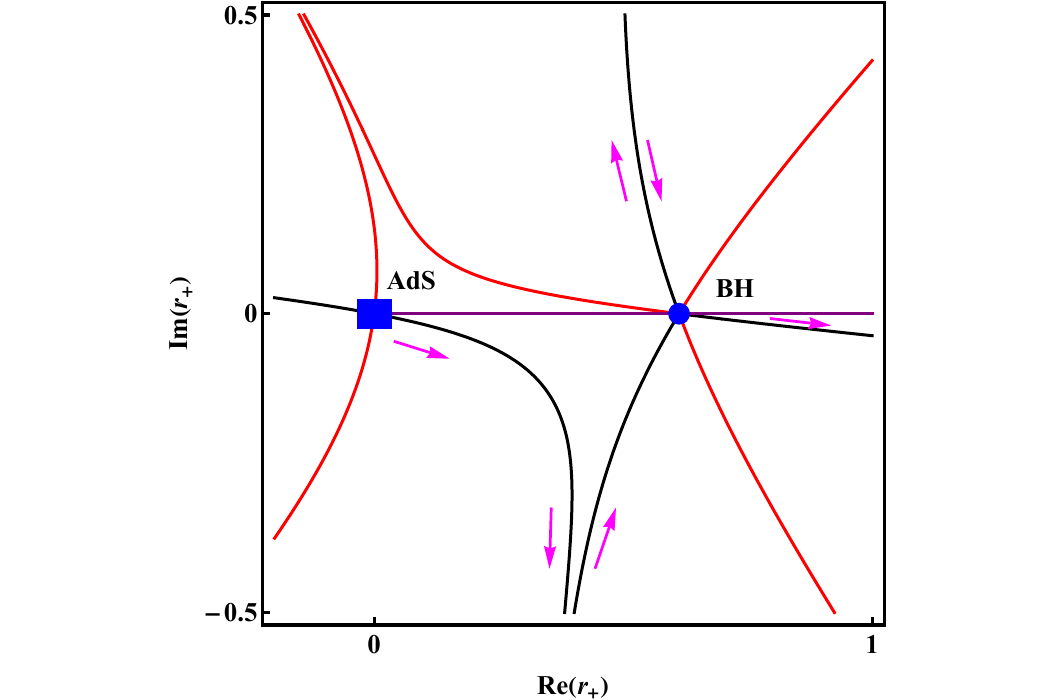}
}
\caption{Steepest descent-ascent contour for $AdS_5$ in the degenerate case ($\beta=\beta_{\rm max}$) with real  rotating $G=|G|e^{i\ep}$. Steepest descent ($\mathcal{J}_\sigma$) curves are plotted in black, and steepest ascent ($\mathcal{K}_\sigma$) curves are plotted in red. The arrows indicate the direction of the integration cycle. Note that due to the coalescing of saddles ($r_+^+=r_+^-$), there are extra descent/ascent lines emanating from the saddle (blue dot). However, the contribution from that branch drops out.}
\label{fig:stokes_jump_deg_case}
\end{figure}
where only thermal a.d.s and large black hole contribute to the partition function in the high temperature limit. When $\beta=\beta_{\rm max}$, one encounters Stokes ray between thermal ads ($r_+=0$) and the degenerate saddle ($r_+=r_+^+=r_+^-$). In this case, one also observes a similar homology jump in the thimbles, see fig (\ref{fig:stokes_jump_deg_case}). Having obtained the intersection numbers($n_\sigma$), the partition function can be computed in a straightforward way by going to the cubic order \cite{Feldbrugge:2017kzv} (see, sec. \ref{sec:stability_direction} for detailed discussion).

As one crosses $\beta=\beta_{\rm max}$, one also observes the homological jump in the Lefschetz thimbles structure as the nature of saddles changes.
 Such situations are often referred to as singularities, and the Stokes ray always passes through them. To see how the thimble structure changes as we move through the singularity, we define a new variable $\mathcal{\varepsilon}=1-\frac{\beta}{\beta_{\rm max}}$, following \cite{Witten:2010cx} (sec. 3.4). The critical points become near $\beta=\beta_{\rm max}$
\begin{equation}
  \label{eq:singulaity}  r_+=0,\cdots,0\,;\quad  r_+^\pm=\frac{2 \pi  l^2}{\beta_{\rm max}d}\pm\frac{2 \sqrt{2} \pi  l^2\sqrt{\varepsilon }}{\beta_{\rm max} d} \, .
\end{equation}
The orientation of the thimble changes as one switches from $\varepsilon<0$ to $\varepsilon>0$ . The resultant homology jump can be expressed as
\begin{equation}
\label{eq:homology_jump_critical_point}
\mathcal{J}(0)\rightarrow\tilde{\mathcal{J}}(0)-\tilde{\mathcal{J}}(r_+^-)+\tilde{\mathcal{J}}(r_+^+),\quad \mathcal{J}(r_+^+)\rightarrow -\tilde{\mathcal{J}}(r_+^-)+\tilde{\mathcal{J}}(r_+^+),\quad \mathcal{J}(r_+^-)\rightarrow \tilde{\mathcal{J}}(r_+^+),
\end{equation}
where $\mathcal{J}$ and $\tilde{\mathcal{J}}$ are the thimbles for $\varepsilon<0 (\beta>\beta_{\rm max})$ and $\varepsilon>0 (\beta<\beta_{\rm max})$. For $\varepsilon>0$, one observes a Stokes ray, and to break it, we complexify $G$ with $\ep>0$. See figures. (\ref{fig:bt_gr_bmax_AdS6}) and (\ref{fig:stokes_jump}) for orientations of thimbles for $\varepsilon<0$ and $\varepsilon>0$, respectively.

\subsubsection{Direction of steepest descent near saddles}

\label{sec:stability_direction}
In this section, we will briefly analyse the direction of steepest descent curves ($\mathcal{J}_\sigma$), which gives information about local minima near the saddle.
To find the direction of descent near the saddles, we expand $\mathcal{I}(r_+)$ around the saddles ($r_+^s$) as
\begin{equation}
\label{eq:expansion_around_saddle}
\mathcal{I}(r_+)-\mathcal{I}(r^s_+)=\left. \frac{\partial^2 \mathcal{I}}{\partial r_+^2}\right|_{r_+^s}+ \left.\frac{\partial^3 \mathcal{I}}{\partial r_+^3}\right|_{r_+^s}+\cdots.
\end{equation}
To ensure there is a minimum along the real and positive $r_+$ axis, one requires the first non-vanishing term in the r.h.s. in eq. (\ref{eq:expansion_around_saddle}) to be negative definite. For the real black hole saddles (Euclidean instanton) analyzing the second derivative of $\mathcal{I}$, we get
\begin{equation}
    \label{eq:double_derivative}
   \left. \frac{\partial^2 \mathcal{I}}{\partial r_+^2}\right|_{r_+^\pm}=-\frac{(d-1)\omega l^2\pi}{2Gd\beta}(r_+^\pm)^{d-4}\left[1-\frac{\beta^2}{\beta_{\rm max}^2}\pm\sqrt{1-\frac{\beta^2}{\beta_{\rm max}^2}}\right],\quad \beta<\beta_{\rm max}.
\end{equation}
Hence, at the large black hole ($r_+^+$), it is negative, while at the small black hole ($r_+^-$), it is positive.
Such behaviour is expected as a small black hole is known to be thermodynamically unstable, whereas a large black hole is stable. Since these are real saddles, the direction of minima is along the real axis only. While a small black hole is unstable along the real axis, it admits a direction of steepest descent when analytically continuing $r_+$ to the complex plane, see fig (\ref{fig:stokes_jump}). At $\beta=\beta_{\rm max}$, one encounters degeneracy as the right-hand side in eq. (\ref{eq:double_derivative}) vanishes. Evaluating at the cubic order, at the degenerate saddle, we get
\begin{equation}
    \label{eq:deg_case}
 \Biggl.\frac{\partial^3\mathcal{I}}{\partial r_+^3}\Biggr|_{r_+^+=r_+^-}=-(d-1) \left(\frac{d-2}{d}\right)^{\frac{d-4}{2}} l^{d-4},
\end{equation}
where, $-$ sign signifies the presence of steepest descent along the real-positive $r_+$ direction. Interestingly, there are also two different steepest descent directions in the complex $r_+$-plane, see fig. (\ref{fig:stokes_jump_deg_case}).

Now consider the case of the thermal AdS saddle at $r_+=0$ of degeneracy $d-3$. Presence of degeneracy implies 
\begin{equation}
    \label{eq:deg_thermal_ads}
    \frac{\partial^2 \mathcal{I}}{\partial r_+^2}=\cdots=\frac{\partial^{d-3} \mathcal{I}}{\partial r_+^{d-3}}=0,\,\,\hspace{7mm} \frac{\partial^{d-2} \mathcal{I}}{\partial r_+^{d-2}}\neq 0.
\end{equation}
 For $d=4$, the saddle is non-degenerate, while for higher $d$, degeneracy appears. In the presence of degeneracy, the saddle point approximation breaks down, and one needs to go into higher orders. The partition function when evaluated at $r_+=0$ saddle gives
\begin{equation}
    \label{eq:deg_Z[beta]}
    \begin{split}
    &Z_0[\beta]\sim\int_0^{\infty} dr_+ \exp\left[\mathcal{I}(0)+\frac{1}{(d-2)!}\frac{\partial^{d-2} \mathcal{I}}{\partial r_+^{d-2}}r_+^{(d-2)}\right] \\
    &=\int_0^{\infty} dr_+ \exp\left[-\frac{\beta\omega(d-1)}{16\pi G}r_+^{(d-2)}\right]= \Gamma \left(\frac{d-1}{d-2}\right) \left(\frac{\beta  (d-1) \omega }{16 \pi G}\right)^{\frac{1}{2-d}}.
    \end{split}
\end{equation}
Since the integral is convergent along the real $r_+$ axis, there is a steepest descent along this direction. When $d>4$, there are other curves (apart from the real $r_+$ line) in the complex $r_+$ plane along which the integral also converges. To see this we write $r_+=|r_+|e^{i\varphi}$, near the saddle convergence remains unaffected if $\cos(\varphi(d-2))=1$. When $d=4$, there are only two directions ($\varphi=\pm \pi$, along the real line) along which the integral converges. However, for $d=5$ the directions (near the saddle) along which the integral converges are $\varphi=0,\pm2\pi/3$, see fig. (\ref{fig:stokes_jump}). Similarly, for higher $d$, there will be multiple paths in the complex $r_+$ plane along which the integral converges absolutely apart from the real ($\varphi=0$) contour, precisely a total of $(d-2)$ $\mathcal{J}_\sigma$'s. Explicitly, the directions emanating from the saddle at $r=0$ are 
\begin{equation}
    \label{eq:deg_sadd_direc}
    \varphi_n=\frac{2\pi n}{(d-2)},\quad n=0,1,2,\cdots,(d-3).
\end{equation}

\subsection{Thermodynamic phase-transition and instability}
\label{sec:phase_transition}
\begin{figure}
    \centering
    \includegraphics[width=0.6\linewidth]{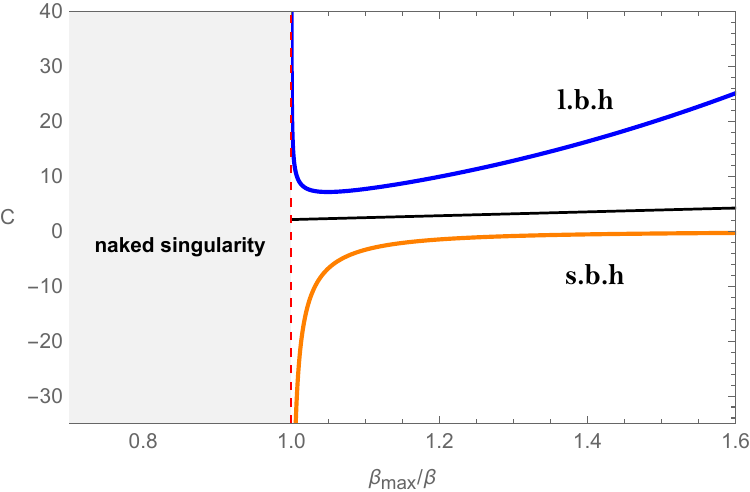}
    \caption{Behaviour of specific heat ($C$) for $\beta>\beta_{\rm max}$ and for $\beta<\beta_{\rm max}$ at various saddles.}
    \label{fig:placeholder}
\end{figure}

In this section, we wish to understand the thermodynamic phase transitions and instabilities from the viewpoint of Picard-Lefschetz analysis \cite{Witten:2010cx,Kanazawa:2014qma,Fujimori:2021oqg}. We observe that any
discontinuous changes to the orientation of Lefschetz thimbles/intersection number are related to the phase transition and thermodynamic instability of the black holes. To justify the statement, let us examine the behaviour of specific heat ($C$), which is given by
\beq
\label{eq:spec_heat_def}
C =  T \biggl(\frac{\partial S}{\partial T}\biggr) \, .
\eeq
 Utilizing the expression for entropy $S$ given by eq. (\ref{eq:entropy_and_mass}) and the temperature given by the regularity condition of black hole $T(r_{+}) = f'(r_{+})/(4\pi)$, the specific heat for the black holes with regular horizon $r_{+}$ is given by:
\beq
\label{eq:sp_heat_bh}
C(r_+) = \frac{\omega (d-1) r_{+}^{(d-2)}}{4 G} \frac{T(r_{+})}{T'(r_{+})}\, .
\eeq
The specific heat of the black hole saddles given in eq. (\ref{eq:roots_canonical}) is
\begin{equation}
    \label{eq:specific_heat}
    C(r_+^\pm)=\pm\frac{\omega (d-1) \beta_{\rm max} (r_{+}^{\pm})^{d-1}}{4G\sqrt{(\beta_{\rm max})^2 - \beta^2}} \,.
\end{equation}
At high temperatures ($\beta<\beta_{\rm max}$), one finds the specific heat $C$, of small black holes ($r_+^-$) is negative, while for large black holes ($r_+^+$) it is positive, see fig. (\ref{fig:placeholder}). This indicates the large black hole is thermodynamically stable, while the small black hole is unstable. We also observed that the thimbles ($\mathcal{K}(\rm s.b.h)$) emanating from a small black hole exhibit a jump in orientation across the Stokes ray, resulting in a discontinuous change in the intersection number ($n_\sigma$). This turns out to be true for the more general charged case as well, where a negative specific heat saddle corresponds to a jump in $n_\sigma$, (see sec. \ref{sec:instablity_jump} and \ref{sec:canonical_sp_heat}). As a consequence, their contribution to the partition function drops out upon homology averaging. At low temperatures ($\beta > \beta_{\rm max}$), $C$ becomes complex, signifying the solutions describing complex black holes or naked singular geometries.  A similar conclusion is also observed earlier in \cite{Marolf:2022ybi},  where the authors also noticed that when black hole saddles have positive specific heat, they contribute to the partition function with non-zero weight in the semi-classical limit.

Let us also point out that the specific heat diverges (dictating the continuous phase transition) when $T'(r_{+}) = 0$, which happens when the saddles coalesce at $\beta=\beta_{\rm max}$, see fig. (\ref{fig:placeholder}). As discussed earlier, when one crosses this point, the orientation of the Lefschetz thimbles ($\mathcal{J}$) changes discontinuously, see eq. (\ref{eq:homology_jump_critical_point}). Hence, we find that any discontinuous jump in the orientation of thimbles is associated with unphysical behaviour of specific heat and leads to thermodynamic instability.

\section{Charged AdS Schwarzschild black hole}
\label{sec:charge_ads}

We next consider the case of charged AdS geometries to test the observations made in the previous section on pure-AdS: naked-singular geometries, although being dominant in partition function in some cases, are eventually discarded in the Picard-Lefschetz analysis and small size sub-dominant blackhole geometries despite being relevant, drops out from the partition function due to homology-averaging. 

The charged-AdS Schwarzschild with asymptotic boundary $S_\beta^1\times S^{d-1}$ has the following metric
\bea
\label{eq:char_AdS_line}
ds^2 & = & f(r) d\tau^2 + \frac{dr^2}{f(r)} + r^2 d\Omega_{d-1}\, , \hspace{5mm}\tau \sim \tau + \beta \, , \;  \text{where} \\
\label{eq:f(r)_def_char}
f(r) & = & 1 + \frac{r^2}{l^2} - \frac{\mu}{r^{d-2}} + \frac{q^2}{r^{2d-4}} \,  ,
\eea
where $\mu$ describes the mass and $q$ denotes the charge. The non-zero vector potential component is given by:
\beq
\label{eq:A_tau}
A_{\tau} = i (\Phi - \frac{q}{\gamma r^{d-2}} ) \, ,
\hspace{5mm}
{\rm where}
\hspace{5mm}
\gamma = \sqrt{2(d-2)/(d-1)} \, .
\eeq
The physical mass ($M$) and charge ($Q$) are given by
\beq
\label{eq:phy_mass_char}
M=\frac{(d-1)\omega\mu}{16\pi G} \, ,
\quad Q=\frac{\gamma(d-1)\omega q}{8\pi G} \, .
\eeq
The geometries that correspond to black hole solutions, consist of inner and outer horizons at $r_-$ and $r_+$, respectively. These can be found using $f(r)=0$, where smaller positive real root correspond to inner horizon and larger positive real root correspond to outer horizon. As $r_+$ satisfies $f(r_{+}) = 0$, this immediately gives a relation between $\mu$ and $q$
\beq
\label{eq:mu_def}
\mu = r_{+}^{d-2} \left[1 + \frac{r_{+}^2}{l^2} + \frac{q^2}{r_{+}^{2d-4}} \right] \, .
\eeq

\begin{figure}
    \includegraphics[width=0.40\linewidth]{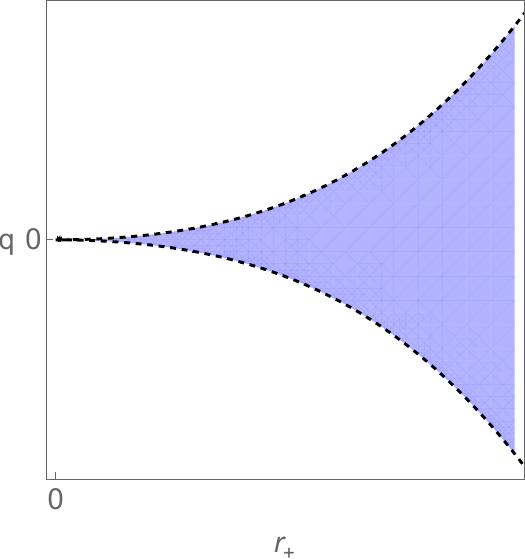}
    \caption{ In the blue shaded region, one obtains black holes with a horizon/no naked singularities, satisfying eq. (\ref{eq:no_naked_singularity}). We use $d=4, l=1$ for plotting. This is the region one obtains at the real and positive $r_+$-saddle. The dashed lines describe extremal black holes.}
    \label{fig:region_of_integration}
\end{figure}

Note that the presence of a real and positive $r_+$ automatically ensures another real and positive root ($r_-$, inner horizon). This is a consequence of Descartes' rule of signs that constrains $f(r)=0$ to have either two or zero real and positive roots. For a generic black hole geometry, to have a non-singular horizon and avoid the presence of {\it naked singular} geometries in the partition function, one requires
\begin{equation}
    \label{eq:no_naked_singularity}
    \frac{d}{(d-2)l^2}r_+^{2d-2}+r_+^{2d-4}\geq q^2 \, ,
\end{equation}
to be satisfied by a real positive $r_{+}$. The inequality in eq. (\ref{eq:no_naked_singularity}) saturates when both the horizons coincide and one gets extremal black hole, see fig (\ref{fig:region_of_integration}). The inequality ensures the positivity of $\beta$ at the horizon ($r_+$) and also put a bound on the mass ($\mu\geq \mu_e$), where $\mu_e$ is the mass of extremal black hole. In supersymmetric theories, one also gets a bound on the mass which is $\mu\geq 2q$, with $\mu=2q$ is a BPS state \cite{London:1995ib,Chamblin:1999tk}. Now, since $\mu_e>2q$ for finite $l$, extremal solution is always non-supersymmetric. When $\mu=2q$ (supersymmetric solution), one gets from eq. (\ref{eq:mu_def})
\begin{equation}
    \label{eq:susy_solution}
    \left(1+\frac{q}{r_+^{d-2}}\right)^2+\frac{r_+^2}{l^2}=0.
\end{equation}
Clearly, eq. (\ref{eq:susy_solution}) has no real $r_+$ solutions for real $q$ and hence will describe a naked singularity \cite{Chamblin:1999tk}. Such solutions will also violate the inequality, \ref{eq:no_naked_singularity}.
We point out that all solutions violating eq. \ref{eq:no_naked_singularity} will correspond to a naked singular geometry; however, satisfying it doesn't always guarantee the absence of a naked singularity. For instance, real negative saddles will satisfy the condition (\ref{eq:no_naked_singularity}). However, they all correspond to a naked singularity, which is in conflict with the cosmic censorship. We will address this issue more systematically by investigating the {\it relevance} of such geometries that appear as dominant saddle geometries of the gravitational partition function in the semiclassical limit.
We analyze the partition function for the charged case in two different settings. The first corresponds to fixing the inverse temperature $\beta$ and the electrostatic potential $\Phi$ at spatial infinity, known as the grand canonical ensemble. The other case is where a fixed charge ($Q)$ and $\bt$ asymptotically, known as the canonical ensemble.

\subsection{Fixed Potential - Grand Canonical Ensemble}
\label{fix_pot}

When fixing the Potential ($\Phi$) at infinity, the grand partition function is given by:
\beq
\label{eq:Z_char_AdS_par}
Z(\beta, \Phi) = \int_0^\infty dS \int_{-\infty}^{+\infty}dQ \,\, \exp[S-\beta(E-\Phi Q)] \, ,
\eeq
where
\begin{equation}
    \label{eq:phi_q}
   S(r_+)=\frac{\omega r_+^{d-1}}{4G},\quad E(r_+,q)= \frac{(d-1)\omega}{16\pi G}\left(r_+^{d-2}+\frac{r_+^d}{l^2}+\frac{q^2}{r_+^{d-2}}\right).
\end{equation}
Similar to the AdS-Schwarzschild case, it will be convenient to switch from $(S,Q)$ to the $(r_+,q)$ variables and perform the saddle-point analysis. Ignoring the Jacobian, we get
\begin{equation}
\label{eq:grand_partition_mini}
Z(\beta, \Phi) =\int_{\mathbb{D}} 
{\rm d} r_+ {\rm d}q \,\, 
\exp{(\mathcal{I}_{\rm CAdS})}
\end{equation}
where, the region of integration ($\mathbb{D}$) spans $0< r_+<\infty$ and $-\infty<q<+\infty$ with $r_+=q=0$ point included and the action is given by
\beq
\label{eq:cha_AdS_act}
\mathcal{I}_{\rm CAdS} = \frac{\omega}{16\pi G} \left[4\pi r_{+}^{d-1} - (d-1)\beta \left( r_{+}^{d-2} + \frac{r_{+}^{d}}{l^2} + \frac{q^2}{r_{+}^{d-2}} - 2 \gamma q \Phi\right)\right] \, .
\eeq
In region $\mathbb{D}$, the integrand has no essential singularity. To start with, we allow all possible charges and take the integration contour for $q$ to run from $-\infty$ to $+\infty$, as there is no classical instability present (unlike the rotating solutions \cite{Hawking:1999dp}). However, as we explain below at saddles, $q$ and $r_+$ get constrained and satisfy eq. (\ref{eq:no_naked_singularity}).
The integral in eq. (\ref{eq:grand_partition_mini}) is a two-dimensional integral. Since the argument inside the exponent is quadratic in $q$ (and hence, there is only one saddle for it), one can first perform the integration over $q$ and retain the leading order term in $G$, reducing to an effective one-dimensional integral. This is appropriate to do for our analysis in the semi-classical limit. The effective one-dimensional integral for $r_+$ is then given by:
\bea
\label{eq:grand_partition}
Z[\beta,\Phi] && = \int_0^\infty {\rm d} r_+ 
\,\, \exp \left[\frac{\omega r_+^{d-1}}{4G}-\beta\frac{\omega(d-1)}{16\pi G}\left\{\left(1-\gamma^2\Phi^2\right)r_+^{d-2}+\frac{r_+^d}{l^2}\right\}\right] 
\notag \\
&& =\int_0^\infty {\rm d} r_+ 
\,\, e^{\mathcal{I}(r_{+},\Phi)}\, .
\eea
Note that after performing the integration over $q$, the partition function eq. (\ref{eq:grand_partition}) admits $\Phi\rightarrow-\Phi$ symmetry. As expected when $\Phi=0$, eq.(\ref{eq:grand_partition}) reduces to eq. (\ref{eq:can_partion_ads_bh}). We now proceed with the Picard–Lefschetz analysis, which follows in the same manner as in the AdS Schwarzschild case. The dominant contribution to the charged partition function is given by saddle points, which are computed using the following equation:
\beq
\label{eq:char_sads_eqn}
\frac{\partial \mathcal{I}}{\partial r_+} = 0\, ,
\eeq
that gives the relation satisfied by the $r_{+}$ as:
\beq
\label{eq:rp_sad}
r_{+}^{(d-3)} \left[ r_{+}^2 - \frac{4\pi l^2}{d\beta} r_{+} + l^2 \left(\frac{d-2}{d}\right)(1-\g^2 \Phi^2)\right] = 0 \, ,
\eeq
The saddles are thermal-AdS at $r_{+}^{(0)} = 0$ with $q^{(0)} =0$ having a degeneracy of $(d-3)$, along with  two more saddles denoted $r_{+}^{-}$ and $r_{+}^{+}$. These saddles have the following expression
\beq
\label{eq:char_BH_sol}
r_{+}^{\pm} = \frac{1}{d \bt} \Bl(
2\pi l^2 \pm l \sqrt{4\pi^2 l^2 - d (d-2) \beta^2 (1- \g^2 \Phi^2)} 
\Br)\, ,
\eeq
with associated charge $q^{\pm} = \g \Phi (r_{+}^{\pm})^{d-2}$. This is obtained from the regularity condition on vector potential $A_{\tau}(r_{+}) = 0$ at the horizon. Defining ($\beta_{\rm max}^\Phi$, $\Phi_{\rm max}$)- a special value of external parameters where one observes qualitative changes in the saddles and Lefschetz thimbles
\begin{equation}
\label{eq:betamaxphimin}
\beta_{\rm max}^\Phi=\frac{2\pi l}{\sqrt{d(d-2)\left\{1-\gamma^2\Phi^2\right\}}},\,\,\,\hspace{4mm}\Phi_{\rm max}=\frac{1}{\gamma}=\sqrt{\frac{(d-1)}{2(d-2)}} \, ,
\end{equation}
the roots can be rewritten as
\begin{equation}
\label{eq:roots_grand_canonical}
r_+^{\pm}=\frac{2\pi l^2}{d\beta}\left(1\pm\sqrt{1-\frac{\beta^2}{(\beta_{\rm max}^{\Phi})^2}}\right)
= \frac{2\pi l^2}{d\beta} \bl(
1 \pm \tanh \theta
\br) \, ,
\end{equation}
where we define 
\beq
\label{eq:tanhTH_charge}
\tanh \theta = \sqrt{1-\frac{\beta^2}
{(\beta_{\rm max}^{\Phi})^2} } \, .
\eeq
The saddles in eq. (\ref{eq:roots_grand_canonical}) share the same structure as the saddles for the AdS Schwarzschild case, given in eq. (\ref{eq:roots_canonical}) with the only exception that $\beta_{\rm max}^\Phi$ can become imaginary depending on $\Phi$. These saddles are regular for all $\bt$ and $\Phi$. The Kretschmann scalar for these saddle geometries can be computed as before (see appendix \ref{sec:curvature_scalar}), and it is given by
\begin{equation}
    \label{K_explicit}
    \begin{split}
        K(r)=\frac{A}{r^{4d-4}}+\frac{B}{r^{3d-2}}+\frac{C}{r^{2d-2}}+\frac{D}{r^{2d}}+E
    \end{split}
\end{equation}
where,
\begin{equation}
    \label{K_explicit_component}
    \begin{split}
       & A=2 \left(8 d^4-52 d^3+127 d^2-139 d+58\right) q^4,\hspace{5mm} B=-4 (d-1)^2 \left(2 d^2-7 d+6\right) \mu  q^2\, ,\\
         & C=\frac{4 \left(d^2-5 d+6\right) q^2}{l^2},\hspace{4mm} D=(d-2) (d-1)^2 d \mu ^2,\hspace{5mm} E=\frac{2 d (d+1)}{l^4}\, ,\\
    \end{split}
\end{equation}
and $\mu$ is mentioned in eq. (\ref{eq:mu_def}).
It is evident that at $r=0$, the Kretschmann scalar blows up for all saddle geometries. If $\theta$ becomes complex, which happens for real $\bt_{\rm max}^\Phi$ ($\Phi<\Phi_{\rm max}$) with $\beta>\beta_{\rm max}^\Phi$, the saddles $r_+^\pm$ become complex. These saddles lack a horizon, exposing the singularity at $r=0$. These correspond to naked-singular geometries. Real solutions for $r_+$ occur under three circumstances: (1) $\Phi>\Phi_{\rm max}$, leading to imaginary $\bt_{\rm max}^\Phi$ giving real $\theta$ for values of $\bt$, (2) Extremal case when $\Phi=\Phi_{\rm max}$, implying $\bt_{\rm max}^\Phi \to \infty$ leading to two real solutions for all $\bt$, (3) $\Phi<\Phi_{\rm max}$ with $\beta<\beta_{\rm max}^\Phi$. In the first case when $\Phi>\Phi_{\rm max}$, one has $r_+^+>0$ and $r_+^-<0$. The case where both $r_+^\pm$ are real and non-negative correspond to second and third case above. 

\begin{figure}
    \centering
\includegraphics[width=12cm, height=8cm]{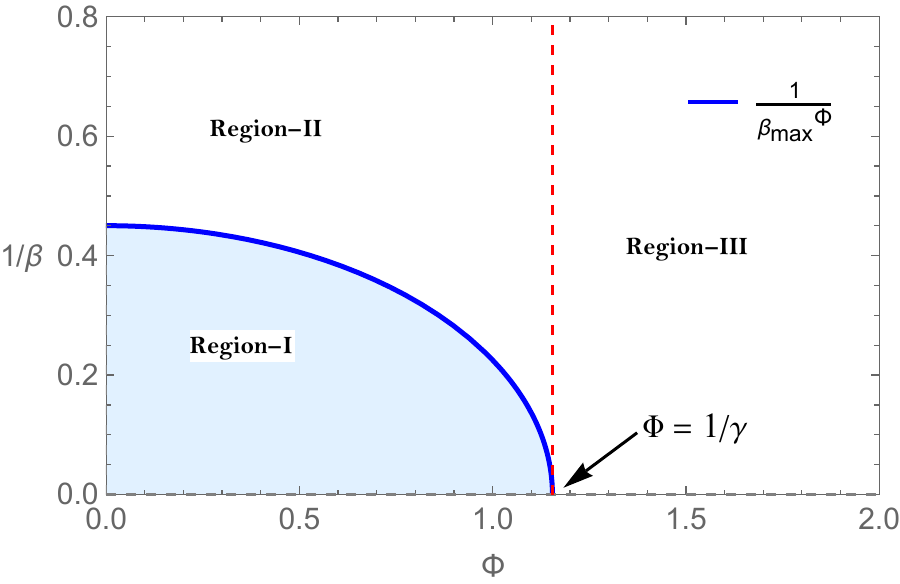}
    \caption{Phase diagram for charged $AdS_5$ black hole. The blue shaded region describes naked singular geometries. Region-II describes a black hole with a non-singular horizon.}
    \label{fig:phase_diagram}
\end{figure}

We summarize these in table \ref{tab:nature_horizon} and pictorially in fig. (\ref{fig:phase_diagram}).
\begin{table}[hbt!]
    \centering
    \begin{tabular}{|c|c|c|}
    \hline
        \text{Range of $\beta$} &\text{Range of $\Phi$}  &\text{nature of the $r_+$ horizon}\\
        \hline
        $\beta<\beta_{\rm max}^\Phi$ & $\Phi<\Phi_{\rm max}$ & two real and positive $r_+$ solution\\
         \hline
         $\beta>\beta_{\rm max}^\Phi$ & $\Phi<\Phi_{\rm max}$ & two complex conjugate $r_+$ solution\\
         \hline
         $0<\beta<\infty$ & $\Phi_{\rm max}<\Phi$ & real, positive and negative $r_+$ solution\\
          \hline
         $0<\beta<\infty$ & $\Phi=\Phi_{\rm max}$ & one real positive $r_+$ solution\\
         \hline
    \end{tabular}
    \caption{Various ranges of the free parameters $(\beta,\Phi)$ and the corresponding nature of the solutions}
    \label{tab:nature_horizon}
\end{table}
The exponent at the thermal AdS solution is given by $\mathcal{I}(r_{+}^{(0)},\Phi) =0 $ and at the other two saddles it reads as:
\beq
\label{eq:on_act_char_BH}
\mathcal{I}(r_{+}^{\pm},\Phi) =  \frac{\omega (r_{+}^{\pm})^{d-1}}{(d-2) 8\pi G l^2} \left[ (d-1)\beta r_{+}^{\pm} - 2\pi l^2\right] \, .
\eeq
Similar to the AdS Schwarzschild case, the spacetime geometries at the charged saddle points $r_{+}^{\pm}$ are regular, i.e. $\beta = 4\pi/f'(r_{+})$ is satisfied, implying there is no conical singularity at the horizon. Further, when $r_+^\pm$ is a positive saddle, utilizing $q = \g \Phi (r_+^\pm)^{d-2}$, one satisfies the condition for the no-naked singularity given in eq. (\ref{eq:no_naked_singularity}). 

Just like in the case of AdS–Schwarzschild, one encounters a similar puzzle in the grand canonical ensemble that is the complex saddles dominate at extremely low temperatures, $\beta \gg \beta_{\rm max}^{\Phi}$, for $\Phi < \Phi_{\rm max}$. The analysis proceeds analogously to the Schwarzschild case, with $\beta_{\rm max}$ replaced by $\beta_{\rm max}^{\Phi}$. Furthermore, for the grand canonical case, in the regime $\Phi > \Phi_{\rm max}$, there exists a negative saddle. At extreme low temperatures $\bt \to \infty$, it remains subdominant with respect to the positive black hole saddle. However, its dominance relative to the thermal AdS varies with the dimension: the negative saddle dominates over the thermal AdS in even $d$, while in odd $d$, the thermal AdS phase remains dominant. Similar to earlier case, we now proceed to analyze these saddle points to determine their contributions to the partition function across different parameter regimes of $\beta$ and $\Phi$, as summarized in the table. (\ref{tab:nature_horizon}), utilizing Picard-Lefschetz (PL) methods.

\subsubsection{Complex saddles - $\beta>\beta_{\rm max}^\Phi$ and $\Phi<\Phi_{\rm max}$}
\label{comp_c_sads}

When $\Phi<\Phi_{\rm max}$ and $\beta > \beta_{\rm max}^\Phi$ (low temperature), we obtain thermal-AdS along with two complex saddles, similar to the AdS-Schwarzschild case. In a similar argument discussed in sec. \ref{sec:cbh}, these complex saddles, although sometimes dominant, do not contribute to the partition function, according to the PL method, see fig \ref{fig:bt_gr_bmax_AdS6}. It can be quickly understood as follows: in the low-temperature regime, the complex saddle points yield a nonzero value of $\mathcal{H}(r_{+}^{\pm},\Phi)$ and fails to satisfy the relevance condition $\mathcal{H}(r_+^\pm,\Phi)=0$ in any dimension $d\geq 3$. Only the thermal saddle contributes as it lies on the integration contour. This implies that the complex saddles corresponding to naked-singular geometries, despite being dominant, fail to become ``relevant'' saddles as the deformed integration contour, according to the PL method, doesn't pass through them. 
\begin{figure}[h]
    \centering
    \includegraphics[width=15cm, height=10cm]{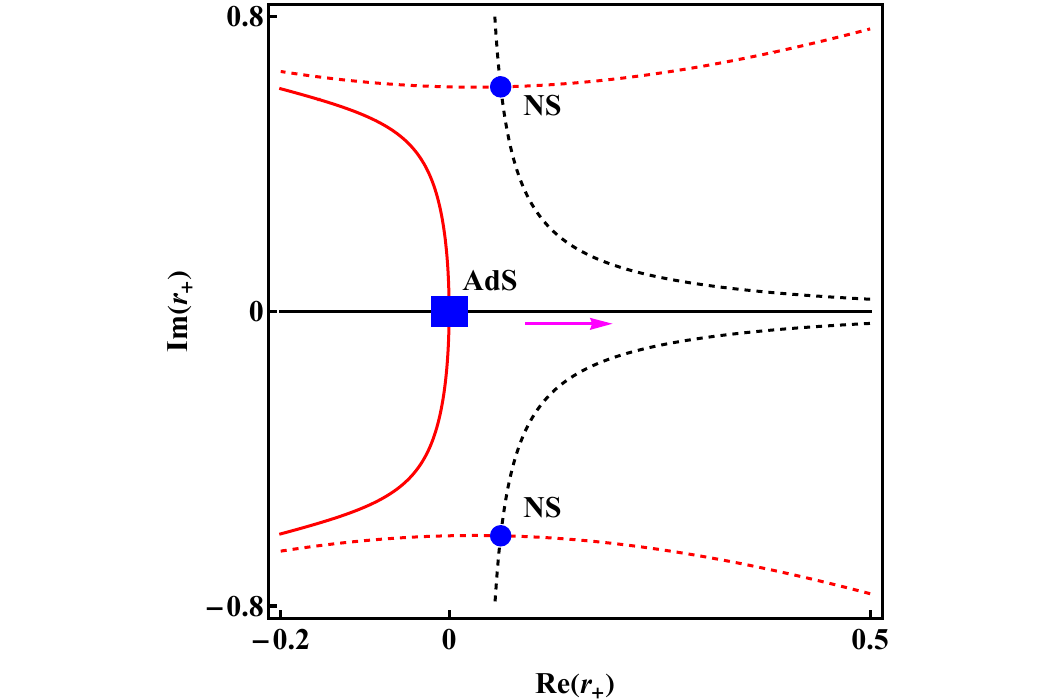}
    \caption{
    PL plot for charged $AdS_{5}$ blackhole in the grand canonical ensemble when $\Phi<\Phi_{\rm max}$ and $\beta > \beta_{\max}^{\rm \Phi}$. For illustration, we choose the parameters as: $l=1, \Phi = \Phi_{\rm max}/2, \beta = 10\beta_{\rm max}^{\Phi}$ and $G=1$  with $\Phi_{\rm max}=\sqrt{3}/2$ and  $\beta_{\rm max}^{\Phi} = \sqrt{2/3}\pi$ . The steepest ascent associated with the Empty $AdS$ saddle (square box) intersects the original integration contour, making it relevant, while the steepest ascent (red dotted curve) corresponding to complex saddle/naked singularities (blue dots) doesn't intersect, thereby irrelevant to the partition function.}
    \label{fig:phi_ls_phmax}
\end{figure}
%

\subsubsection{Negative real saddle - $\Phi > \Phi_{\rm max}$}
\label{neg_c_sad}

\begin{figure}[htbp]
\centering
\subfigure[ $\ep =\pi/10$ \label{fig:phi_gr_phmax_pos}]{
    \includegraphics[trim={2.8cm 0 2.8cm 0cm}, clip, scale =0.6]{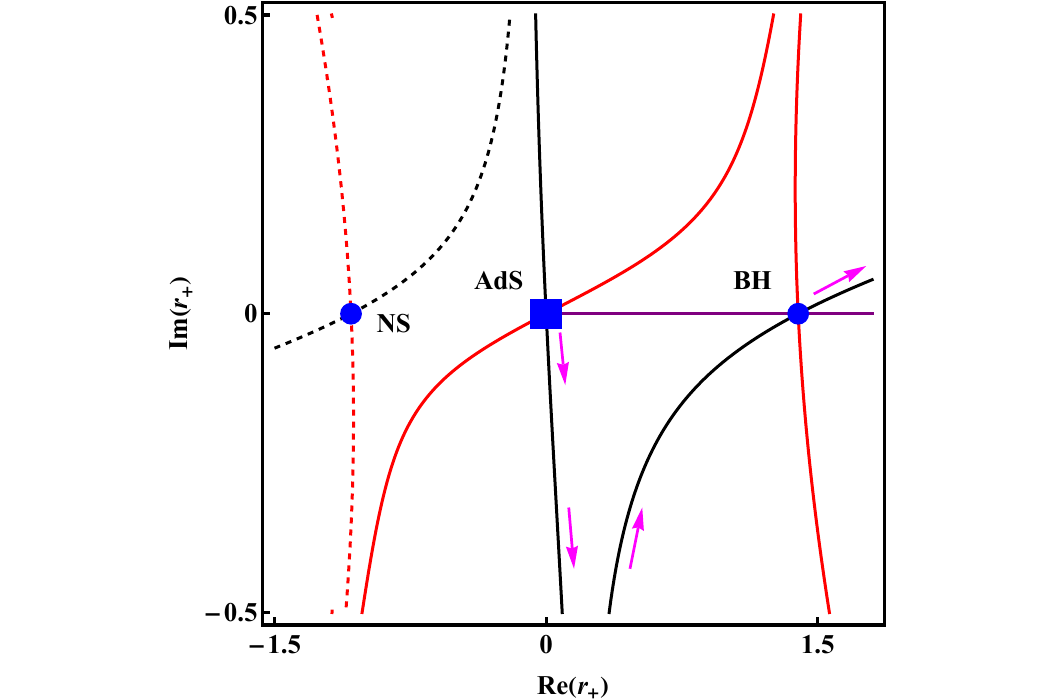}
}
\hspace{0.5cm}
\subfigure[$\ep =-\pi/10$\label{fig:phi_gr_phmax_neg}]{
    \includegraphics[trim={2.8cm 0 2.8cm 0cm}, clip, scale =0.6]{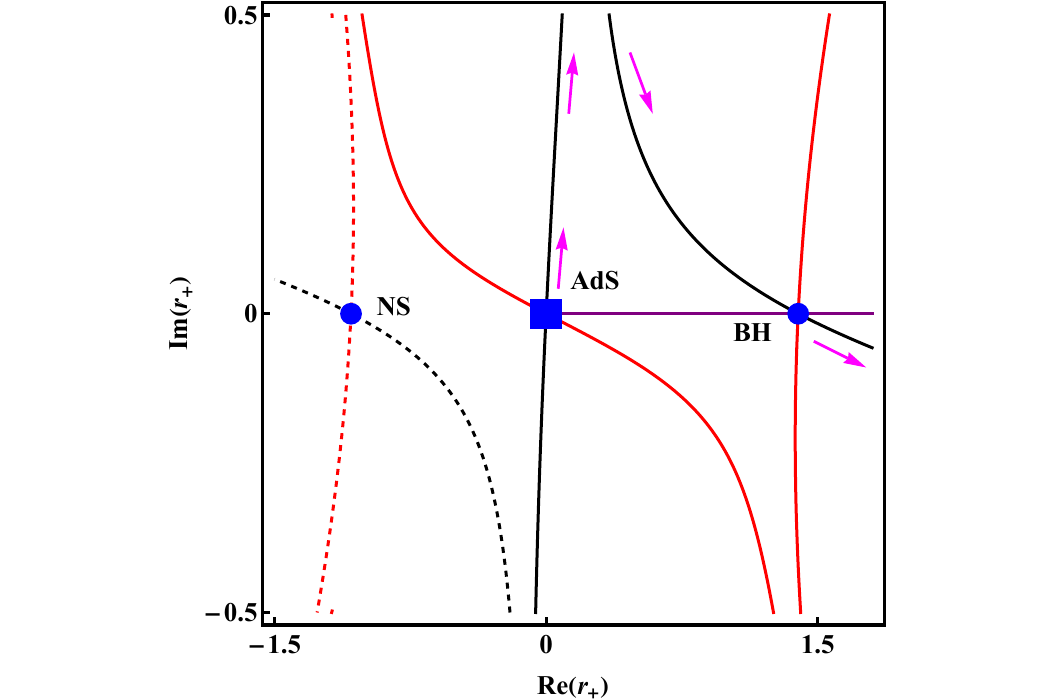}
}
\caption{ PL plot illustrating the breaking of stokes degeneracy for $\Phi>\Phi_{\rm max}$ in case of $AdS_{5}$. We choose the parameters as: $l=1, \Phi = 2\Phi_{\rm max}, \beta = 10$ and $G=e^{i\ep}$, with $\Phi
_{\rm max} = \sqrt{3}/2$. 
}
\label{fig:PL_phi_gr_phmax}
\end{figure}

When $\Phi > \Phi_{\rm max}$, at any temperature, we have thermal AdS, a real positive $(r_{+}^{+})$ and a negative saddle $(r_{+}^{-})$. The positive saddle corresponds to a black hole with a horizon. However, the negative saddle corresponds to a naked singularity. Since all the saddles lie on the real axis, we have $\mathcal{H}(r_{+}^{s}) = 0$ for each of them and hence saddles lie on a Stokes ray. In a way similar to the AdS-Schwarzschild case, we complexify the Newton constant ($G=|G|e^{i\ep}$) and perform the PL analysis (see fig. \ref{fig:PL_phi_gr_phmax}). In this regime $\Phi > \Phi_{\rm max}$, we find that both the thermal-AdS saddle and the positive black hole saddle are relevant for either sign of $\ep$, while the negative saddle doesn't contribute since the associated steepest ascent doesn't intersect the original integration contour. Hence, we reach the same conclusion that {\it naked singular} geometries do not contribute to the partition function.

%

\subsubsection{Positive real saddles - $\beta \leq \beta_{\rm max}^\Phi$ and $\Phi < \Phi_{\rm max}$}
\label{real_c_sad}

For $\Phi<\Phi_{\rm max}$, at high temperatures ($\beta<\beta_{\rm max}$), one obtains two real black hole saddles along with the thermal AdS. Since $\mathcal{H}=0$ for all the saddles, the saddles lie on a Stokes ray. By complexifying $G$, one can lift the Stokes ray and perform the PL analysis, as shown in fig. \ref{fig:stokes_jump}. The analysis is similar to the AdS-Schwarzschild case with $\beta < \beta_{\rm max}$; therefore, we don't repeat it. In a similar way, the contribution from the small black hole drops out from the partition function after homology averaging. Only thermal-AdS and large black-hole contribute. When $\beta=\beta_{\rm max}^\Phi$, the black hole saddles coalesce into a single degenerate saddle at $r_{+}^+=r_+^- = 2\pi l^2/(d\beta_{\rm max}^{\Phi})$. For such situation PL analysis is similar to the AdS-Schwarzschild case with $\beta = \beta_{\rm max}$, see fig. \ref{fig:stokes_jump_deg_case}. 

\begin{figure}[h!]
\centering
\subfigure[$\ep =\pi/10$\label{fig:cs5_pos}]{
    \includegraphics[trim={2.8cm 0 2.8cm 0cm}, clip, scale =0.6]{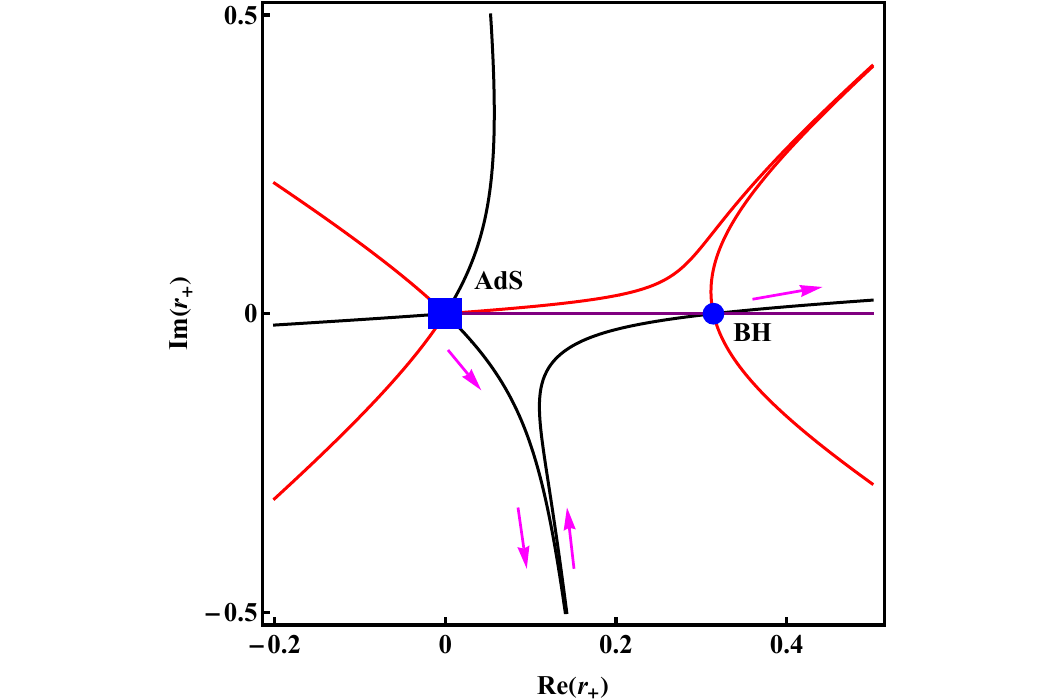}
}
\hspace{0.5cm}
\subfigure[$\ep =-\pi/10$\label{fig:cs5_neg}]{
    \includegraphics[trim={2.8cm 0 2.8cm 0cm}, clip, scale =0.6]{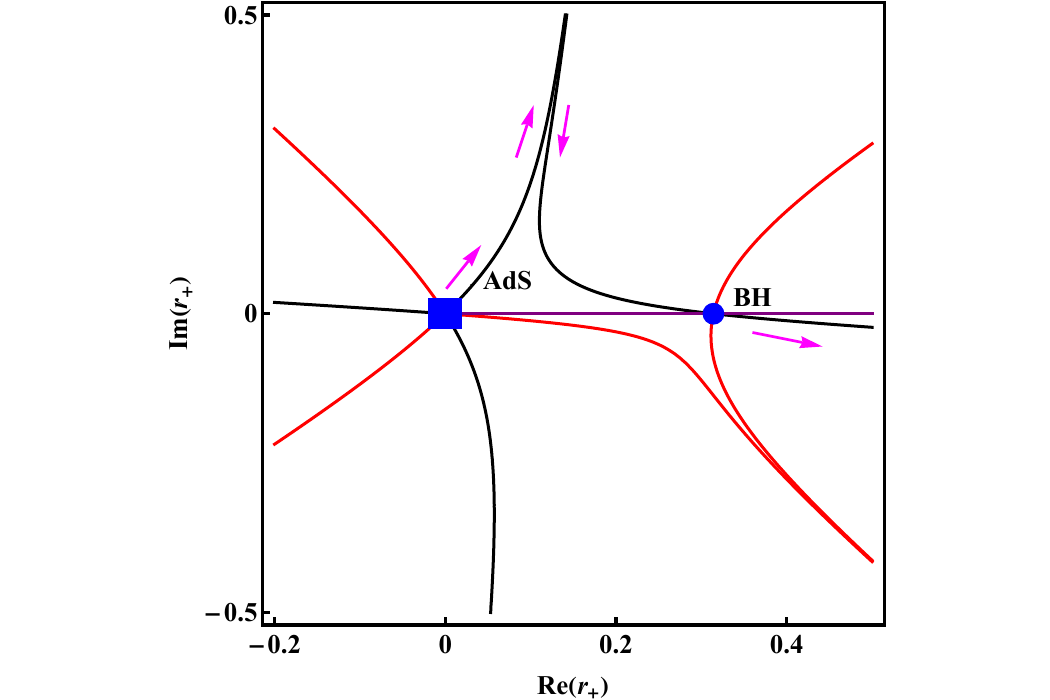}
}
\caption{The PL demonstration of extremal case $\Phi = \Phi_{\rm max}=1/\g$ for $AdS_{5}$ in grand canonical ensemble. For this, we set the following choice of parameters: $\Phi=\Phi_{\rm max}=\sqrt{3}/2, \beta =10$ and $G=e^{i\ep}$.}
\label{fig:PL_cs5}
\end{figure}
%

\subsubsection{Extremal case - $\Phi =\Phi_{\rm max}$}
\label{Ext_case}

For the extremal case, one finds the saddle solutions are the thermal AdS saddle (now with the degeneracy factor of $d-2$, as one of the black hole saddles coincides with thermal ads) and a black hole saddle with horizon at $r_{+} =  4\pi l^2/(d\beta)$ which exists for all temperatures ($0<\beta<\infty$). Since saddles are real, they lie on the Stokes ray, and as before, one can complexify $G$ and proceed with PL analysis (see fig. \ref{fig:PL_cs5}).  We find that both the saddles contribute to the partition function for either sign of $G$ - rotation.

%

\subsubsection{Thermodynamic instability and homology jump}\label{sec:instablity_jump}

\begin{figure}[htbp]
\centering
\subfigure[\label{fig:spe_heat_d_4}]{
    \includegraphics[trim={2.6cm 0 2.6cm 0cm}, clip, scale =0.6]{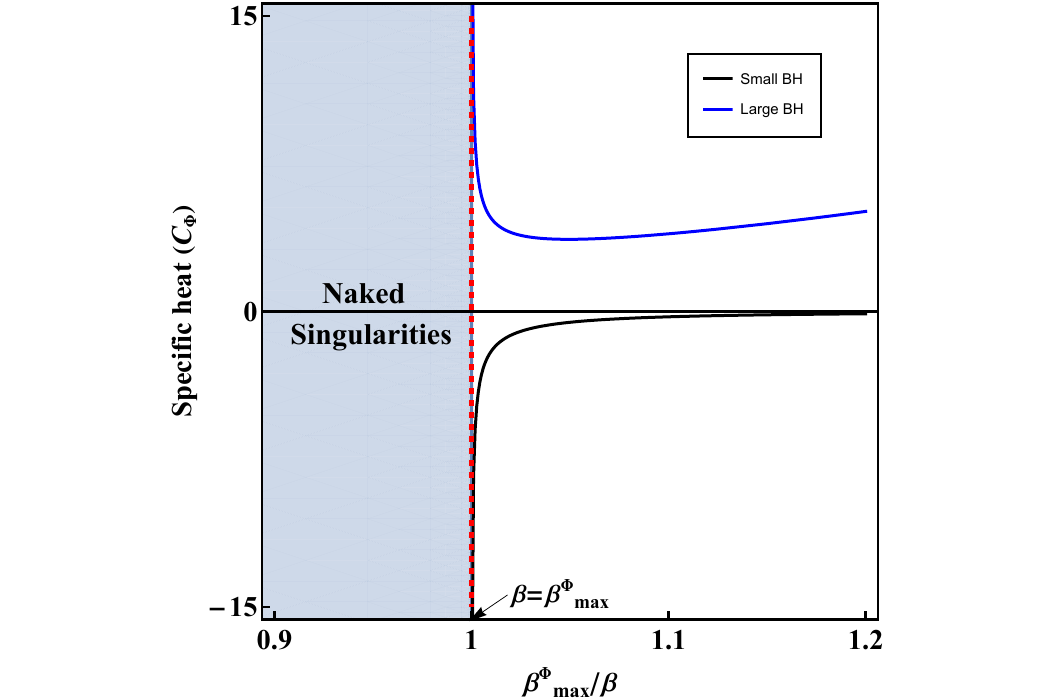}
}
\hspace{0.2cm}
\subfigure[\label{fig:sus_d_4}]{
    \includegraphics[trim={2.6cm 0 2.6cm 0cm}, clip, scale =0.6]{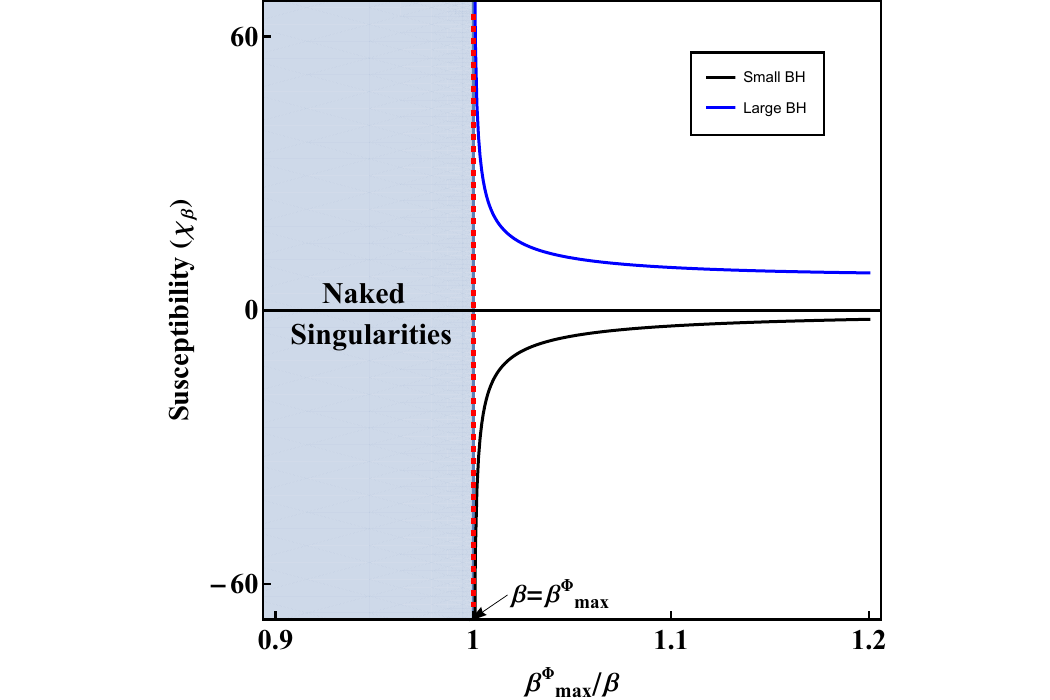}
}
\caption{This figure shows the behavior of specific heat capacity and susceptibility  $AdS_{5}$ black holes for $\Phi <\Phi_{\rm max}$ in the case of grand canonical ensemble at various temperatures. For illustration we choose $l=1, G= 1 $ and $\Phi =0.9 \Phi_{\rm max}$  with $\Phi_{\rm max} = \sqrt{3}/2$.}
\label{fig:thermo_quant}
\end{figure}

In this section, we examine the specific heat ($C_\Phi$) at constant potential ($\Phi$) and susceptibility ($\chi_{\beta}$) of a charged black hole, extending the analysis done in sec. (\ref{sec:phase_transition}) for the Schwarzschild black hole. We wish to understand how the unphysical behaviour of these thermodynamic quantities is related to the discontinuous change in the orientation of the thimbles/intersection number. The specific heat of the black hole at constant potential ($\Phi$) is given by:
\beq
\label{eq:spec_heat_def_charge}
C_{\Phi} =  T \biggl(\frac{\partial S}{\partial T}\biggr)_{\Phi} \, .
\eeq
Using the expression of $S(r_+)$ given in eq. (\ref{eq:phi_q}) and regularity at saddles $T(r_{+}) = f'(r_{+})/(4\pi)$, we get
\begin{equation}
    \label{eq:specific_heat_grand}
    C_\Phi(r_+^\pm)=\pm\frac{\omega (d-1) \beta_{\rm max}^{\Phi} (r_{+}^{\pm})^{d-1}}{4G\sqrt{(\beta_{\rm max}^{\Phi})^2 - \beta^2}} \,.
\end{equation}
In the regime $\beta<\beta_{\rm max}^\Phi$ and $\Phi < \Phi_{\rm max}$, $C_\Phi$ is positive for large black hole ($r_+^+$) and negative for small black hole $(r_+^-)$. Hence, the large black hole is thermodynamically stable, while the small black hole is unstable. At low temperatures $\beta > \beta_{\rm max}^{\Phi}$ and $\Phi < \Phi_{\rm max}$, $C_{\Phi}$ becomes complex, signifying that the corresponding solutions describe complex black holes or {\it naked singular} geometries, which are unphysical. This scenario has been demonstrated in the fig. (\ref{fig:spe_heat_d_4}). The specific heat $(C_\Phi)$ diverges, when $\beta=\beta_{\rm max}^\Phi$. For the small black hole branch, it diverges to $-\infty$ while for the large black hole branch, it diverges to $+\infty$. As discussed earlier, as one crosses the critical point $\beta=\beta_{\rm max}^\Phi$, the orientation of the Lefschetz thimbles changes discontinuously.

Similar to the AdS-Schwarzchild, charged small black holes are always subdominant than AdS and the large black hole, and hence always suffer from the Stokes jump. The Lefschetz thimbles associated with small black holes undergo a homology jump across the Stokes ray. Upon taking the homology average, their contribution is eliminated from the partition function. Similarly, in the regime $\Phi > \Phi_{\rm max}$, the specific heat $C_{\Phi}$ associated with the positive black hole saddle remains positive, indicating thermodynamic stability. In contrast, the negative black hole saddle exhibits a negative specific heat for odd $d$ and a positive specific heat for even $d$. When $d$ is odd, the negative saddle is subdominant with respect to both the AdS and the positive saddle and therefore exhibits a similar homology jump in the Lefschetz thimbles. As discussed earlier, the negative saddle corresponds to a {\it naked singular} geometry and, according to the Picard–Lefschetz technique, does not contribute to the partition function.

Let us now briefly examine the behaviour of another thermodynamic quantity- susceptibility. The susceptibility is given by:
\beq
\label{eq:sus_def}
\chi_{\beta} = \frac{\partial Q}{\partial \Phi}\biggr|_{\beta} \, .
\eeq
The susceptibility at the saddle points is given by:
\beq
\label{eq:sus_def_sad}
\chi_{\beta}(r_{+}) =  \frac{\omega (d-2) r_{+}^{d-3}}{4\pi G} \frac{d(d-1)\bt r_{+}^2 - 2(2d-3) \pi l^2 r_{+} + \bt l^2 (d-2)^2}{d \bt r_{+} - 2\pi l^2}  \, .
\eeq
It is easy to see from the above equation for $d\geq 3$, at high temperatures, the susceptibility for the small black hole $r_{+}^{-} < 2\pi l^2/(d\beta)$ is negative, while large black holes $r_{+}^{+} > 2\pi l^2/(d\beta)$ is positive. This indicates again that the small black hole is thermodynamically unstable, while the large black hole is stable. The susceptibility diverges when both black holes coalesce to one with horizon radius $r_+=2\pi l^2/d\beta$, which corresponds to $\beta=\beta_{\rm max}^\Phi$ where $C_\Phi$ also diverges. This situation is illustrated in the figure \ref{fig:sus_d_4}. At low temperatures, the susceptibility at the complex saddle solution is complex. However, the susceptibility of negative solution alter sign as we go through different dimensions. To see this explicitly, utilizing the saddle equation for negative saddle $r_{+} = - |r_{+}|$, and recast the $\chi_{\bt}$ in eq. (\ref{eq:sus_def_sad}) as:
\beq
\label{eq:sus_def_neg_sad}
\chi_{\beta}(r_{+}) = (-1)^{d} \frac{\omega (d-2) |r_{+}|^{d-3}}{4\pi G} \frac{d\bt |r_{+}|^2 + 2 \pi l^2 |r_{+}| + \bt \g^2 l^2 (d-2)^2 \Phi^2}{d\bt |r_{+}| + 2\pi l^2}  \, ,
\eeq
which manifestly shows the susceptibility of negative saddle is negative when $d$ is odd and positive for $d$ even, similar to the behavior of specific heat.


\subsection{Fixed charge: Canonical ensemble}
\label{fix_char}

We now focus on the situation where we fix the charge of the black hole instead of the electrostatic potential at infinity, a.k.a the canonical ensemble. For the fixed charge case, the partition function is given by 
\beq
\label{eq:par_char_can}
Z(\beta, q) = \int^\infty_{0^{+}} dS  \,\, \exp(S-\beta E) \, ,
\eeq
where $S(r_{+})$ and $E(r_{+},q)$ are entropy and energy mentioned in eq.(\ref{eq:phi_q}) respectively. In terms of $r_{+}$ variable (ignoring Jacobian), we get
\beq
\label{eq:Z_char_can_par}
Z(\beta, q) = \int_{\mathbb{D}} dr_{+}  \,\, \exp(\tilde{I}_{\rm CAdS}(r_{+})) \, .
\eeq
The integration contour $\mathbb{D}$ runs along $0^{+} < r_{+} < \infty$, that excludes the $r_{+} =0$ point, as it corresponds to the essential singularity of the exponent when $q\neq 0$, which also denotes a physical consequence that empty AdS ($r_+=0$) is no longer a solution of Einstein's equation in the presence of non-zero charge. The exponent for the canonical ensemble is given by 
\begin{equation}
    \label{eq:fix_charge}
  \tilde{I}_{\rm CAdS}(r_{+}) = \frac{\omega}{16\pi G} \left[4 \pi r_+^{d-1}-(d-1)\beta \biggl(r_+^{d-2}+\frac{r_+^d}{l^2}+\frac{q^2}{r_+^{d-2}}\biggr)\right].
\end{equation}
The dominant saddle configurations that extremize the exponent $\tilde{I}_{CAdS}(r_{+})$ of the integral in eq. (\ref{eq:Z_char_can_par}) are given by the saddle equation:
\begin{equation}
    \label{eq:sadd_eq_fix_charge}
    \frac{d}{(d-2)l^2}r_+^{2d-2}-\frac{4\pi}{(d-2)\beta}r_+^{2d-3}+r_+^{2d-4}-q^2=0.
\end{equation}
It is worth pointing out that in an arbitrary dimension $d$, any positive solution to the above equation corresponds to a regular, no–naked singular black hole. Since, at these positive saddles, the regularity condition $\beta = 4\pi/f'(r_{+})$, indeed the saddle point equation itself and the no–naked singularity condition
\begin{equation}
    \label{eq:no_naked_singularity_saddle}
    \frac{d}{(d-2)l^2}r_+^{2d-2}+r_+^{2d-4}=\frac{4\pi}{(d-2)\beta}r_+^{2d-3}+q^2>q^2 \,\,\text{for}\,\, r_{+}>0
\end{equation}
are satisfied. While, a negative saddle $r_{+} = - |r_{+}|$, although satisfies regularity condition, it corresponds to a naked geometry, as 
\begin{equation}
    \label{eq:no_naked_singularity_saddle_sim}
    \frac{d}{(d-2)l^2}|r_+|^{2d-2} + |r_+|^{2d-4}= q^2 - \frac{4\pi}{(d-2)\beta}|r_+|^{2d-3} < q^2 \,,
\end{equation}
the no-naked singularity condition is violated, and the same goes for the complex saddle points. Computing the analytic expressions of saddles in an arbitrary dimension $d$ is impossible as eq. (\ref{eq:sadd_eq_fix_charge}) becomes a higher-order polynomial. In the following, we focus on the $d=3$ ($AdS_{4}$) case, and $d=4,5$ in the extremely low temperature limit $\bt \to \infty$, where some analytic traceability is possible. Furthermore, we employed the Picard-Lefschetz technique to determine the relevance of saddle points.

\subsubsection{$AdS_4$}
\label{AdS_4}

\begin{figure}[hbtp]
    \centering
    \includegraphics[width=15cm, height=10cm]{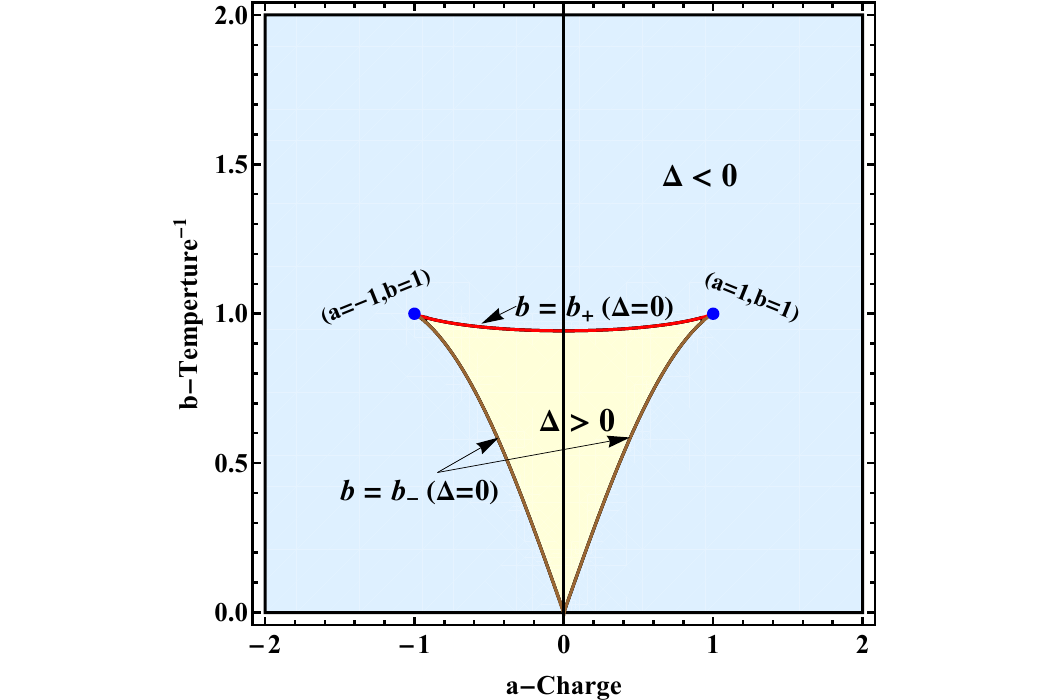}
    \caption{This plot illustrates the different regimes of the parameter $a$ (charge) and $b$ (inverse temperature), along with the nature of the discriminant ($\Delta$). 
    The blue region has complex roots, while the yellow region has all real roots, see
    table (\ref{tab:can_d_3_nature_of_horizon}).}
    \label{fig:a_b_regimes}
\end{figure}

We now proceed to analyze the canonical ensemble for $AdS_4$, for which the saddle point equation reads as
\beq
\label{eq:sad_eq_d_3}
3 \beta r^{4}_{+} - 4 \pi l^2 r^{3}_{+} + \beta l^2 r_{+}^2 - \beta l^2 q^2 = 0 \, .
\eeq
The above equation is quartic in nature and can be solved explicitly to determine the saddle points. However, significant insight about the nature of the saddles and their relevance can be obtained by simply analyzing the discriminant without explicitly computing the saddles. For convenience, let us introduce two variables, namely:
\beq
\label{eq:cri_var}
q_{c} = \frac{l}{6} \quad \text{and} \quad \beta_{c} = \sqrt{\frac{3}{2}}\pi l \, ,
\eeq
which correspond to the critical point satisfying the conditions $\partial\beta/\partial r_+=\partial^2\beta/\partial r_+^2=0$. We rescale the charge and inverse temperature parameters as $q = a\, q_{c}$ and $\beta = b\,\beta_{c}$, respectively. The discriminant of the saddle eq. (\ref{eq:sad_eq_d_3}) in these parameters is given by:
\beq
\label{eq:dis_d_3}
\Delta(a,b) = - \mathbb{C} a^2 b^2 q_{c}^{12} \beta_{c}^{6} \left[b^4 (a^2 + 3)^2 - 8b^2(3a^2+1) + 16a^2\right] \, ,
\eeq
where $\mathbb{C}$ is a positive real number. In the following, we analyze the nature of the roots for different parameter regimes of $a$ and $b$, see fig. \ref{fig:a_b_regimes}. 

\begin{figure}[hbtp]
\centering
\subfigure[$\,|q|>q_{c}$\label{fig:a_gr_1_b_ls_1}]{
    \includegraphics[trim={2.8cm 0 2.5cm 0cm}, clip, scale =0.55]{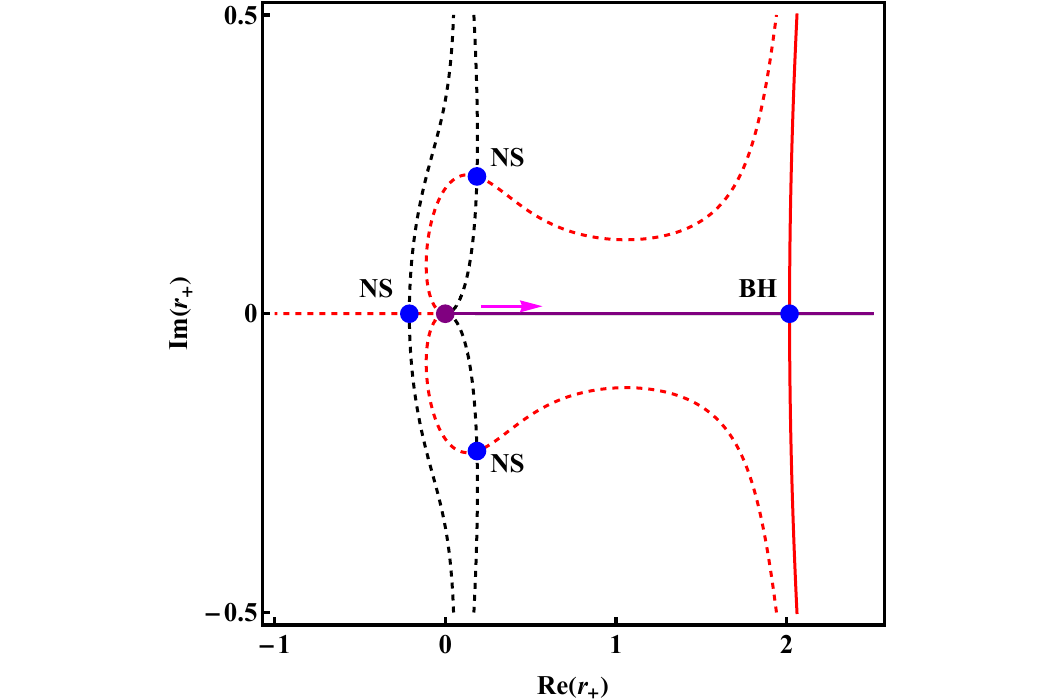}}
\hspace{0.5cm}
\subfigure[$\,|q|<q_{c}$\label{fig:a_gr_1_b_ls_bm}]{
    \includegraphics[trim={2.8cm 0 2.5cm 0cm}, clip, scale =0.55]{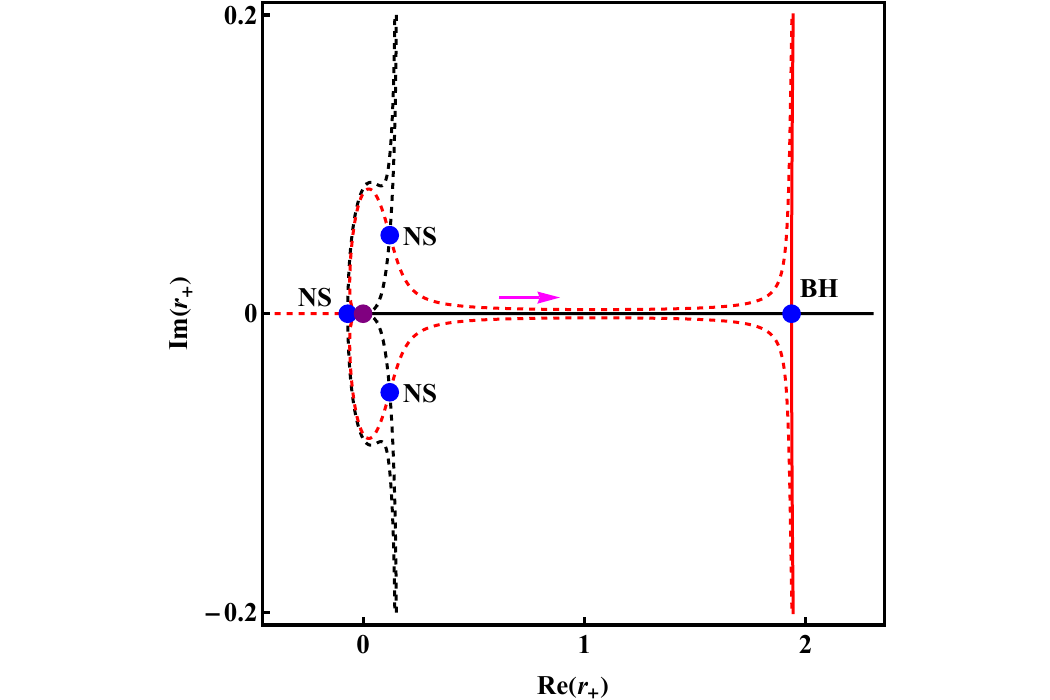}
}
\caption{The figures show PL plot for the case $\D<0$, with $\beta < \beta_{c}$, $l=1,b=1/2$ and $G=1$ with $a=2$ for the \textbf{left} panel, while $a=1/2$ for the \textbf{right} panel.  Blue dots represent the saddle points, the red/black lines denote the steepest descent/ascent thimbles. The purple dot at $r_{+} = 0$ is the singularity, and the purple line represents the original integration contour. Dashed line are used to denote flows lines associated with irrelevant saddles, while solid lines for relevant saddles. Clearly, the positive black hole (BH) saddle is relevant, and the negative/complex saddles correspond to {\it naked singularities} (NS) are irrelevant by PL.}
\label{fig:b_ls_1}
\end{figure}

\begin{figure}[ht]
\centering
\subfigure[$\, \ep =\pi/10$ \label{fig:a_gr_1_b_gr_1_pos}]{
    \includegraphics[trim={2.8cm 0 2.5cm 0cm}, clip, scale =0.55]{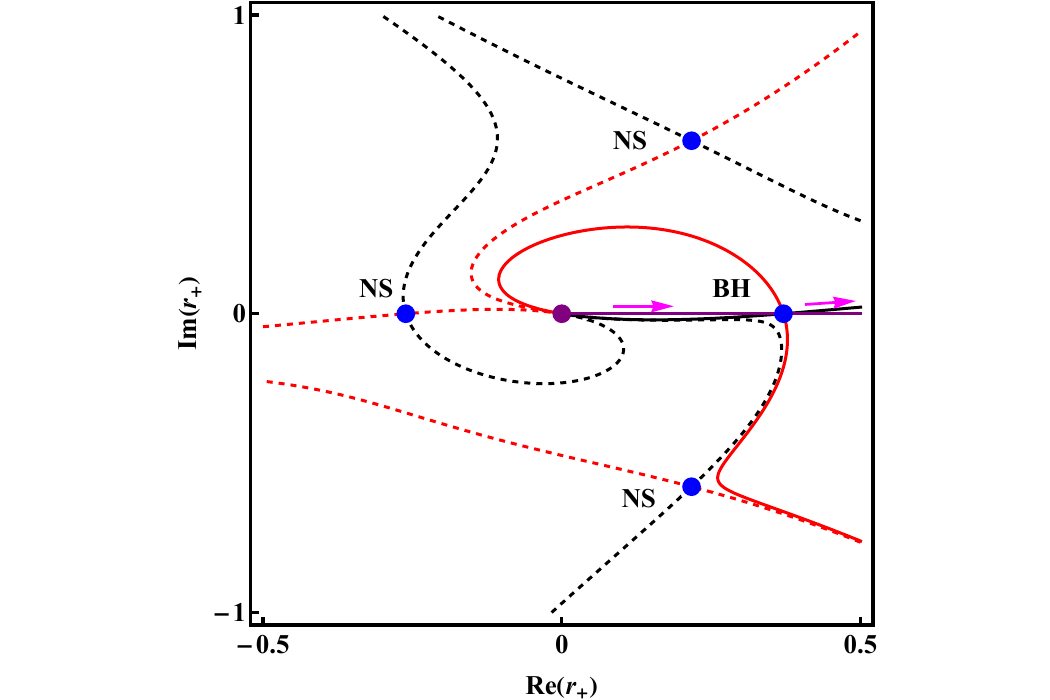}
}
\hspace{0.5cm}
\subfigure[$\, \ep =-\pi/10$\label{fig:a_gr_1_b_gr_1_neg}]{
    \includegraphics[trim={2.8cm 0 2.5cm 0cm}, clip, scale =0.55]{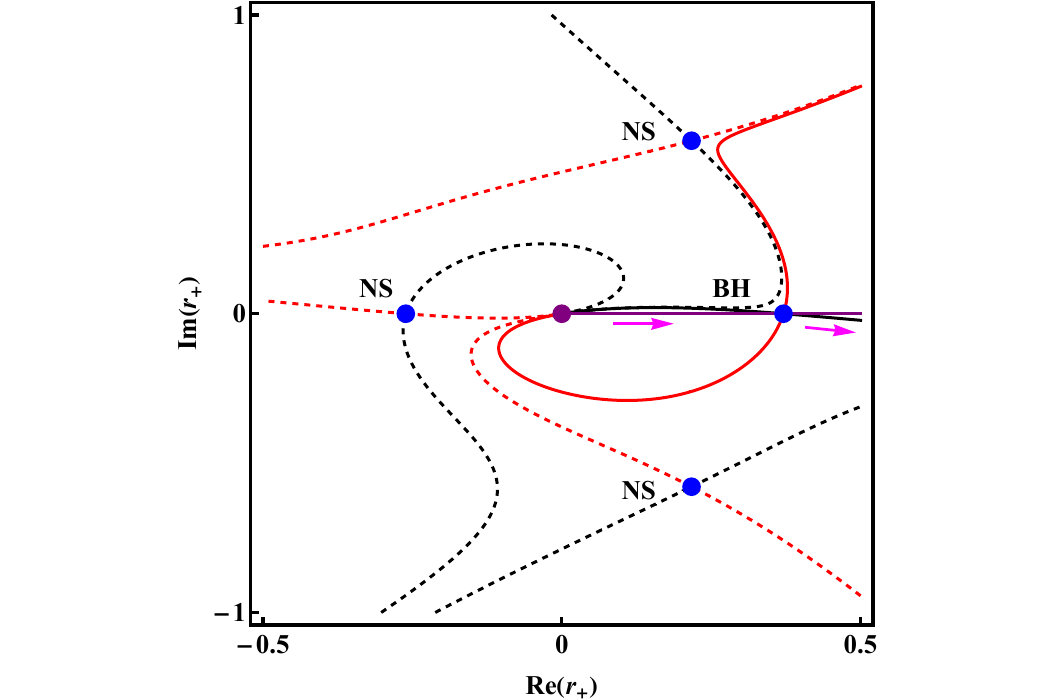}
}
\caption{A similar plot to the Fig. (\ref{fig:b_ls_1}), showing the lifting of stokes degeneracy for the case $|q| > q_{c}$ with $b>1$ where $\D<0$. We set $l=1, a =2, b=2 $ and $G=e^{i\ep}$.}
\label{fig:a_gr_1_b_gr_1}
\end{figure}

For $|a|\geq 1$, the discriminant remains negative for all values of $b \neq 1$, (see figure \ref{fig:a_b_regimes}), indicating the presence of one positive real saddle point, one negative real saddle point, and a pair of complex conjugate saddles. According to the Picard–Lefschetz method, the complex conjugate pair of saddles—corresponding to naked singular geometries— are irrelevant and do not contribute to the partition function, as they fail to satisfy the necessary condition, $\mathcal{H}(r_{+}^{s}) = 0$. The negative real saddle (also naked singularity)— and the positive saddle, both satisfy the necessary condition, $\mathcal{H}(r_{+}^{s}) = 0$. However, in this regime, it does not lead to any Stokes degeneracy (see figure \ref{fig:b_ls_1}) for $0<b<1$, but does for $1<b$. Similar to the earlier cases, we lift the Stokes degeneracy by complexifying the Newton constant. Lifting the degeneracy, we find that the negative saddle is irrelevant and does not contribute to the partition function. Meanwhile, the positive real saddle satisfies the necessary condition and lies on the defining contour, making it a relevant saddle according to the Picard–Lefschetz prescription. Therefore, for $|a|\geq 1$ at all temperatures the partition function gets contributions from only the regular and no-naked singular geometry. This situation is being illustrated in the figures (\ref{fig:b_ls_1}) and (\ref{fig:a_gr_1_b_gr_1}). 
%

For $|a|<1$, the discriminant can be positive or negative depending on $b$ (see fig \ref{fig:a_b_regimes}). Particularly, one notices that when $0 < b < b_{-} $ or $b > b_{+}$, the discriminant remains negative, leading to the same conclusion as in the $|a|>1$ case. The quantities $b_{\pm}$ are the roots of the discriminant $\D$, given by
\beq
\label{eq:bs_def}
b_{\pm} = \frac{2}{3+a^2}\bigl[1 + 3 a^2 \pm (1 - a^2)^{3/2}\bigr]^{1/2}\,
\eeq
with the associated temperatures defined as: $\beta_{\pm} = b_{\pm}\beta_{c}$. Here, $0<b_{-}<1$ and $2\sqrt{2}/3 < b _{+} < 1$ as $0 < |a| < 1$. At the critical point $|a|=1$ and $b=1$, or equivalently $q = |q_{c}| $ and $\beta = \beta_{c}$, where $\D$ vanishes, we have a positive triply degenerate saddle solution $r_{c}$, given by
\beq
\label{eq:critcal_horizon_rad}
r_c = l/\sqrt{6}\, ,
\eeq
and one negative real solution $r_{\rm neg} = -l/(3\sqrt{6})$. Applying PL analogous to the earlier case, we obtain that the negative saddle is irrelevant as it lies outside defining contour. Whereas the triply degenerate positive saddle satifies the necessary condition and lies on the defining contour as shown in the figure (\ref{fig:a_eq_1_b_eq_1}). Therefore, the triply degenerate positive black holes is relevant and contribute to the partition function. Similarly, along the curves $b = b_{+}$ or $b = b_{-}$ with $|a| < 1$, the discriminant vanishes. However, in these cases, unlike the critical case, one finds a real negative saddle, a real positive saddle, and a doubly degenerate positive saddle. All of them satisfying the necessary condition $\mathcal{H}(r_{s}^{+}) = 0$, with $\mathfrak{h}(r_{+}^{s}) \neq \mathfrak{h}(r_{+}^{s^{\prime}})$ for $s\neq s'$, denoting the stokes degeneracy situation. Upon complexifying the Newton constant and applying the Picard–Lefschetz analysis, one finds that both distinct positive black hole saddles are relevant, whereas the negative horizon solution is irrelevant. The doubly degenerate saddle ($r_{d}$) in either the cases $b=b_{+}$ or $b=b_{-}$ is given by
\beq
\label{eq:dbly_deg_sad}
r_{d} = \frac{r_c}{b_{\pm}} (1 - \sqrt{1-b_{\pm}^2} ) \, ,
\eeq
respectively. Furthermore, the other two solutions in terms of $r_{d}$ read as follows:
\beq
\label{eq:other_sads}
\begin{split}
r_{1} & = (2\pi l^2 - r_{d}) - \sqrt{(2\pi l^2 - r_{d})^2 + b_{\pm} \beta_c l^2 q^2/r_{d}^2} \, ,\\
r_{2} & = (2\pi l^2 - r_{d}) + \sqrt{(2\pi l^2 - r_{d})^2 + b_{\pm} \beta_c l^2 q^2/r_{d}^2} \, .
\end{split}
\eeq
\begin{figure}[htbp]
\centering
\subfigure[$\, \ep =\pi/10$\label{fig:a_ls_1_b_eq_bm_pos}]{
    \includegraphics[trim={2.8cm 0 2.5cm 0cm}, clip, scale =0.55]{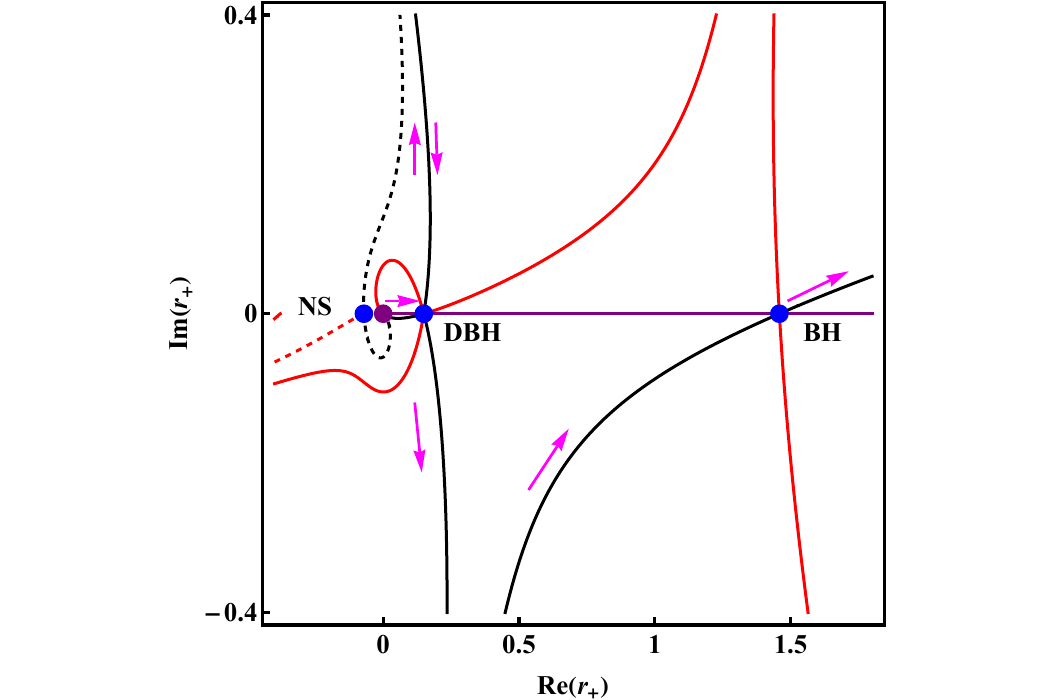}
}
\hspace{0.5cm}
\subfigure[$\, \ep =-\pi/10$\label{fig:a_ls_1_b_gr_bp_neg}]{
    \includegraphics[trim={2.8cm 0 2.5cm 0cm}, clip, scale =0.55]{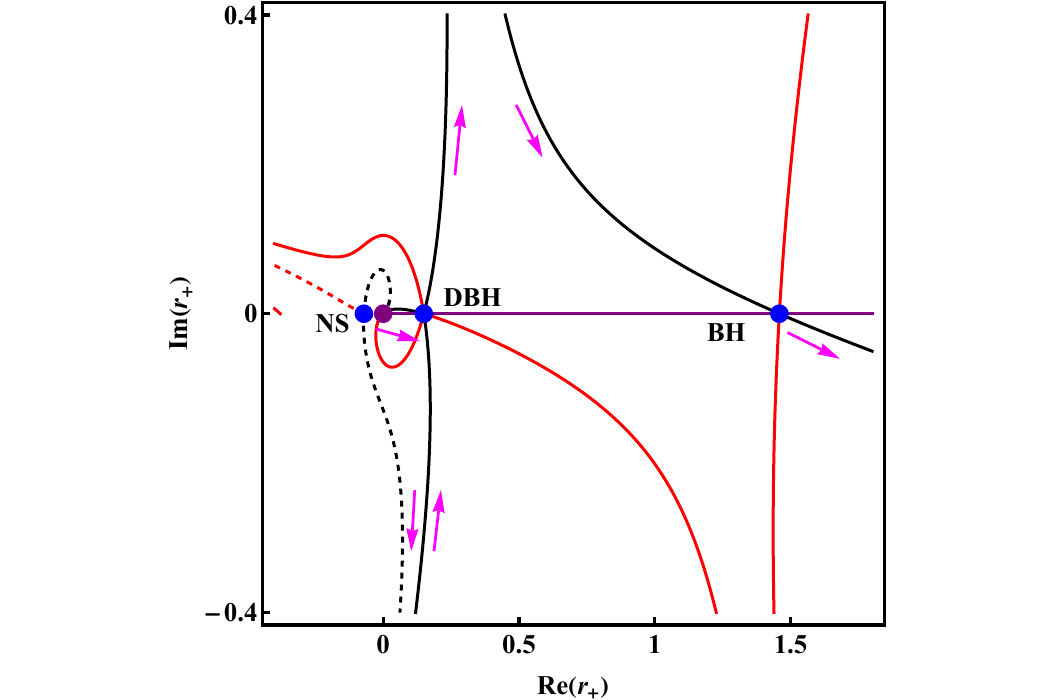}
}
\caption{ The Picard Lefschetz plot for the degenerate case where $\D =0$, for $|q| < q_c$ with $\bt=\bt_{-}$. For this case, we have a negative saddle (NS), a positive saddle (BH) and a distinct double degenerate positive saddle (DBH). The Stokes degeneracy in this case is lifted by complexifying the Newton constant $G =e^{i\ep}$. The parameters for this plot are as follows: $a=1/2, b=b_{-} = 0.6456$. }
\label{fig:a_ls_1_b_eq_bm}
\end{figure}
\begin{figure}[hbtp]
\centering
\subfigure[$\, \ep =\pi/10$\label{fig:a_eq_1_b_eq_1_pos}]{
    \includegraphics[trim={2.8cm 0 2.5cm 0cm}, clip, scale =0.55]{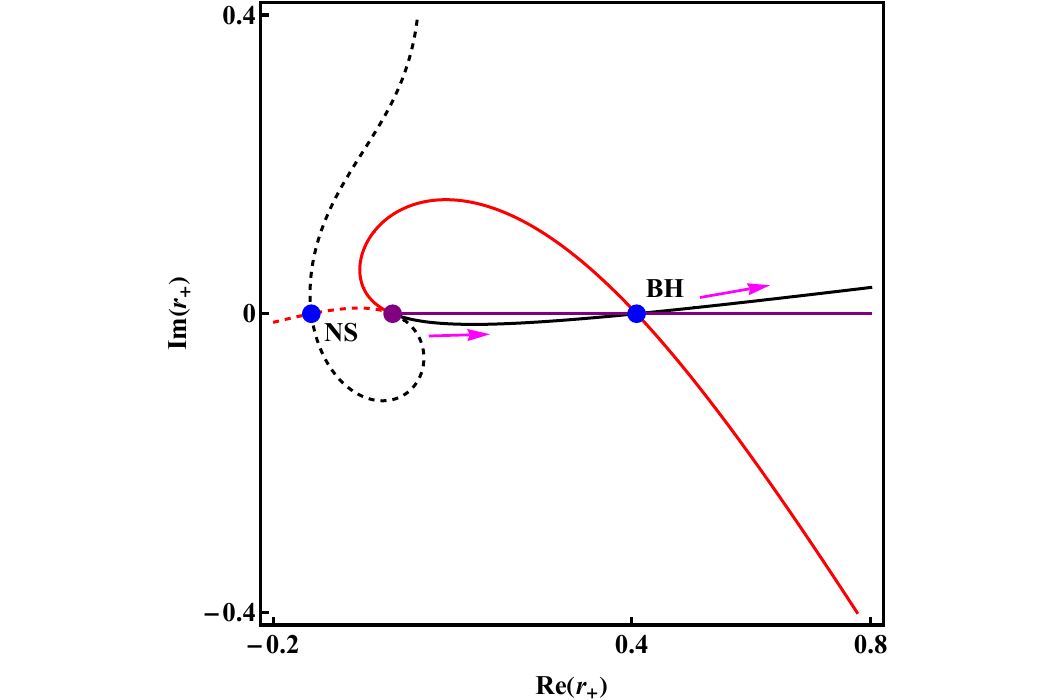}
}
\hspace{0.5cm}
\subfigure[$\, \ep =-\pi/10$\label{fig:a_eq_1_b_eq_1_neg}]{
    \includegraphics[trim={2.8cm 0 2.5cm 0cm}, clip, scale =0.55]{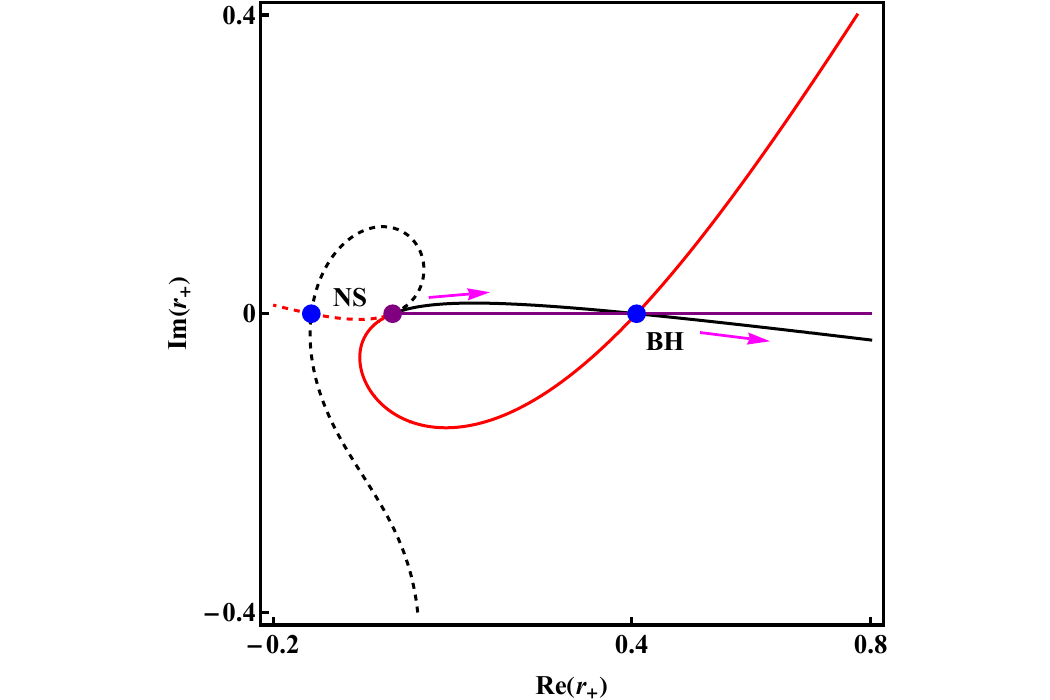}
}
\caption{ A similar plot to figs. (\ref{fig:a_ls_1_b_eq_bm})  at the critical point $|q| = q_c$ with $\beta = \beta_c$ for which we have $\D=0$. For this case, we get one negative saddle (NS) and another triply degenerate positive black hole saddle (BH). The parameters for this plot are as follows: $a=1/2, b=1$. }
\label{fig:a_eq_1_b_eq_1}
\end{figure}
It is clearly evident that $r_{1}$ is the negative saddle, while $r_2$ is the positive saddle. This Picard Lefschetz analysis for $b=b_{-}$ is shown in the figure (\ref{fig:a_ls_1_b_eq_bm}).
\begin{figure}[htbp]
\centering
\subfigure[$\, \ep =\pi/10$\label{fig:a_ls_1_bm_b_bp_pos}]{
    \includegraphics[trim={2.8cm 0 2.5cm 0cm}, clip, scale =0.55]{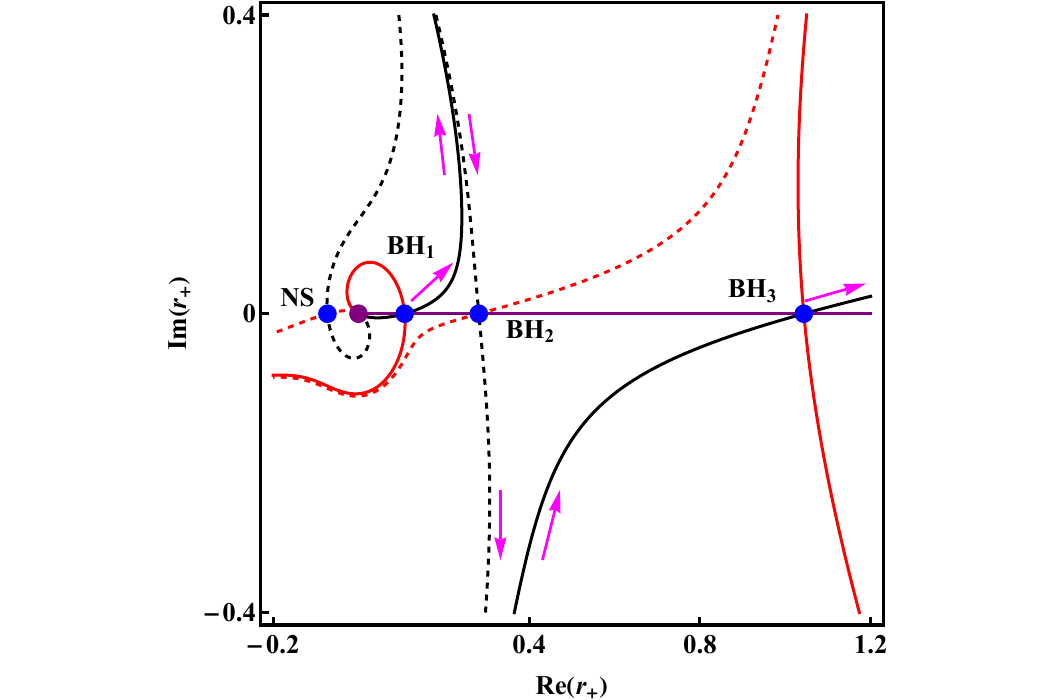}
}
\hspace{0.5cm}
\subfigure[$\, \ep =-\pi/10$\label{fig:a_ls_1_bm_b_bp_neg}]{
    \includegraphics[trim={2.8cm 0 2.5cm 0cm}, clip, scale =0.55]{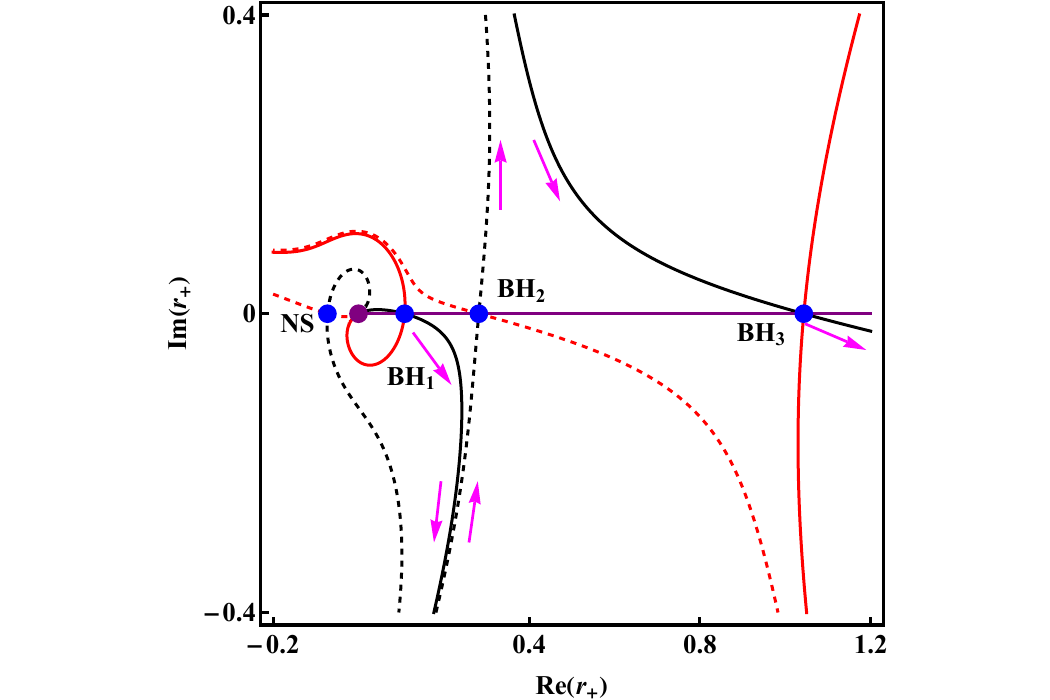}
}
\caption{ PL plot for the $|q|<q_{c}$ and $\bt_{-}<\bt <\bt_{+}$ case, where the discriminant $\D>0$. In this case, we have a negative saddle denoted by NS, while the three distinct positive saddles $r_{+,1},r_{+,2},r_{+,3}$ are represented by BH$_1$, BH$_2$ and BH$_3$, respectively. The flows associated with the negative saddle and the positive saddle that suffer Stokes jump are represented using dashed lines, while flow lines associated with contributing saddles are denoted by solid lines. For this illustration we choose $a=1/2, b=0.8$, and $G=e^{i\ep}$ where $b_{-}\approx0.6456$ and $b_{+}\approx 0.95325$.  
}
\label{fig:a_ls_1_bm_b_bp}
\end{figure}
For $b_{-}< b <b_{+}$, one can convince themselves from the plot that $\Delta > 0$, implying that the equation admits four real saddle solutions—either one positive and three negative saddles, or three positive and one negative saddle, depending on the value $b$ takes. However, according to Descartes' rule of signs, which states that in a regime where all the roots are real, it can have only one negative saddle and three positive saddles, one finds that the latter case occurs. Again we encounter the Stokes degeneracy, which is resolved by complexifying the Newton constant. Let $r_+^{-}$ denote the negative saddle and $r_{+,1}$, $r_{+,2}$, and $r_{+,3}$ denote the three positive solutions satisfying $r_{+,1} < r_{+,2} < r_{+,3}$ for $|a|<1$ and $b_{-}< b <b_{+}$ case. Employing the Picard–Lefschetz technique, we find that all positive saddles are relevant for either sign of $\ep$, see fig. (\ref{fig:a_ls_1_bm_b_bp}). However, the intermediate saddle $r_{+,2}$ undergoes a Stokes jump and drops upon taking homology averaging as discussed in sec. (\ref{Homology}). 
\begin{table}[H]
    \centering
    \begin{tabular}{|c|c|c|}
    \hline
        \text{Range of $q$} &\text{Range of $\beta$}  &\text{nature of the $r_+$ horizon}\\
        \hline
        $|q|\geq q_c$ & $\beta \neq \beta_{c}$ & one complex conjugate pair, one positive and a negative solution \\
         \hline
         $|q| < q_{c}$ & $\beta < \beta_{-}$ or $\beta > \beta_{+}$ & one complex conjugate pair, one positive and a negative solution \\
         \hline
         $|q|<q_c$ & $\beta_{-} < \beta < \beta_{+}$ & one negative and three positive solutions\\         
         \hline
        $|q| < q_c$ & $\beta = \beta_{-}$ or $\beta = \beta_{+}$  & one negative, one positive and a doubly degenerate positive solution\\
        \hline
         $|q| = q_c$ & $\beta = \beta_c$ & one negative and a triply degenerate positive solution\\         
         \hline
    \end{tabular}
    \caption{Various ranges of the free parameters $(q,\beta)$ and the corresponding nature of the solutions in case of canonical ensemble.}
    \label{tab:can_d_3_nature_of_horizon}
\end{table}
To summarize, we find that in various regimes of $(q,\beta)$, as mentioned in table. \ref{tab:can_d_3_nature_of_horizon} complex/negative saddles describing naked singularity are always irrelevant by PL and don't contribute to the partition function. Only real and positive solutions contribute to the partition function, with the single exception of the $r_{+,2}$ saddle, whose contribution vanishes upon homology averaging.

%

%

%

%


%

\subsubsection{$AdS_5$ and $AdS_6$}
\label{AdS_56}

\begin{figure}[htbp]
\centering
\subfigure[$AdS_{5}$\label{fig:AdS_5}]{
    \includegraphics[trim={2.8cm 0 2.5cm 0cm}, clip, scale =0.55]{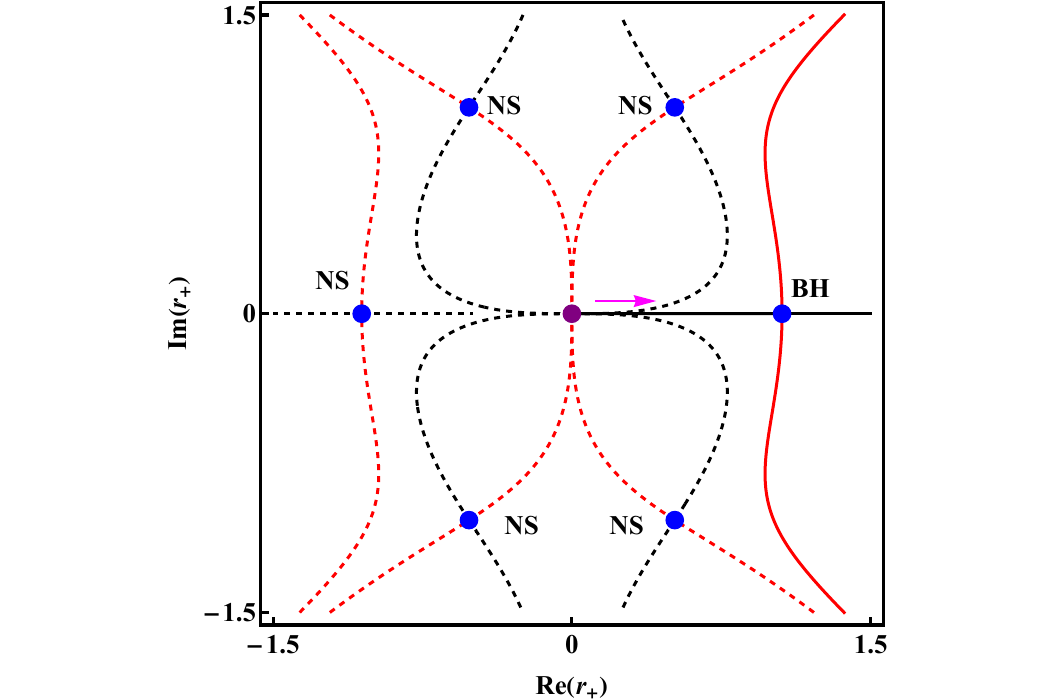}
}
\subfigure[$AdS_{6}$\label{fig:AdS_6}]{
    \includegraphics[trim={2.8cm 0 2.5cm 0cm}, clip, scale =0.55]{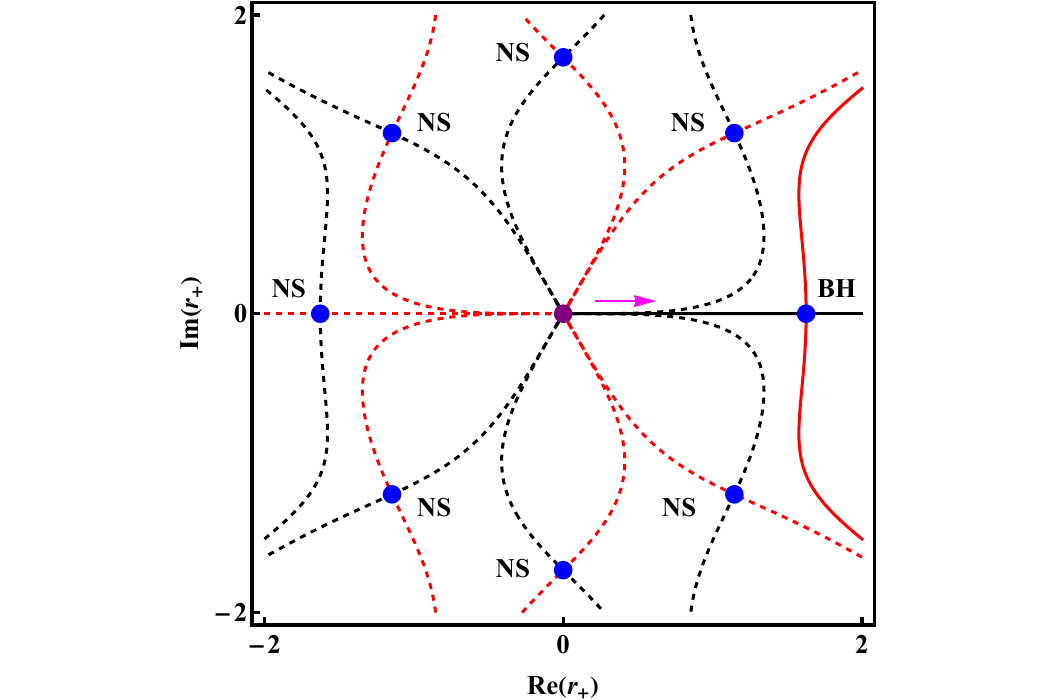}
}

\caption{This figure demonstrates the PL analysis of $AdS_{5}$ and $AdS_{6}$ in the large $\beta$ limit. We set the parameters as: $l=1,\bt=10^{5}$ and $G=1$ for both case, while $q=2$ for $AdS_{5}$ and $q=10$ for $AdS_{6}$ case.}
\label{fig:AdS_56}
\end{figure}

In the limit $\beta \to\infty $ the odd term $r_+^{2d-3}$ in the saddle eq. (\ref{eq:sadd_eq_fix_charge}) can be dropped, that excites for an analytic understanding about the saddle geometries of $AdS_5$ and $AdS_6$ for arbitrary $q$. In this limit, the saddle equation reduces to
\begin{equation}
    \label{eq:saddle_eq_d_45}
    \begin{split}
        \frac{2}{l^2}r_+^{6}+r_+^{4}-q^2=0 \quad \text{and}\quad \frac{5}{3l^2}r_+^{8}+r_+^{6}-q^2=0\,.
    \end{split}
\end{equation}
for $d=4$ and $d=5$ respectively. Defining a new variable $x=r_+^2$, these equations can be made cubic and quartic in $x$ given by:
\begin{equation}
    \label{eq:saddle_eq_d_4_d_5}
    \begin{split}
        \frac{2}{l^2}x^{3}+x^{2}-q^2=0 \quad \text{and}\quad \frac{5}{3l^2}x^{4}+x^{3}-q^2=0.\quad .
    \end{split}
\end{equation}
for $d=4$ and $d=5$ respectively. 
\begin{figure}[htbp]
\centering
\subfigure[\,\,\, $q<q_{\rm crit}$ ]{
\includegraphics[width=0.3\textwidth]{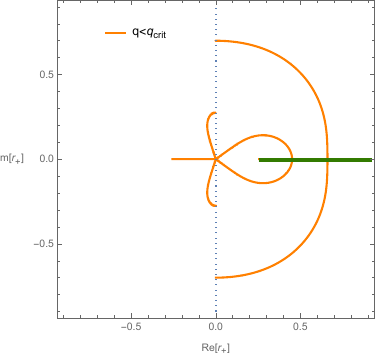}
}
\subfigure[ \,\,$q=q_{\rm crit}$ ]{
  \includegraphics[width=0.3\textwidth]{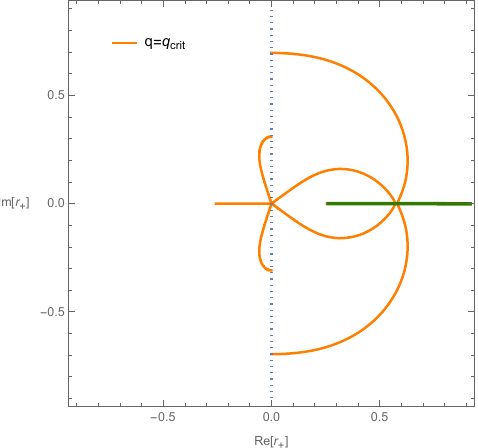}
}
\subfigure[$\,\,q>q_{\rm crit}$]{
\includegraphics[width=0.3\textwidth]{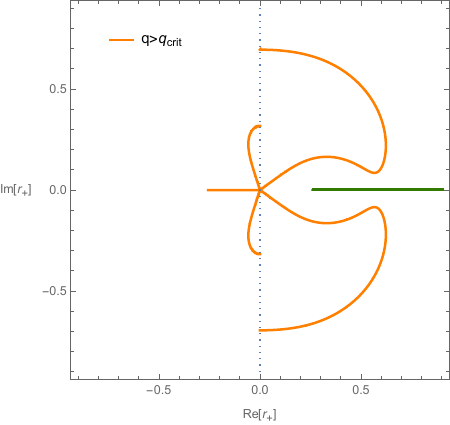}
}
\caption{Nature of the $r_+$- saddles for various ranges of $q$ in $AdS_5$ as we vary $\beta$ (keeping real and positive), $q_{\rm crit}=l^2/3\sqrt{15},l=1$. For each point on the curve, one satisfies the regularity criterion $\beta=4\pi/f'(r_+)$. All the geometries lying on the orange curve are naked singular geometries (disallowed by PL), whereas points on the green curve describe a black hole with a horizon (allowed by PL). 
}
\label{fig:stokes_jump_1}
\end{figure} 
Similar to the $AdS_{4}$, we proceed with the same approach of analyzing the discriminant of the above equation, which serves as a good candidate to identify the nature of the saddle point solutions (The nature of the saddles for arbitrary $\beta > 0$ with various regimes of $q$ is illustrated in Fig.(\ref{fig:stokes_jump_1})). For the quartic equation (the second equation in \ref{eq:saddle_eq_d_4_d_5}), the discriminant $\D$ is found to be negative for any value of $q$. This implies that, in the extremely low temperature limit, i.e., $\beta \to \infty$, $AdS_{6}$ has one real positive saddle, a real negative saddle and the rest are complex conjugate pairs. Implementing the Picard Lefschetz technique, we obtain that only the real positive saddle is relevant and contributes to the partition function. However, unlike $AdS_{6}$ case, the sign of $AdS_{5}$ discriminant depends on the value of charge $q$. To be precise, the discriminant of the cubic equation $\D$ (first equation in \ref{eq:saddle_eq_d_4_d_5}) is given by:
\beq
\D = l^4 q^2 (l^4 - 27 q^2)/4 \, .
\eeq
For $q^2 < l^2/27$, the above discriminant is negative, while for $q^2 > l^2/27$ it is positive and vanishes for $q^2 = l^2/27$. Irrespective of the discriminant sign, there always exists one real positive saddle, one real negative saddle, and the rest are complex conjugate saddles. The nature of  Among these, only the real positive saddle is the relevant and contributing saddle to the partition function. These situations are illustrated in the figure (\ref{fig:AdS_56}).

For finite $\beta$ , one expects more real roots and positive roots to appear. In general, the saddle eq. (\ref{eq:sadd_eq_fix_charge}) can have a maximum of three real and positive roots ($\tilde{r}_{+,1},\tilde{r}_{+,2},\tilde{r}_{+,3}$). Similar to the $AdS_4$, these real roots will lie at $\tilde{r}_{+,1}<\tilde{r}_{+,2}<\tilde{r}_{+,3}$. In an exactly similar argument, the saddle $\tilde{r}_{+,2}$ will be subdominant, and the Lefschetz thimble ($\mathcal{J}(\tilde{r}_{+,2})$) corresponding to that saddle will show the Stokes jump.
\subsubsection{Homology averaging of Lefschetz Thimbles}
\label{Homology}
\begin{figure}[hbt!]
    \centering
    \includegraphics[width=11.5cm, height=8.5cm]{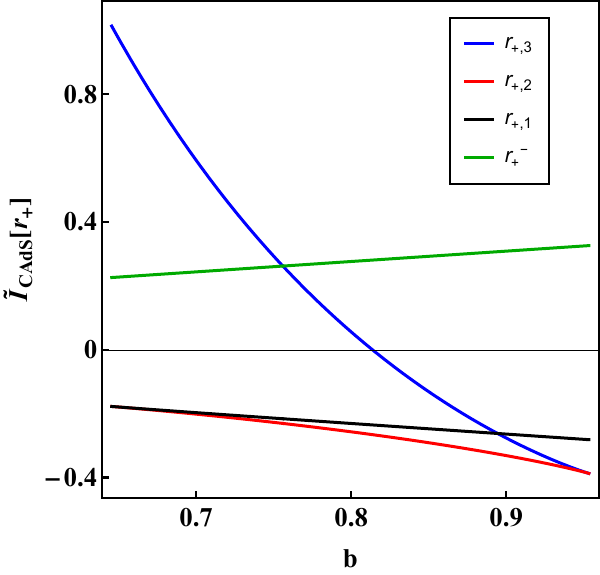}
    \caption{Plot of the canonical partition function exponent for various saddle points in the case $q = 1/2$, shown for different values of $b$ in the range $b_{-} < b < b_{+}$, where $b_{-} \approx 0.6456$ and $b_{+} \approx 0.95325$. The plot clearly demonstrates that the intermediate saddle $r_{+,2}$ (red curve) is always subdominant compared to the other saddles.
   }
    \label{fig:can_exp}
\end{figure}

Similar to the earlier analysis of the AdS Schwarzschild black hole in sec. (\ref{sec:real_saddle_relevance}), in this section we discuss about the Stokes jump and homology averaging for the canonical ensemble. We notice when one fixes the charge; for the range of parameters $|q|<q_c$ and $\beta_{-} < \beta < \beta_{+}$, there are four saddles, specifically: one negative ($r_+^-$ ) and three positive solutions ($r_{+,1}<r_{+,2}<r_{+,3}$). Picard-Lefschetz analysis reveals the existence of a Stokes line among these saddles. To break this Stokes degeneracy, we complexify $G=|G|e^{i\epsilon}$ which leads to the discontinuous jump in the thimbles for $\ep>0$ and $\ep<0$ as shown in the figure (\ref{fig:a_ls_1_bm_b_bp}). The homology jump in thimbles associated with the saddles can be expressed as
\begin{equation}
\label{eq:homology_jump_four_saddle}
\left(\begin{array}{c}
     \mathcal{J}(r_+^-)  \\
      \mathcal{J}(r_{+,1})\\
    \mathcal{J}(r_{+,2}) \\
    \mathcal{J}(r_{+,3})
\end{array}\right)\rightarrow\left(\begin{array}{c|ccc}
        1 & 0 &1 &0\\
        \hline
       0  & 1 &1 &0\\
       0  &0 &1 &0\\
        0 &0 & 1&1
\end{array}\right)\left(\begin{array}{c}
     \mathcal{J}(r_+^-)  \\
      \mathcal{J}(r_{+,1})\\
    \mathcal{J}(r_{+,2}) \\
    \mathcal{J}(r_{+,3})
\end{array}\right),\quad \text{for}\,\quad \ep^+ \rightarrow\ep^-\, .
\end{equation}
Since the negative saddle $r_+^-$ is irrelevant and doesn't contribute to the integral cycle, we consider only the lower $3\times 3$ matrix in eq. (\ref{eq:homology_jump_four_saddle}, which is exactly the same as obtained earlier in eq. (\ref{eq:stokes_jump_matrix_J}).  Hence, the analysis will proceed in the same way as in the case of the high-temperature AdS Schwarzschild case. Note that the thimble $\mathcal{J}(r_{+,2})$ remains unchanged across the jump, as it is subdominant compared to the other saddles (see figure \ref{fig:can_exp}). Therefore, the contour of integration $\mathbb{D}\in (0^+,\infty)$ can be expressed homologically two different ways as
\begin{equation}
\label{eq:homology_jump_canonical}
    \begin{split}
\mathbb{\mathbb{D}}=&\mathcal{J}(r_{+,1})-\mathcal{J}(r_{+,2})+\mathcal{J}(r_{+,3}) \hspace{3mm} \text{for $\ep >0$} \\
    =&\mathcal{J}(r_{+,1})+\mathcal{J}(r_{+,2})+\mathcal{J}(r_{+,3}) \hspace{3mm} \text{for $\ep <0$}.
    \end{split}
\end{equation}
It is important to note that there is no $1/2$ factor in front of $\mathcal{J}(r_{+,1})$ as we consider the entire thimble associated with $r_{+,1}$. This difference comes because of an essential singularity at $r_+=0$ in the action when $q\neq 0$ and the thimble ends there. The intersection number for the intermediate black hole saddle ($r_{+,2}$) is different for the two signs of $\epsilon$. In an exactly similar argument presented in sec. (\ref{sec:real_saddle_relevance}), upon doing the homology averaging, the contribution from the $r_{+,2}$ saddle (sub-leading) drops out. 

Furthermore, we find other stokes degeneracy situations: $|q|\leq q_c$ and $\beta = \bt_{\pm}$ or $\beta = \beta_{c}$, where we observe the merging of saddle points. The homology analysis for this case, goes exactly similar way as explained in Sec. (\ref{sec:real_saddle}). One finds the Lefschetz thimbles undergo a discontinuous change in their thimble structure across the degenerate point. However, for either choice of complex deformation angle we find all the positive black hole solutions are relevant.

\subsubsection{Thermodynamic stability analysis}
\label{sec:canonical_sp_heat}

Similar to earlier cases, we examine the specific heat at fixed charge to analyse the thermodynamic stability of black holes. The specific heat at constant charge for the black hole saddles reads as
\beq
\label{eq:spe_heat_can}
C_{q}(r_+)= \frac{\omega (d-1)\pi}{2G} \frac{l^2 r_{+}^{d}}{d(d-1)\bt r_{+}^2 - 2\pi(2d-3)l^2 r_{+} + (d-2)^2 l^2 \bt} \, .
\eeq
Analogous to the grand canonical ensemble, the specific heat at fixed charge becomes complex when $r_{+}$ is complex, corresponding to unphysical naked singular geometries. For real positive black hole saddles, the sign of the denominator determines thermodynamic stability: a positive denominator yields $C_q > 0$, corresponding to a stable black hole, whereas a negative denominator leads to $C_q < 0$, indicating instability. The $C_{q}$ diverges when the denominator vanishes, which happens for the critical point $|q|=q_c$ and $\beta =\beta_c$.

Meanwhile for real negative black hole saddles, $r_{+} = -|r_{+}|$, the specific heat at constant charges becomes:
\beq
\label{eq:spe_heat_can_neg_bh}
C_{q} = (-1)^d \frac{\omega (d-1)\pi}{2G} \frac{l^2 |r_{+}|^{d}}{d(d-1)\bt |r_{+}|^2 + 2\pi(2d-3)l^2 |r_{+}| + (d-2)^2 l^2 \bt} \, .
\eeq
As we are interested in $d\geq 3$, it is evident from the above expression that the negative black hole saddle exhibits a negative specific heat when $d$ is odd and a positive specific heat when $d$ is even, analogous to the regime, $\Phi > \Phi_{\rm max}$ in the grand canonical. Particularly, for $d=3$, the expression for the specific heat reads as:
\beq
\label{eq:spe_heat_can_d_3}
C_{q}= \frac{\omega\pi}{G} \frac{l^2 r_{+}^{3}}{6\bt r_{+}^2 - 6\pi l^2 r_{+} +  l^2 \bt} \, .
\eeq
From the above discussion, it is clear that the specific heat associated with the negative real saddle point, which appears in various regimes, is negative, whereas the specific heat of the complex saddles is complex. Now, we proceed to examine the behaviour of specific heat at the positive real saddles in various regimes by studying the sign of the denominator of $C_q$.

\begin{figure}
    \centering
    \includegraphics[trim={2.6cm 0 2.6cm 0cm}, clip, scale=0.75]{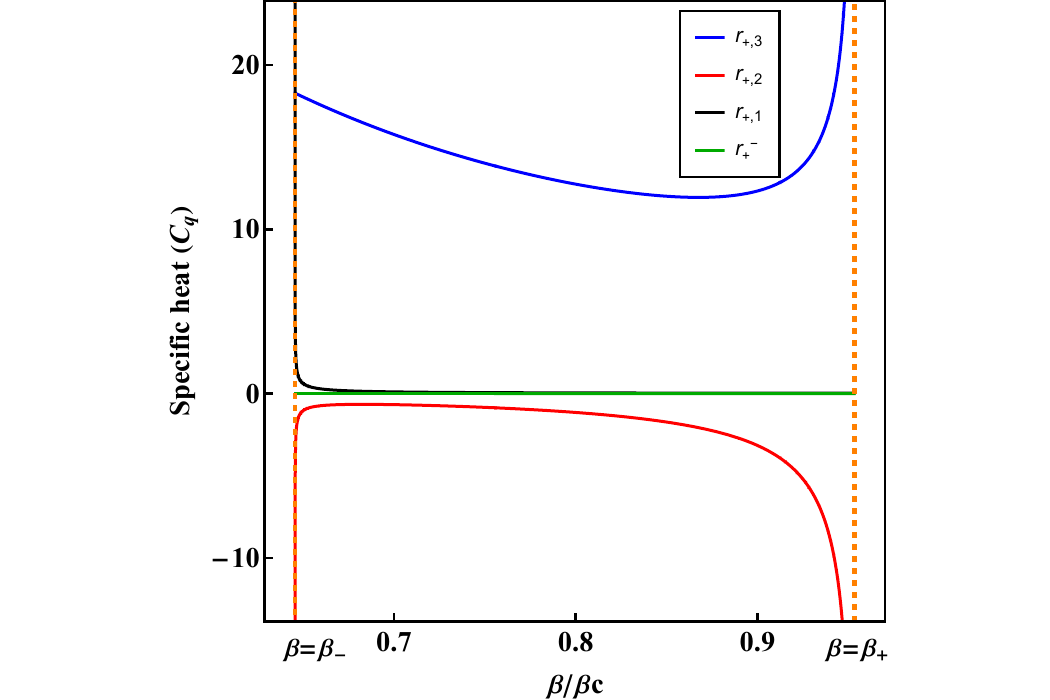}
    \caption{A plot of specific heat capacity at fixed charge of $AdS_{4}$ black holes for $|q| <q_c$ and $\bt_{-}<\beta < \beta_{+}$. For illustration we choose $l=1, G= 1 $ and $ q =  q_c/2$. As explained in the text, the subdominant saddle $r_{+,2}$ has negative specific heat.}
    \label{fig:can_spec}
\end{figure}

Simply, by examining the nature of saddle solutions and with a little bit of algebra, one may determine the sign of specific heat at the positive saddle solution. For $|q|\geq q_c$, at any temperature except at the critical temperature and $|q|< q_c$ with either $\beta <\beta_{-}$ or $\beta>\beta_{+}$, one finds a positive real black hole that has positive specific heat, indicating thermodynamic stability and is the sole relevant contributing saddle to the partition function. For $|q| < q_c$ we have three positive black hole solutions with horizon radii $r_{+,1}$, $r_{+,2}$, and $r_{+,3}$, satisfying $r_{+,1} < r_{+,2} < r_{+,3}$, all of which are relevant. We notice that the specific heat of small $r_{+,1}$ and large $r_{+,3}$ black holes are positive, while intermediate black holes $r_{+,2}$ acquire a negative specific heat (see figure \ref{fig:can_spec}), indicating thermodynamic instability \cite{Chamblin:1999tk, Ladghami:2024wkv}. Nevertheless, the contribution of $r_{+,2}$ to the partition function drops out upon performing the homology averaging, as explained in sec. \ref{Homology}. Hence, we reach the same conclusion as obtained before for AdS-Schwarzschild and the charged black hole in the grand ensemble. 

Along the curves $\beta=\beta_{-}$ or $\beta=\beta_{+}$, with $|q| < q_c$, the specific heat at the non-degenerate positive saddle is positive, whereas the specific heat diverges at the doubly degenerate saddle point. Similarly, at the critical case $|q| = q_c$ and $\beta = \beta_c$, $C_q$ diverges at the triply degenerate positive saddle. At these points where saddles merge/specific heat diverges, the structure of Lefschetz thimbles changes discontinuously. 

\section{KSW allowability of Complex Black Holes}
\label{d_KSW}

\begin{figure}[htbp]
\centering
\subfigure[$AdS_{4}$ \label{fig:3d}]{
        \includegraphics[trim={0.4cm 0 1.1cm 0cm}, clip, scale=0.35]{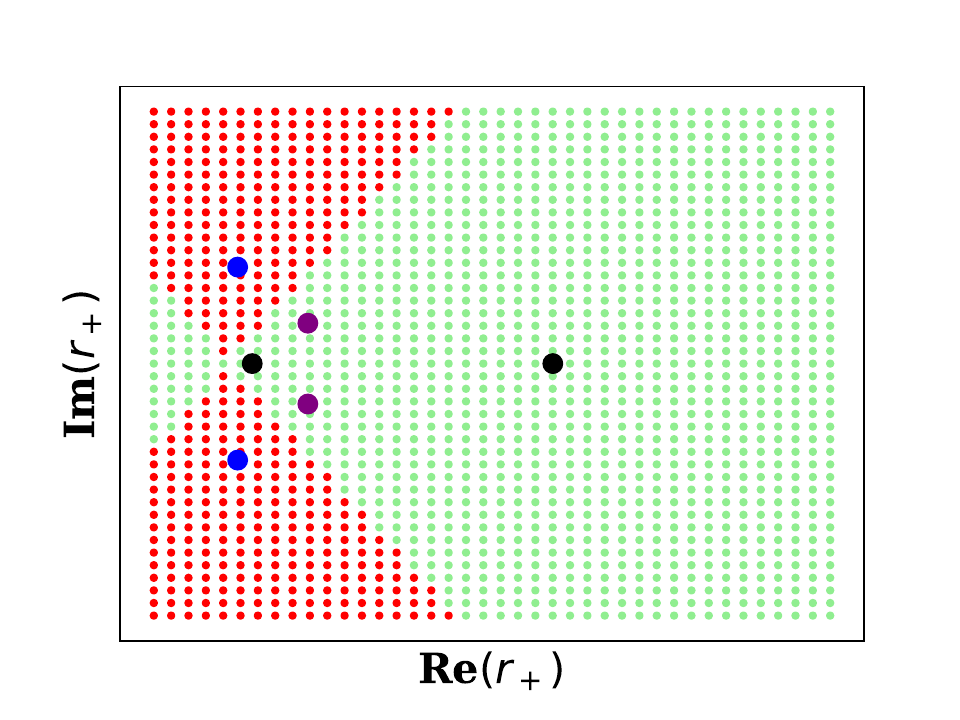}
}
\hspace{-0.5cm}
\subfigure[$AdS_{5}$  \label{fig:4d}]{
    \includegraphics[trim={1.1cm 0 1.1cm 0cm}, clip, scale=0.35]{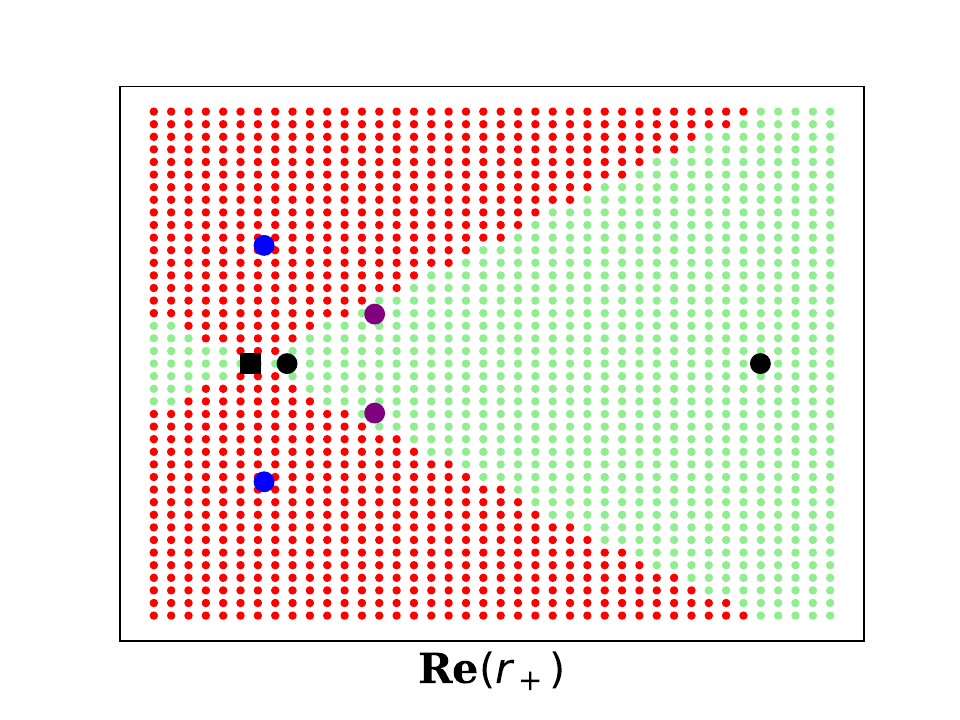}
}
\hspace{-0.5cm}
\subfigure[$AdS_{26}$  \label{fig:25d}]{
        \includegraphics[trim={1.1cm 0 1.1cm 0cm}, clip, scale=0.35]{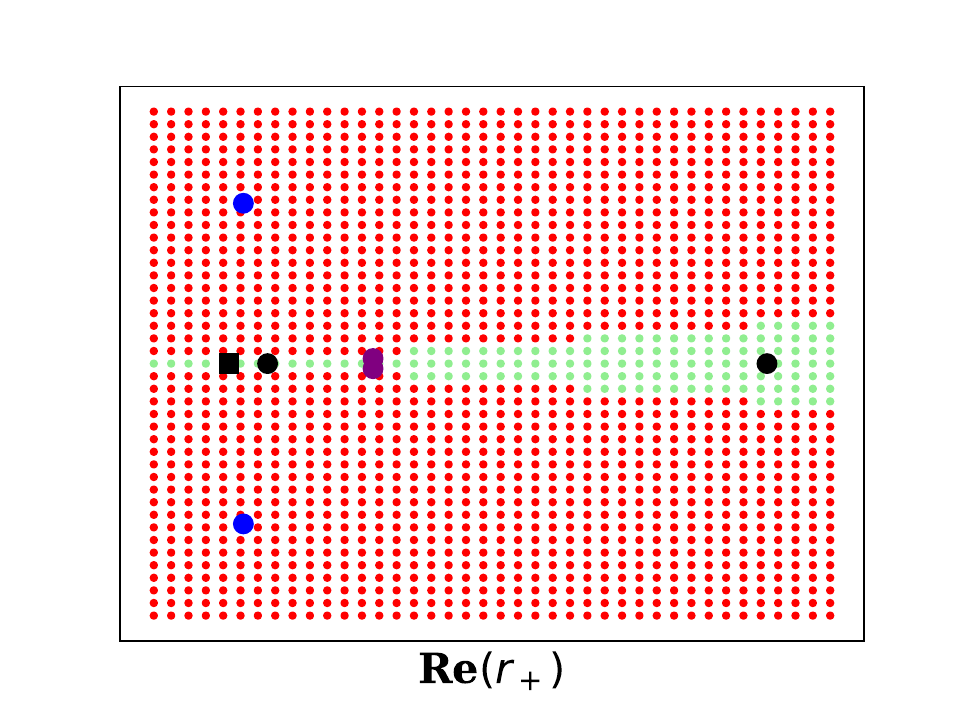}
}
\caption{KSW plots for different asymptotically AdS-Schwarzschild spacetime. The green dots represent the weak KSW allowable points in the $r_{+}$ plane, while red denotes the KSW disallowed complex geometries. The black square and black dots denote the thermal AdS and positive real black holes ($\bt=\bt_{\rm max}/2$), respectively, that weak KSW allowed. The purple dots are the weak KSW allowed complex naked singular geometries at $\beta$ slightly above $\beta_{\rm max}$, while blue dots represent the KSW disallowed complex saddles ($\bt=10\bt_{\rm max}$).}
\label{fig:KSW_all}
\end{figure}

\begin{figure}[htbp]
\centering
\hspace{-0.5cm}
\subfigure[ $\Phi < \Phi_{\rm max}$ \label{fig:phi_ls_phim}]{
    \includegraphics[trim={1.1cm 0 1.1cm 0cm}, clip, scale=0.35]{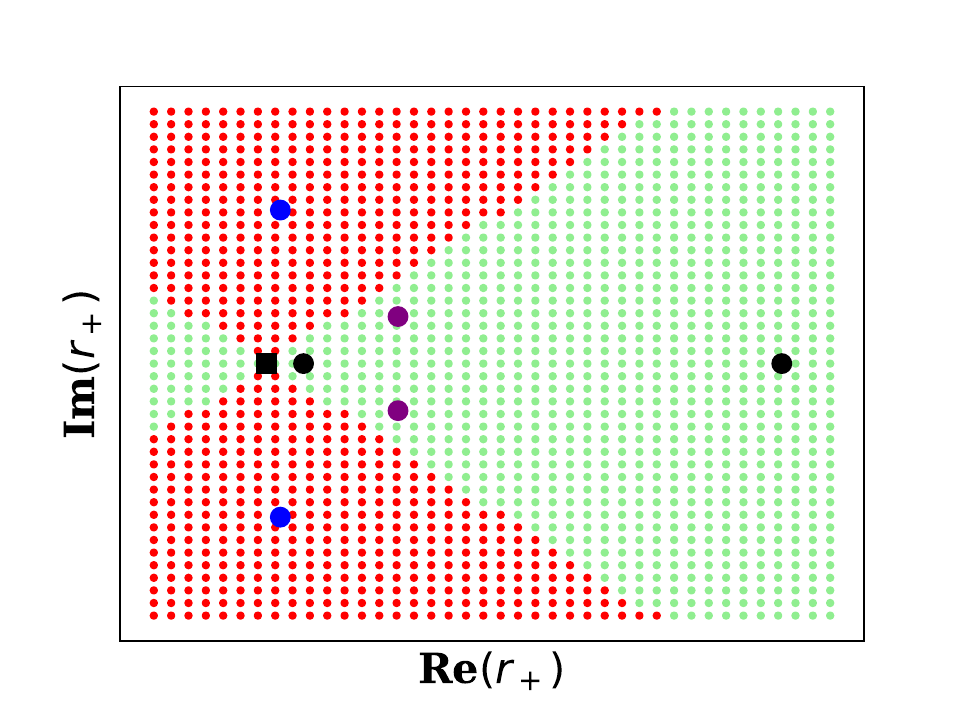}
}
\hspace{-0.5cm}
\subfigure[$\Phi = \Phi_{\rm max}$ \label{fig:fig:phi_eq_phim}]{
    \includegraphics[trim={1.3cm 0 1.3cm 0cm}, clip,scale=0.35]{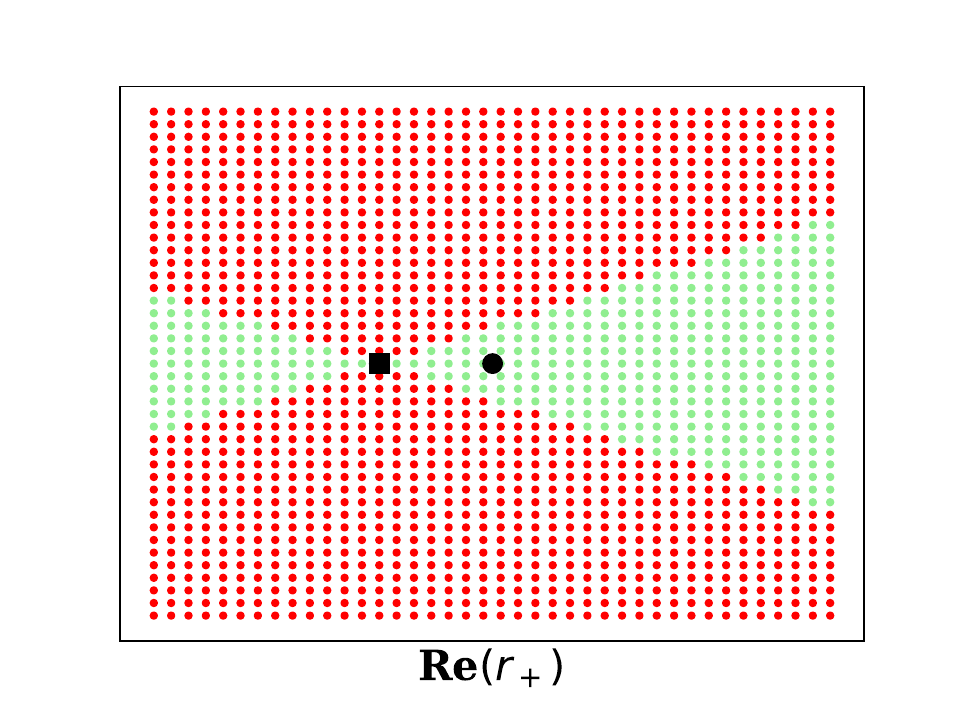}
}
\hspace{-0.5cm}
\subfigure[$\Phi > \Phi_{\rm max}$ \label{fig:fig:phi_gr_phim}]{
    \includegraphics[trim={1.3cm 0 1.3cm 0cm}, clip,scale=0.35]{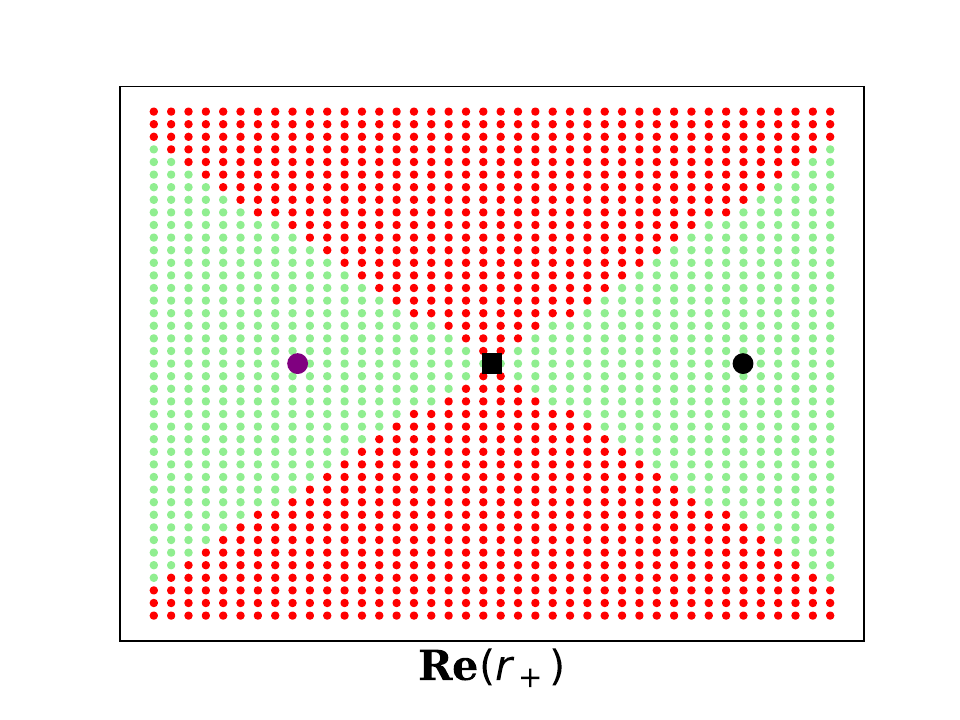}
}
\caption{A similar plot to fig.(\ref{fig:KSW_all}) corresponding to various regions of grand canonical AdS$_5$ case. The purple saddle in (c) corresponds to negative naked singular geometry. Following values of $\Phi=a\Phi_{\rm max}$ are utilized in plotting: a) $a=1/2$, b) $a=1$ and c) $a=2$.}
\label{fig:KSW_grand_d_4}
\end{figure}

\begin{figure}[htbp]
\centering
\hspace{-0.5cm}
\subfigure[ $|q|>q_c$ \label{fig:q_gr_qc}]{
    \includegraphics[trim={1.1cm 0 1.1cm 0cm}, clip, scale=0.35]{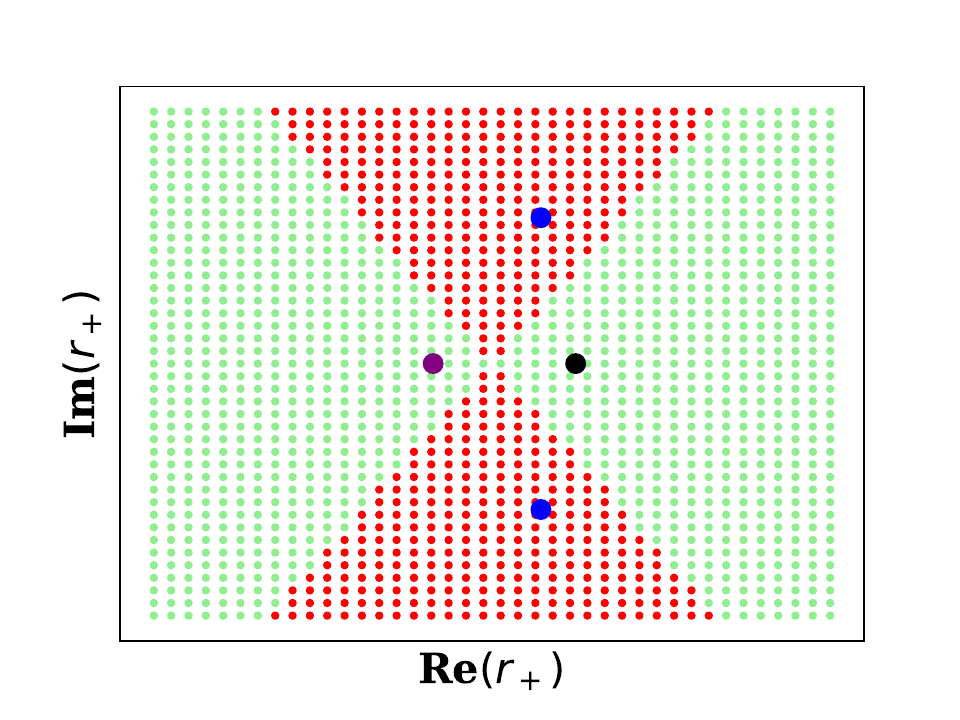}
}
\hspace{-0.5cm}
\subfigure[ $|q|=q_c$ and $\bt = \bt_c$\label{fig:q_eq_qc}]{
    \includegraphics[trim={1.3cm 0 1.3cm 0cm}, clip,scale=0.35]{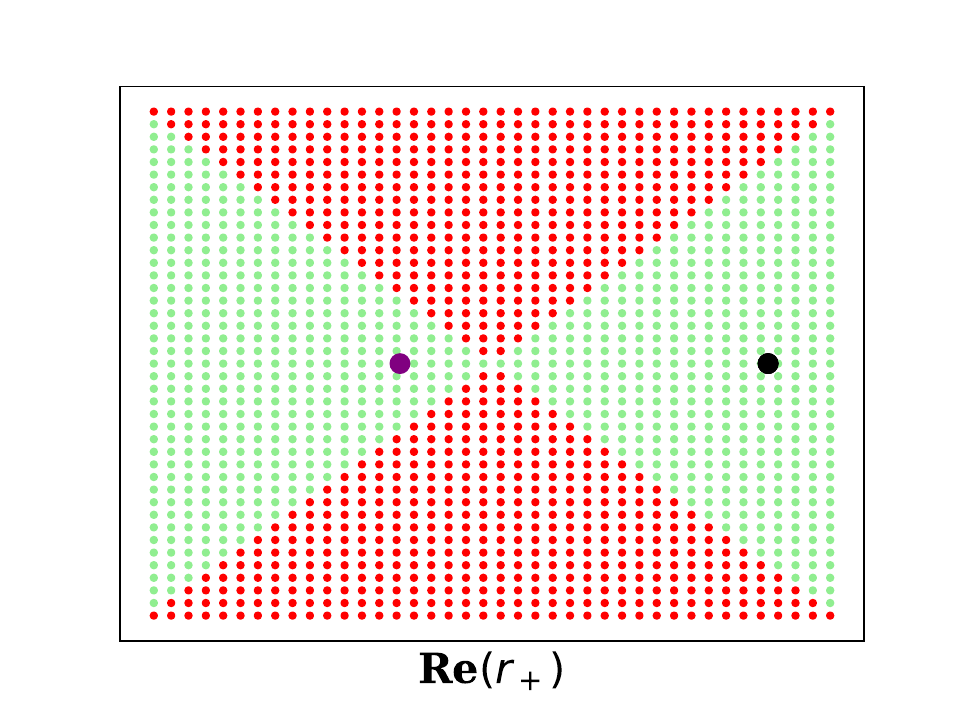}
}
\hspace{-0.5cm}
\subfigure[ $|q|<q_c$ and $\beta_{-}< \beta < \beta_{+}$ \label{fig:q_ls_qc}]{
    \includegraphics[trim={1.3cm 0 1.3cm 0cm}, clip,scale=0.35]{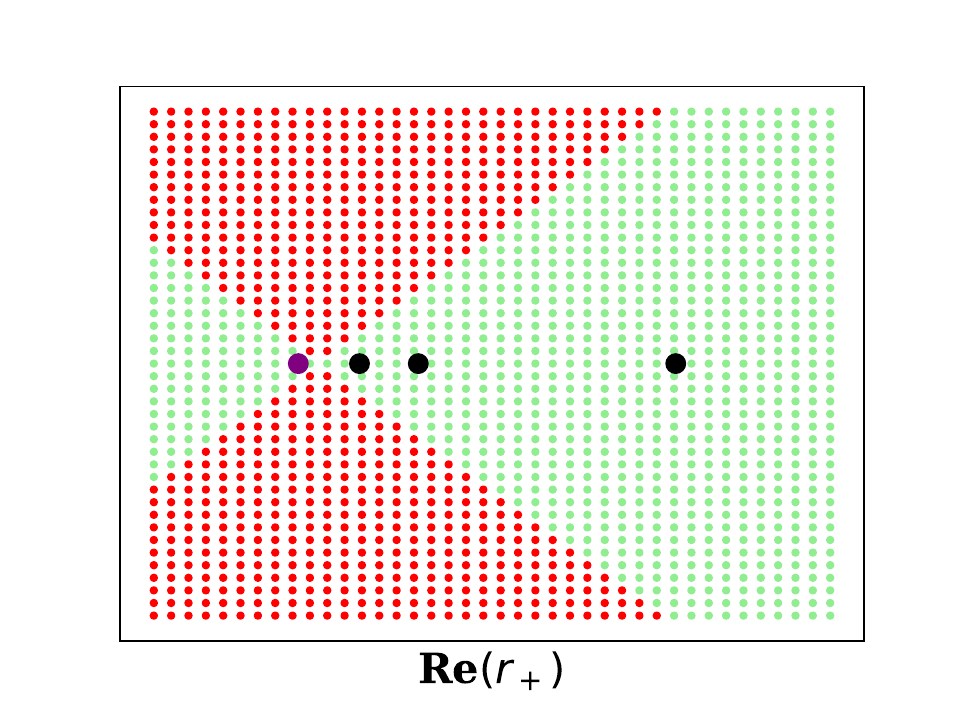}
}
\caption{A similar plot to figs.(\ref{fig:KSW_all}) and (\ref{fig:KSW_grand_d_4}) corresponding to various regions of canonical AdS$_4$ case. The purple saddle in all plots correspond to negative naked singular geometry. Following values of $q=a q_c$ and $\beta = b \beta_c$ are utilized in plotting: a) $a=2,b=2$, b) $a=1,b=1 $ and c) $a=1/2, b=1/2$. }
\label{fig:KSW_can_d_3}
\end{figure}

In this section, we study the KSW (Louko–Sorkin–Kontsevich–Segal–Witten) allowability of generic complex geometry/saddles for the asymptotically AdS spacetimes. We are interested in seeing whether this could help in resolving the puzzle pointed out in the earlier part of the paper. The KSW is a diagnostic tool to identify ``allowable" complex spacetime metrics on which sensible QFTs can be defined \cite{Witten:2021nzp, Louko:1995jw, Kontsevich:2021dmb}. While this seems like  a simple requirement, in practice, the KSW  puts strong constraints on the metrics which are physically ``allowable" \cite{Jonas:2022uqb, Lehners:2023pcn, Hertog:2023vot, Hertog:2024nbh, Liu:2024zti}. In the following, we examine the compatibility of the complex saddles contributing to the partition function with the KSW criterion. The line-element for the Charged-AdS$_{d+1}$ spacetime is given by:
\beq
\label{AdS_line}
ds^{2} = \frac{dr^2}{f(r)} + f(r) d\tau^2 + r^2 d\Omega_{d-1} \, , \quad \text{with}\,\,\, r_{+} \leq r < \infty \, ,
\eeq
with $f(r)$ given the relation (\ref{eq:f(r)_def_char}). 
Let us consider a generic complex curve $r(u)$ in complex $r$ plane which  monotonically increases along real axis, parameterized by a real parameter $u$ such that $r(0) = r_{+}$ and $r(1) = \infty$. In terms of the $u-$ variable, the line element reads as:
\beq
\label{AdS_line_u}
ds^{2} = \frac{r'(u)^2}{f(r(u))} du^2 + f(r(u)) d\tau^2 + r(u)^2 d\Omega_{d-1} \, , \quad \text{with}\, 0 \leq u \leq 1 \, .
\eeq
Therefore, a charged AdS black hole with horizon radius $r_{+}$ is said to be KSW allowable if the components of its line element satisfy \footnote{This condition remains valid as long as there is no Stokes degeneracy in the system. However, in the presence of such degeneracy, one must complexify the Newton constant to lift it. As a consequence, the KSW condition receives a modification that depends on the complex rotation angle $\theta$ of the Newton constant, leading to the modified criterion \cite{Ailiga:2025fny}: $\Sigma (u) < \pi -2\theta$. Nevertheless, for any choice of $\theta$, the real line is always allowed, hence the real/Euclidean saddles.
}
\beq
\label{KSW_con}
\Sigma (u) = \left|\arg\left( \frac{r'(u)^2}{f(r(u))} \right)\right|  + \left|\arg(f(r(u)) \right| +  (d-1) \left| \arg(r(u)^2) \right| < \pi \, \quad \text{for all \, } u \in [0,1].
\eeq
Since the above criterion is pointwise, it must be satisfied at every point along the contour parameterized by $u$. In principle, to determine whether a given saddle is KSW allowable, one should evaluate the extremal curve test as described in \cite{Hertog:2023vot, Ailiga:2025fny}. However, carrying out such an extremal curve test is technically challenging. Nevertheless, one can still obtain conclusive information regarding the disallowable black hole geometries by employing the weak KSW criterion, which corresponds to the non-radial part of the full KSW condition—namely, it involves only the angular and temporal sectors of the metric while neglecting the radial component. The weak KSW criterion is given by \footnote{ As a consequence of the absolute value in the criterion, if a complex saddle $r_{+}$ is allowed (or disallowed), then its complex conjugate $r_{+}^{*}$ is also allowed (or disallowed).}
\beq
\label{weak_KSW_con}
\Sigma_{\text{weak}} (u) =  \left|\arg(f(r(u)) \right| + (d-1) \left| \arg(r(u)^2) \right| < \pi \, \quad \text{for all \, } u \in [0,1].
\eeq
Moreover, for a spacetime, it is necessary but not sufficient to satisfy the weak KSW for it to be an allowable. As for any generic complex curve, since the boundary condition is fixed, if a spacetime violates the weak KSW condition at $u=0$ i.e $r=r_{+}$, one may therefore conclusively regard that geometry as disallowed. While the weak KSW condition does not guarantee the allowability of a geometry, it does offer partial information about geometries that are disallowed. Hence, a generic spacetime with a complex horizon $r_{+}$ is KSW disallowed if 
\beq
\label{weak_KSW_dis_con}
\Sigma_{\text{weak}} (0) =  (d-1) \left| \arg(r_{+}^2) \right| \geq \pi \, ,
\eeq
that is, 
\beq
\label{KSW_dis}
 \left| \arg(r_{+}^2) \right| \geq \frac{\pi}{(d-1)} \, .
\eeq
Utlilizing the above condition, the KSW allowability scenario using weak condition is illustrated for Schwarzschild case in figure (\ref{fig:KSW_all}), for grand canonical case in figure (\ref{fig:KSW_grand_d_4}) and canonical case in figure (\ref{fig:KSW_can_d_3}).    The region covered with red dots corresponds to the complex geometries that violate the KSW criterion, while the region covered with green dots corresponds to points that obey the weak KSW criterion. It is worth noting that the disallowed region increases with the number of spacetime dimensions. Such behaviour is also observed in earlier literature \cite{Jonas:2022uqb}.

Let us now consider the large $\beta$ limit of eq. (\ref{weak_KSW_dis_con}) and see if the weak KSW can resolve the puzzle as posed in sec. (\ref{sec:cbh}). In this limit, the $\arg(r_+^2)$ evaluated at the complex saddles is given by (in the case of a charged black hole in the grand canonical ensemble)
 \begin{equation}
     \label{eq:large_beta_KSW}
|\arg((r_+^\pm)^2)|\biggr|_{\beta\gg \beta_{\rm max}}\sim \biggl|\pi -\frac{2\beta_{\rm max}^\Phi}{\beta}-\frac{1}{3}\left(\frac{\beta_{\rm max}^\Phi}{ \beta }\right)^3-\mathcal{O}\left(\frac{\beta_{\rm max}^\Phi}{ \beta }\right)^5\biggr|
 \end{equation}
Since, in the large $\beta$ limit, r.h.s in eq. (\ref{eq:large_beta_KSW}) approaches $\pi$, implying that weak KSW is sufficient to resolve the puzzle raised in sec. (\ref{sec:cbh}). It excludes all the complex black holes in the large $\beta$ limit. Moreover, one finds that these complex black holes become weak KSW allowed only for $1 < \beta/\beta^{\Phi}_{\rm max} < \sec[{\pi/(2d-2)}]$, for $\beta$ slightly above $\beta_{\rm max}^\Phi$. 
 %
%
%

%

\section{Conclusion}
\label{conc}

In this paper, we explore the issues raised recently in the work \cite{Mahajan:2025bzo} where the authors focused their attention on analysing subtleties of the Euclidean gravitational AdS partition function. They noticed that such partition function in the saddle-point approximation can be reduced to one-dimensional integrals which can be studied using complex analysis and Picard-Lefschetz theory. An interesting puzzle that was seen is the occurrence of complex saddles, which can become dominant in the partition function and are naively expected to govern its behaviour. However, if that is the case, then this would be in tension with AdS/CFT, which is a serious concern. The authors in \cite{Mahajan:2025bzo} then used Picard-Lefschetz methods to realize that such complex saddles which although become dominant in certain cases and are in conflict with AdS/CFT, actually either become irrelevant or drop out from path-integral under homology-averaging, thereby resolving the conflict. 

In this paper, we explore these features, by considering the Euclidean partition function for the case of ${\rm AdS}_{d+1}$-Schwarzschild and ${\rm AdS}_{d+1}$-charged spherically symmetric geometries. As Euclidean gravitational path-integral suffers from the conformal factor problem, following the ideas proposed in \cite{Marolf:2022ybi}, Lorentzian gravitational path-integral is utilized to define a Euclidean one via an integral transform, which has been seen to reproduce the results of standard Euclidean-gravitational path-integral at the saddle-points. The thermal partition function is then a finite-dimensional integral, which becomes our starting point. These one-or two-dimensional integral is then applied to ${\rm AdS}_{d+1}$-Schwarzschild and charged ${\rm AdS}_{d+1}$ geometries, respectively, where they are analysed using Picard-Lefschetz methods. 

We first consider the case of ${\rm AdS}_{d+1}$-Schwarzschild geometries, which follows from Einstein's equation. The thermal partition function $Z(\bt)$ is computed by utilizing the expression of Entropy and Energy as a function of $r_+$, which is the horizon's radius. The one-dimensional integral mentioned in eq. (\ref{eq:EPI_grav}) over entropy reduces to one-dimensional integral over $r_+$ (modulo Jacobians). This one-dimensional integral is studied in the saddle-point approximation. For the case of ${\rm AdS}_{d+1}$-Schwarzschild, there are three saddles: one appearing at $r_+=0$ (with degree $d-3$), the other two follow from quadratic equation and are denoted by $r_+^\pm$. The Kretschmann scalar shows that all the saddle geometries associated with non-zero $r_{+}^{\pm}$ are singular at $r=0$. It is seen that for all $\bt$, the saddle with $r_+=0$ corresponds to thermal-AdS. For $\bt>\bt_{\rm max}$ (low-temperature), the two saddles $r_+^\pm$ form a complex-conjugate pair and correspond to naked-singular geometries. For $\bt<\bt_{\rm max}$ (high-temperature), $r_+^\pm$ are real saddles with $r_+^+$ corresponding to larger black hole geometry and $r_+^-$ corresponding to smaller black hole. 

We then ask the question about the contribution of such geometries in the gravitational partition function. A naive expectation would be that all saddles will contribute with the dominance coming from those saddles with the hightest $\rm Re({\cal I}(r_+))$. If this is the case then one gets into conflict with AdS/CFT where it well known that thermal-AdS reproduces results consistent with the dual CFT in the low temperature regime \cite{Witten:1998zw, Witten:1998qj}. This puzzle was tackled in a recent work for ${\rm AdS}_5$ \cite{Mahajan:2025bzo}, where the authors realized that Picard-Lefschetz method when utilized in evaluating the one-dimensional integral stated in eq. (\ref{eq:EPI_grav}), helps in picking only the ``relevant saddles'' which might not be the dominant saddles. For the case of ${\rm AdS}_{d+1}$-Schwarzschild, complex saddles corresponding to naked singular geometries which become dominant for certain values of $\bt>\bt_{\rm max}$ (see fig. \ref{fig:dominance_check}) are {\it irrelevant} in the partition function. Small black holes which are although {\it relevant} but sub-dominating for $\bt<\bt_{\rm max}$ drops out of the path-integral after homology averaging. These sub-dominant small-sized black holes also have negative specific heat. It is seen that methods of Picard-Lefschetz systematically picks the {\it relevant} saddles by deforming the original integration contour to pass along the Lefschetz thimbles of the {\it relevant} saddles. 

We also showed analytically that complex saddles corresponding to naked singularities for ${\rm AdS}_{d+1}$-Schwarzschild are always {\it irrelevant}. This is easy to see by noting that the $\rm Im({\cal I}(r_+))$ is conserved along the steepest ascent and descent contours. As $\rm Im({\cal I}(r_+))$ at these saddles is always non-zero, steepest ascent contours from these saddles will never intersect the original integration contour lying along the positive real line, where the $\rm Im({\cal I}(r_+))$ is zero. This analytically proves that naked-singularities, which are solutions to Einstein's equation and can be dominating in the partition function for certain values of $\bt>\bt_{\rm max}$, are not {\it relevant} saddles. 

Small-sized black hole saddles for the case of ${\rm AdS}_{d+1}$-Schwarzschild are always {\it relevant}, as the $\rm Im({\cal I}(r_+))=0$. However, making an unambiguous decision regarding their relevance is not easy, as the flow lines from all the saddles overlap with each other. This is due to a symmetry mentioned in eq. (\ref{eq:real_symmetry}) present in the system. Only after the symmetry is broken and degeneracy is lifted, the {\it relevance} of saddles can be determined. Complex deformation of Newton's constant $G=|G|e^{i\ep}$ is one way to lift the degeneracies, a process first introduced in \cite{Ailiga:2025fny}. Once the degeneracies are lifted, it is observed that the steepest ascent contours from all the saddles intersect the original integration contour, and hence, they all contribute to the partition function. The process of lifting degeneracies via complex deformation of $G$, however, shows that the orientation of thimbles in the complex plane changes depending on the sign of $\ep$. The descent thimble emanating from Small-BH saddle undergoes homology jumps as the sign of $\ep$ flips. The contribution from these sub-dominant saddles having negative specific heat drops out from the partition function in homology-averaging.

We then studied the case of charged ${\rm AdS}_{d+1}$ black holes, addressing the same issues investigated for ${\rm AdS}_{d+1}$-Schwarzschild BHs. We first focus on the grand canonical case. Here, we observe that complex saddles and saddles with negative real $r_+$ correspond to naked singularities. As before, they are dominating 
in regimes where $\beta > \beta_{\rm max}$ and $\Phi<\Phi_{\rm max}$ (regime of complex saddle), and for $\Phi > \Phi_{\rm max}$ (regime of negative real saddle). As before, if the partition function is computed naively at low temperature, by taking into account the contributions from all saddles, we will run into conflict with the AdS/CFT results. However, when the situation is analysed using Picard-Lefschetz, then it is seen that these saddles corresponding to naked singular geometries don't contribute to the partition function as they are {\it irrelevant} (which is the case for negative/complex saddles). For low temperature, the {\it relevant} saddles are either thermal-AdS or real positive BH, which are the only ones contributing to the partition function thereby resolving the conflicts. 

We then discuss the high-temperature phase for the charged AdS in a similar way. Here, we also notice that in the various regimes of potential ($\Phi$), the saddles are real (there are no complex saddles in this regime). These real saddles as before have $\rm Im({\cal I}(r_+))=0$, thereby implying the steepest ascent from them intersect the integration contour. However, as all the saddles satisfy this criterion, there is overlapping of flow-lines making it impossible to decide upon the {\it relevance} of each saddle unambiguously. Here again we need to employ complex deformation of Newton's constant $G=|G|e^{i\ep}$ in order to lift the degeneracy. All the positive real saddles are relevant once the degeneracy is resolved. Depending on the sign of $\ep$, the thimbles structure in the complex plane undergo a homology jump. The sub-dominating saddle at $r_+^-$ having negative specific heat drops out of the partition function on homology averaging despite being a {\it relevant} saddle. 

Moving on to the canonical ensemble for the charged AdS, we notice a similar pattern. However, due to algebraic limitations, we could only do a study of ${\rm AdS}_{4,\, 5, \, 6}$. In ${\rm AdS}_4$ case, we analyse for arbitrary temperature (see table \ref{tab:can_d_3_nature_of_horizon} for various regimes), but for ${\rm AdS}_{5, \, 6}$ only low-temperature analysis was possible analytically. However, one can argue that the same will hold for arbitrary dimensions. Here again we realize that saddles corresponding to naked-singular geometries (which are either complex saddles or saddles with $r_+<0$) don't contribute to the partition function despite being dominant saddles in low-temperature regime. They are {\it irrelevant} saddles according to PL method. For high temperature, sub-dominant saddles (intermediate black holes) are {\it relevant}. However, they drop out of the partition function under homology averaging. Only thermal-AdS and/or saddles with positive specific heat contribute to the partition function. 

In this paper, we studied the Euclidean partition function for various ${\rm AdS}$ geometries in arbitrary dimensions and noticed a pattern. Naked singular geometries (which correspond to either complex saddles or saddles with negative $r_+$) despite being solution to Einstein's equation and could be dominant saddles in certain regimes of parameter space, drop out of the partition function. Based on Picard-Lefschetz analysis it is seen that they are {\it irrelevant}. Sub-dominant saddles with positive $r_+$ come with negative specific heat. Although being {\it relevant} in the Picard-Lefschetz analysis, they drop out under homology averaging thereby not contributing to the partition function. Only thermal-AdS saddle and stable-BH are {\it relevant} and contribute to the partition function for the cases investigated in this paper. 

We also investigate how these saddles fare in light of the KSW allowability criterion. We make use of a weaker version of KSW to study saddles allowability. It is seen that complex saddles which correspond to naked singularities are always disallowed for large enough $\beta$. Meanwhile, for $\beta$ slightly above $\beta_{\rm max}^{\phi}$, weak KSW allows these complex saddles. We believe a more sophisticated version of KSW will make all such complex saddles corresponding to naked singularities disallowed. Naked singularities corresponding to saddles with negative real $r_+<0$ are KSW allowed. Sub-dominant small sized black holes with $r_+>0$ are also KSW allowed, despite dropping out from the partition function. Thermal-AdS and large-sized black holes are KSW allowed. We believe that this allowability of real saddles corresponding to naked-singularity with negative $r_+<0$ and small-sized BHs with $r_+>0$ is an anomaly and suggesting that KSW criterion is insufficient to rule out such ``unphysical'' geometries. Such situations have been witnessed earlier \cite{Witten:2021nzp}, where it has been noted that there are geometries which are unphysical but are allowed by KSW criterion. Perhaps an upgraded version of the KSW criterion is needed to rule out such configurations. 

An interesting future direction will be to extend the analysis to the complex chemical potential $(\Phi)$ case, which arises in the context of gravitational index computation and plays a central role in counting black-hole microstates \cite{Jones:2025gno}. It would be worthwhile to explore the relevance of such complex solutions utilizing Picard-Lefschetz theory. This extension, however, is expected to be non-trivial, as gravitational action becomes genuinely complex; consequently, the arguments presented in this paper need suitable modifications.

\bigskip
\centerline{\bf Acknowledgments} 

We are thankful to Justin David and Chethan Krishnan for helpful discussions at various stages of the work. We also like to thank Pradipta Pathak for the discussion related to BHs. GN would like to acknowledge the startup support from IISc which allowed purchase of workstation where some of the simulations were performed. 

\appendix

\section{Computation of Kretschmann scalar in $AdS_{d+1}$}
\label{sec:curvature_scalar}

We want to compute the Kretschmann scalar ($K$) defined by
\begin{equation}
    \label{eq:compute_K}
K=g^{\alpha\beta}g^{\mu\nu}g^{\rho\sigma}g^{\gamma\delta}R_{\alpha\mu\rho\gamma} R_{\beta\nu\sigma\delta} \, .
\end{equation}
For the $AdS_{d+1}$ metric given by
\begin{equation}
    \label{eq:ads_metric}
    \begin{split}
   & ds^2=f(r) d\tau^2+\frac{dr^2}{f(r)}+r^2d\Omega^2_{d-1},\hspace{3mm} f(r)=1+\frac{r^2}{l^2}-\frac{\mu}{r^{d-2}}+\frac{q^2}{r^{2d-4}} \, ,\\
   & d\Omega_{d-1}=d\theta_1^2+\sin\theta_1^2d\theta_2^2+\sin\theta_1^2\sin\theta_2^2d\theta_3^2+\cdots +\prod_{i=1}^{d-2}\sin\theta_{i}^2d\theta_{d-1}.
    \end{split}
\end{equation}
The non-zero (independent) components of $R_{\alpha\beta\mu\nu}$ are
\begin{equation}
\label{eq:non_zero_curvature}
\begin{split}
    &R_{\tau r\tau r}=\frac{f''(r)}{2}\, ,\\
     &R_{\tau\theta \tau\theta}=-R_{r\theta r\theta}=\frac{1}{2}r f(r)f'(r)\times \text{(angular factor for corresponding $\theta$)} \, ,\\
    &R_{\theta_1 \theta_2\theta_1\theta_2}=r^2(1-f(r))\times \text{(angular factor for corresponding $\theta_1$ and $\theta_2$)}\, .
    \end{split}
\end{equation}
Hence, the Kretschmann scalar ($K$) is
\begin{equation}
    \label{eq:K}
    \begin{split}
    K=4\left[(g^{\tau\tau})^2(g^{rr})^2(R_{\tau r\tau r})^2+(g^{\tau\tau})^2(g^{\theta\theta})^2(R_{\tau\theta\tau\theta})^2\right.\\
    \left.+(g^{rr})^2(g^{\theta\theta})^2(R_{r\theta r\theta})^2+(g^{\theta_1\theta_1})^2(g^{\theta_2\theta_2})^2(R_{\theta_1\theta_2\theta_1\theta_2})^2\right]
    \end{split}
\end{equation}
The factor $4$ comes from the symmetry properties of $R_{\alpha\beta\mu\nu}=-R_{\beta\alpha\mu\nu}=-R_{\alpha\beta\nu\mu}=R_{\mu\nu\alpha\beta}$. Substituting all the components, we obtain (all the angular factors cancel)
\begin{equation}
    \label{eq:final_K}
    K=f''(r)^2+\frac{2(d-1)}{r^2}f'(r)^2+\frac{2(d-1)(d-2)}{r^4}(1-f(r))^2.
\end{equation}
Substituting the explicit expression of $f(r)$ in the above equation, we get
\begin{equation}
    \label{eq:K_explicit}
    \begin{split}
        K(r)=\frac{A}{r^{4d-4}}+\frac{B}{r^{3d-2}}+\frac{C}{r^{2d-2}}+\frac{D}{r^{2d}}+E \, ,
    \end{split}
\end{equation}
where the coefficients are given by,
\begin{equation}
    \label{eq:K_explicit_component}
    \begin{split}
       & A=2 \left(8 d^4-52 d^3+127 d^2-139 d+58\right) q^4,\hspace{5mm} B=-4 (d-1)^2 \left(2 d^2-7 d+6\right) \mu  q^2 \, ,\\
         & C=\frac{4 \left(d^2-5 d+6\right) q^2}{l^2},\hspace{4mm} D=(d-2) (d-1)^2 d \mu ^2,\hspace{5mm} E=\frac{2 d (d+1)}{l^4} \, .\\
    \end{split}
\end{equation}
From the above expression, it is clearly evident that $K(r=0)$ is singular for all $d\geq 3$. In contrast, for $d=2$ (i.e., AdS$_3$), where $A = B = C = D = 0$, the curvature is singularity free.

\section{Symmetry of flowlines and Stokes Overlap}
\label{sec:symmetry}

In this appendix, we will briefly discuss the symmetry seen in the flowlines and its relation to the Stokes overlap. We start by noting that $\mathcal{I}(r_+)$ is purely real whenever $r_+$ is real. It implies that when analytically continuing $r_+$ to a complex variable, it satisfies 
\begin{equation}
    \label{eq:real_symmetry}
    (\mathcal{I}(r_+))^*=\mathcal{I}(r_+^*).
\end{equation}
To see what eq. (\ref{eq:real_symmetry}) implies about the structure of flowlines in the complex $r_+$-plane, we go to the polar coordinate and write $r_+=R\exp(i\Theta)$. Substituting it in eq. (\ref{eq:real_symmetry}) and noting that $\mathcal{I}(r_+)=\mathfrak{h}(r_+)+i\mathcal{H}(r_+)$, we get
\begin{equation}
    \label{eq:symmetry}
    \mathfrak{h}(R,-\Theta)=\mathfrak{h}(R,\Theta),\,\,\,\,\mathcal{H}(R,-\Theta)=-\mathcal{H}(R,\Theta).
\end{equation}
To see how eq. (\ref{eq:symmetry}) constraints the flowlines in eq. (\ref{eq:floweq}) in the complex $r_+$-plane, we write 
 the steepest descent($-$)/ascent ($+$) thimbles as
\begin{equation}
    \label{eq:polar_descent_line}
\frac{\partial R}{\partial \lambda}=\mp\frac{\partial \mathfrak{h}(R,\Theta)}{\partial R}
\hspace{3mm}
{\rm and}
\hspace{3mm}
\frac{\partial \Theta}{\partial \lambda}=\mp\frac{1}{R^2}\frac{\partial\mathfrak{h}(R,\Theta)}{\partial \Theta},
\end{equation}
where $\lambda$ is a parameter along the flowline and explicitly, $\mathfrak{h}(R,\Theta)$ and $\mathcal{H}(R,\Theta)$ are given by (considering AdS-Schwarzschild black hole)
\begin{equation} 
    \label{eq:h_theta_R}
    \begin{split}
    \mathfrak{h}(R,\Theta)=\frac{\omega}{|G|}R^{d-2}\left[R\cos\{(d-1)\Theta\}-\frac{\beta(d-1)}{4\pi}\left(\cos\{(d-2)\Theta\}+\frac{ R^2}{l^2}\cos\{d\Theta\}\right)\right]\\
    \mathcal{H}(R,\Theta)=\frac{\omega}{|G|}R^{d-2}\left[R\sin\{(d-1)\Theta\}-\frac{\beta(d-1)}{4\pi}\left(\sin\{(d-2)\Theta\}+\frac{ R^2}{l^2}\sin\{d\Theta\}\right)\right].
    \end{split}
\end{equation}
We use ``modulus" in $G$ to indicate that $G$ is real.
From the eqs. (\ref{eq:polar_descent_line}) and (\ref{eq:h_theta_R}), it is clearly evident that the flow lines are invariant under $\Theta\rightarrow-\Theta$ change, or equivalently $(r_+^x,r_+^y)\rightarrow (r_+^x,-r_+^y)$ transformation
\begin{equation}
    \label{eq:transformation_flow_line}
    \mathcal{J}_\sigma\rightarrow \mathcal{J}_\sigma \,\, \text{and}\,\, \mathcal{K}_\sigma\rightarrow  \mathcal{K}_\sigma, \,\,\text{for,}\,\, \Theta\rightarrow-\Theta.
\end{equation}
The real axis ($\Theta=0$) is the axis of symmetry along which
\begin{equation}
    \label{eq:dtheta_dlambda}
  \left.\mathcal{H}(R,\Theta)\right|_{\Theta=0}=0,\hspace{4mm}  \left.\frac{\partial \Theta}{\partial \lambda}\right|_{\Theta=0}=\left.\mp\frac{1}{R^2}\frac{\partial\mathfrak{h}(R,\Theta)}{\partial \Theta}\right|_{\Theta=0}=0 \,\,\, \forall \,\, R\in[0,\infty) \,\,.
\end{equation}
Now if saddles lie on the real axis, then eq. (\ref{eq:dtheta_dlambda}) immediately implies the existence of a Stokes ray (provided $\mathfrak{h}(R,\Theta=0)$ are different at these saddles, which will always be the case if the saddles are non-degenerate). We encounter a similar situation when $\beta\leq\beta_{\rm max}$. Hence, the Stokes ray lies along the real axis/ axis of symmetry transformation. A similar study was also done in cosmology previously in \cite{Ailiga:2025fny}. As discussed in sec. (\ref{sec:real_saddle}), presence of such Stokes degeneracy poses challenges in implementing the Picard-Lefschetz method, and hence it becomes necessary to resolve them.  Various resolutions were considered in \cite{Ailiga:2025fny} to lift such degeneracy; one of them is to rotate Newton's constant by a small phase $G=|G|e^{i\epsilon},|\ep|\ll 1$\footnote{Resolving Stokes degeneracy is also possible by complexifying $\beta$, see \cite{Maldacena:2019cbz}. When complexifying $\beta$, saddles change; however, complexifying Newton's constant wouldn't affect saddles. Complexifying $\beta$ reminds us of using quantum correction for lifting degeneracy, where saddles also receive modifications \cite{Ailiga:2025fny}.}. With this, the Morse function modifies to
\begin{equation}
    \label{eq:modified_morse}
    \mathfrak{h}_\epsilon=\mathfrak{h}+\ep \mathcal{H}.
\end{equation}
Now, the modified steepest descent/ascent equation in the presence of the complexified Newton's constant becomes
\begin{equation}
    \label{eq:polar_descent_line_modified}
\frac{\partial R}{\partial \lambda}=-\frac{\partial \mathfrak{h}_\epsilon(R,\Theta)}{\partial R}
\hspace{3mm}
{\rm and}
\hspace{3mm}
\frac{\partial \Theta}{\partial \lambda}=-\frac{1}{R^2}\frac{\partial\mathfrak{h}_\epsilon(R,\Theta)}{\partial \Theta},
\end{equation}
which is clearly not invariant under $\Theta\rightarrow-\Theta$, as 
\begin{equation}
    \label{eq:h_epsilon}
    \mathfrak{h}_\epsilon(R,-\Theta)=\mathfrak{h}(R,\Theta)-\epsilon\mathcal{H}(R,\Theta)\neq \mathfrak{h}_\epsilon(R,\Theta).
\end{equation}
Hence we conclude
\begin{equation}
    \label{eq:transformation_flow_line_epsilon}
    \mathcal{J}_\sigma^\epsilon \rightarrow \widetilde{\mathcal{J}_\sigma^\epsilon} \quad \text{and} \quad \mathcal{K}_\sigma^\epsilon \rightarrow \widetilde{\mathcal{K}_\sigma^\epsilon}, \quad \text{for} \quad \Theta \rightarrow -\Theta,
\end{equation}
where the tilde over $\mathcal{J}_\sigma$ and $\mathcal{K}_\sigma$ indicates that the thimbles are different after the transformation $\Theta\rightarrow-\Theta$, unlike eq. (\ref{eq:transformation_flow_line}) and superscript $\epsilon$ indicates the flow-lines are computed with complexified $G$. Now suppose we have a Stokes ray between two saddles $\sigma$ and $\sigma'$; then we will have either $\mathcal{J}_\sigma=\mathcal{K}_{\sigma'}$ or, $\mathcal{J}_{\sigma'}=\mathcal{K}_{\sigma}$. However, in the presence of a non-zero phase ($\epsilon$),
\begin{equation}
    \label{eq:eq:breaking_stokes}
\text{$\mathcal{J}_\sigma^\epsilon\neq\mathcal{K}_{\sigma'}^\epsilon$\hspace{4mm} or, \hspace{4mm} $\mathcal{J}_{\sigma'}^\epsilon\neq\mathcal{K}_{\sigma}^\epsilon$},
\end{equation} 
and hence, it will resolve the Stokes ray. To see it
consider the particular flow equation, as mentioned in eq. (\ref{eq:polar_descent_line_modified})
\begin{equation}
    \label{eq:polar_descent_line_modified_1}
\frac{\partial \Theta}{\partial \lambda}=\mp\frac{1}{R^2}\frac{\partial\mathfrak{h}_\epsilon(R,\Theta)}{\partial \Theta}=\mp\frac{1}{R^2}\left(\frac{\partial\mathfrak{h}(R,\Theta)}{\partial \Theta}+\epsilon\frac{\partial\mathcal{H}(R,\Theta)}{\partial \Theta}\right).
\end{equation}
Evaluating at $\Theta=0$ (direction of Stokes ray), we get
\begin{equation}
    \label{eq:flow_eqn_ep}
    \left.\frac{\partial \Theta}{\partial \lambda}\right|_{\Theta=0}=\mp\frac{1}{R^2}\left(\left.\frac{\partial\mathfrak{h}(R,\Theta)}{\partial \Theta}\right|_{\Theta=0}+\left.\epsilon\frac{\partial\mathcal{H}(R,\Theta)}{\partial \Theta}\right|_{\Theta=0}\right)=\epsilon f(R)\neq 0,
\end{equation}
where $f(R)$ is given by,
\begin{equation}
    \label{eq:f(R)}
    \begin{split}
        f(R)=\mp\frac{\omega(d-1)}{G}R^{d-4}\left[R-\frac{\beta}{4\pi}\left((d-2)+\frac{ R^2d}{l^2}\right)\right]\neq 0\,\,\,\forall \,R\,\notin \text{saddles}.
    \end{split}
\end{equation}
In the presence of non-zero $\epsilon$, $\frac{\partial \Theta}{\partial \lambda}$ is non-zero at $\Theta=0$, implying lifting the Stokes ray. The quantity $\frac{\partial \Theta}{\partial \lambda}|_{\Theta=0}$ is interpreted as follows: as we move along the steepest descent/ascent thimbles $\lambda$ changes monotonically, now if the change in $\Theta$ (evaluated at $\Theta=0$) is non-zero, it will imply the thimble deviates and no longer lie along the real axis. Hence, it will necessarily break the Stokes ray. Moreover, since the deviation is proportional to $\epsilon$; changing $\epsilon$ to $-\epsilon$ amounts to changing the direction of $\Theta$ at $\Theta=0$.

Hence, lifting the Stokes ray (which lies along the real axis and hence, is invariant under $\Theta \rightarrow -\Theta$ transformation) is intimately tied to the breaking of the $\Theta\rightarrow-\Theta$ symmetry of the system, \cite{Ailiga:2025fny}. Also note that with the complexified $G$, one satisfies
\begin{equation}
    \label{eq:homo_average}
    \mathfrak{h}_{-\ep}(R,\Theta)=\mathfrak{h}_\ep(R,-\Theta) \quad \text{and},\quad \mathcal{H}_\ep(R,-\Theta)=-\mathcal{H}_{-\ep}(R,\theta),
\end{equation}
where $\mathcal{H}(R,\Theta)=\mathcal{H}(R,\Theta)-\ep\mathfrak{h} (R,\Theta)$ is the phase. Eq. (\ref{eq:homo_average}) ensures that whenever we add contours with $\ep>0$ and $\ep<0$ (homology averaging), one always gets a real answer. In the limit $\ep\rightarrow 0$, Eq. (\ref{eq:homo_average}) reduces to eq. (\ref{eq:symmetry}) from either side.

Let us also mention that complexifying $\beta$ \cite{Maldacena:2019cbz}, even complexifying $l^2$, one can resolve the Stokes degeneracy, which also breaks $\Theta \rightarrow-\Theta$ symmetry. The above discussion extends straightforwardly in the case of charged back holes for both grand canonical and canonical ensemble where the Euclidean action is also real.



\end{document}